\documentclass[twocolumn]{emulateapj}

\usepackage{natbib}
\usepackage{float}
\restylefloat{table}
\usepackage{graphicx}
\usepackage{fmtcount}
\usepackage{verbatim}

\usepackage{epstopdf}
\usepackage{epsfig}

\newcommand{\NumOutliers}{4000} 

\newcommand{\NumMachoLC}{20 million}

\begin{document}
\bibliographystyle{unsrt}

\title{SUPERVISED  DETECTION OF ANOMALOUS LIGHT-CURVES  IN MASSIVE ASTRONOMICAL CATALOGS}
\author{Isadora Nun$^{1}$, Karim Pichara$^{1,2}$, Pavlos Protopapas$^{3,4}$ and Dae-Won Kim$^{5}$ }
\affil{\altaffilmark{1}Computer Science Department, Pontificia Universidad Cat\'olica de Chile, Santiago, Chile}
\affil{\altaffilmark{2}The Millennium Institute of Astrophysics}
\affil{\altaffilmark{3}Institute for Applied Computational Science, Harvard University, Cambridge, MA, USA}
\affil{\altaffilmark{4}Harvard-Smithsonian Center for Astrophysics, Cambridge, MA, USA}
\affil{\altaffilmark{5}Max-Planck Institute for Astronomy, K\"{o}nigstuhl 17, D-69117 Heidelberg, Germany}

\begin{abstract}

The development of synoptic sky surveys has led to a massive amount of data for which resources needed for analysis are beyond human capabilities. In order to process this information and to extract all  possible knowledge, machine learning techniques become necessary. Here we present a new methodology to automatically discover unknown variable objects in large astronomical catalogs. With the aim of taking full advantage of all  information we have about known objects, our method is based on a supervised algorithm. In particular, we train a random forest classifier using known variability classes of objects and  obtain votes for each of the objects in the training set. We then model this voting distribution with a Bayesian network and obtain the joint voting distribution among the training objects. Consequently, an unknown object is considered as an outlier insofar it has a low joint probability. By leaving out one of the classes on the training set we perform a validity test and show that when the random forest classifier attempts to classify unknown light-curves (the class left out), it votes with an unusual distribution among the classes. This rare voting is detected by the Bayesian network and expressed as a low joint probability. 

Our method is suitable for exploring massive datasets given that the training process is performed offline. 
We tested our algorithm on \NumMachoLC{} light-curves from the MACHO catalog and generated a list of anomalous candidates. After analysis, we divided the candidates into two main classes of outliers: artifacts and intrinsic outliers. Artifacts were principally due to air mass variation, seasonal variation, bad calibration or instrumental errors and were consequently removed from our outlier list and added to the training set. After retraining, we selected about \NumOutliers{} objects, which we passed to a post analysis stage by perfoming a cross-match with all publicly available catalogs. Within these candidates we identified certain known but rare objects such as eclipsing Cepheids, blue variables, cataclysmic variables and X-ray sources. For some outliers there were no additional  information. Among them we identified three unknown variability types and few individual outliers that will be followed up  in order to do a deeper analysis.
 

\end{abstract}


\section{Introduction}

Several important discoveries in astronomy  have happened serendipitously while astronomers were examining other effects. For example, William Herschel discovered Uranus on March 13 1781\citep{Herschel1857} while surveying bright stars and nearby faint stars. Similarly, Giuseppe Piazzi found the first asteroid, Ceres, on January 1 1801 \citep{Serio2001} while compiling a catalog of stars positions. Equally unexpected, was the discovery of the cosmic microwave background radiation (CMB) in 1965 by Arno Penzias and Robert Wilson, while testing Bell Labs  horn antenna  \citep{Penzias1965}. 

With the proliferation of data in astronomy and the introduction of automatic methods for classification and characterization, the keen astronomer has been progressively removed from the analysis. Anomalous objects or mechanisms that do not fit the norm are now expected to be discovered systematically: serendipity is now a machine learning task. As a consequence, the astronomer\rq s job is not to be behind the telescope anymore,  but to be capable of selecting and making the interpretation of the increasing amount of data that technology is providing. 

\begin{figure*}[]
\centering
\includegraphics[width=7in]{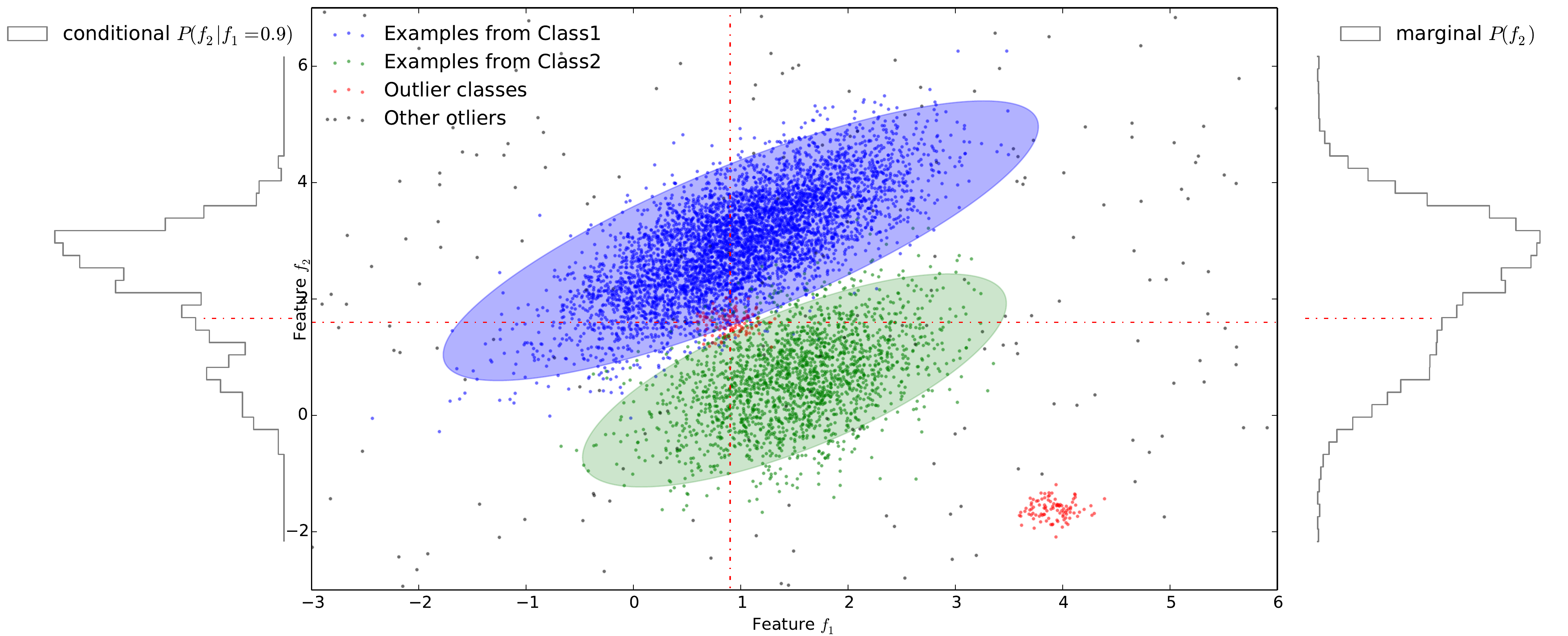}
\label{fig_smiley}
\caption{Simple illustration of the method. In most unsupervised methods the red points in the middle will not be considered as outliers because they are in a region with point density that is not separable. The product of the probabilities or the sum of the distances to the known classes may not be  adequate as an outlier score, and therefore the joint probability is a better measure for outliers. This case occurs when the conditional probability is lower than the marginal probability as it can be seen from this simple illustration.} 
\label{fig:smiley}
\end{figure*}

Outlier detection, as presented here, can guide the scientist on  identifying  unusual, rare or unknown types of astronomical objects or phenomena (e.g. high redshift quasars, brown dwarfs, pulsars and so on). These discoveries might be useful not only to provide new information but to outline observations, which might require further and deeper investigation. In particular, our research detects anomalies in photometric time series data (light-curves). 
For this work, each light-curve is described by 13 variability characteristics (period, amplitude, color, etc.) termed \textit{features} \citep{Kim2011,Pichara2012}, which  have been used  for classification. It is worth noting that the method developed in this paper is not only applicable to time-series data but could also be used for any type of data that need to be inspected for anomalies. In addition to this advantage, the fact that it can be applied to big data, makes this algorithm suitable for almost any outlier detection problem.

Many outlier detection methods have been proposed in astronomy. Most of them are unsupervised techniques, where the assumption is made that there is no information about the set of light-curves or their types
\citep{Connolly2010}. One of these approaches considers a point-by-point comparison of every pair of light-curves in the data base by using  correlation coefficient \citep{Protopapas2008}. Other techniques search for anomalies in lower-dimensional subspaces of the data in order to deal with the massive number of objects or the large quantity of features that describe them \citep{Henrion2012, Connolly2010}. Clustering methods are equally applied in the astronomical outlier detection area aiming to find clusters of new variability classes \citep{Bhattacharyya, Rebbapragada2008}. Unfortunately, these methods either scale poorly with massive data sets and with high dimensional spaces or partially explore  the data therefore missing possible outliers.

In this paper we face these constraints by creating an algorithm able to efficiently deal with big data and capable of exploring the data space as exhaustively as possible. Furthermore, we address this matter from a different point of view as the one presented by \citet{He2006} as \lq\lq the new-class discovery challenge\rq\rq. Contrary to unsupervised methods, it relays on using labeled examples for each known class in the training set, and unlike supervised methods, we assume the existence of some rare classes in the data set for which we do not have any labeled examples. This approach takes advantage of available information but it does not restrict the anomalous findings to a certain type of light-curves.  
 Furthermore, in unsupervised anomaly detection methods, in which no prior information is available about the abnormalities in the data, anything that differs from the whole dataset is flagged as an outlier and consequently many of the anomalies found would simply be noise.
  In contrast to these techniques, supervised methods incorporate specific knowledge into the outlier analysis process, thus obtaining more meaningful anomalies. 
 This is illustrated in Figure.~\ref{fig:smiley}. The blue and green points represent instances in a two dimensional feature space from  known class 1 and class 2 respectively. The shaded areas represent the boundaries learned from a classifier. The grey points represent  isolated outliers and the red points represent outlier classes. In most unsupervised methods the red points in the middle will not be considered as outliers because they are in a region with point density that is not separable. In the naivest supervised methods, anything that is outside the boundaries is considered as an outlier. For the example of the outlier class in the middle, the product of the probabilities or the sum of the distances to the known classes may not be  adequate as an outlier score, and therefore the joint probability is a better measure for outliers. This case occurs when the conditional probability is lower than the marginal probability\footnote{This is not necessary true for all cases} as it can be seen from this simple illustration. The conditional probability shown on the left is smaller than the marginal probability shown on the right. Our model will consider those objects as outliers.
 
In the first stage of our method we build a classifier that is trained with  known classes  (every known object is represented by its features and a label). We then use the classifier decision mechanism to our advantage. 
More precisely, we learn a probability distribution for the classifier votes
on the training set in order to model the behavior of the classifier when the objects correspond to a known variability class. The intuition behind this method is to recognize, and thus to learn, the way the classifier is confused when it comes to voting. By confusion, we refer not only to the hesitation between two or more classes for an object label, but also to the weights it assigns to each of these possibilities.

Therefore, when an unlabeled light-curve is fed into the model, the classifier attempts to label it and, if this classifying behavior is known by the model, the object will have a high probability of occurrence and consequently a low outlierness score. On the contrary, the object will have a higher anomaly score and will be flagged as an outlier candidate insofar as the classifier operates in a different way from the previously known mechanisms. 

Once our outlier candidates are selected, an iterative post-analysis stage becomes necessary. By visual inspection we discriminate artifacts from true anomalies and a) we remove them systematically from our data set and b) create classes of spurious objects that we add to our training set. We then re-run the algorithm and obtain new candidates. These steps are repeated until obtaining no apparent artifacts in our outlier list and a clustering method is finally executed. The purpose of this phase is to group similar objects in new variability classes and consequently to give them an astronomical interpretation. Finally we cross-match the most interesting outliers with all publicly available catalogs with the aim of verifying if there is any additional information about them. In particular we are interested in knowing if they belong to a known class. In the negative case, the outliers will be followed up using spectroscopy to deeply analyze their identity and behavior.

To achieve this, we use random forest (RF) \citep{Breiman2001} for the supervised classification in order to obtain the labeling mechanism for each class on the training set. RF has been extensively and successfully  used in astronomy for catalogation \citep{Pichara2013, Kim2014}. Starting with the RF output, we construct a Bayesian Network (BN) with the purpose of extracting the classifications patterns which we use for our final score of outlier detection.

The paper is organized in the following way: Section \ref{sec:related} is devoted to other methods related to anomaly detections in machine learning and astronomy. In section \ref{sec:background} we detail the background theory including the basic blocks of random forest and Bayesian Networks. Our approach and the pipeline followed in the paper are shown in section \ref{sec:methodology}. Section \ref{sec:data} contains the information about the data used in this work and section \ref{sec:Results} presents the results of the performed tests and the experiments with real data, including re-training and elimination of artifacts. We proceed by explaining in section \ref{sec:post} the post analysis process. Finally, conclusions follow in section \ref{sec:conclusions}.

\section{Related Work}
\label{sec:related}

\subsection{Outlier detection in machine learning}

Vast literature has been published in relation to anomaly/outlier detection problems \citep{Chandola2009, survey2}, but generally they can be classified into  two main classes: supervised and unsupervised methods. 

In unsupervised approaches the examples given to the learner are unlabeled and consequently there is no training set 
in which the data is separated into different classes. In turn, these techniques can be partitioned into three main subcategories: statistical methods, proximity based methods and clustering methods. 

Statistical approaches are the earliest methods used for anomaly detection. These methods detect anomalies as outliers that deviate markedly from the generality of the observations \citep{Grubb1969}, by assuming that a statistical model generates normal data objects and  data that does not follow the model are outliers. In particular, many of these methods use mixture models by applying Gaussian distributions \citep{Agarwal2005, Eskin2000}. The typical strategy considers the calculation of a score and a threshold, both used to identify points that deviate from normal data. For example, \cite{Eskin2000}  proposes an algorithm that fits mixture models, a normal and anomalous, using the Expectation maximization (EM) algorithm and assuming a prior probability $\lambda$ of being anomalous. Then, the author obtains an anomaly score which is based on measuring the variation of the normal distribution when a point is moved to the anomalous distribution. One of the main drawbacks of the statistical approach is that they are generally applied to quantitative data sets or at the very least quantitative ordinal data distributions where the ordinal data can be transformed to suitable numerical values for statistical processing. This limits their range of application and can increase the processing time when complex data transformations are needed as a pre-process. \citep{HodgeAustin2004}.

Clustering-based methods \citep{YangXieLu2006,SonChoYoo2009,Zhang1996} are based on the fact that similar  instances can be grouped into clusters, and that normal data lies on large and dense clusters, while anomalies  belong to small or sparse clusters, or to no cluster at all. Most recent clustering algorithms proposed for anomaly detection are on the context of intrusion detection on networks \citep{YangXieLu2006,SonChoYoo2009}. Unfortunately, clustering algorithms suffer from the curse of dimensionality problem. Often, in large dimensional spaces, distance metrics that are applied to characterize similarity do not provide suitable clusters. Subspace clustering algorithms, a remedy to the dimensionality curse,  have not been commonly used for anomaly detection with exception of some recent works \citep{Seidl2009, Pichara, Pichara2011}. \cite{Seidl2009} perform a subspace clustering algorithm to rank data points according to the size of the clusters and the number of dimensions of each subspace where the points belong.   To identify microclusters containing anomalies, \citet{Pichara}  search for relevant subspaces in subsets of variables that belong to the same factor in a trained BN. Similarly, \citet{Pichara2011} present a semi-supervised algorithm that actively learns to detect anomalies in relevant subsets of dimensions, where dimensions are selected by using a subspace clustering technique that finds dense regions in a sparse multidimensional data set. One of the main drawbacks of these kind of approaches is that they use heuristics to find relevant subspaces and those heauristics may ignore combinations of spaces where anomalies could also lie.

Finally, proximity-based methods follow the intuition that anomalies are records with less neighbors than normal records \citep{Ramaswamy2000,Knorr1998,Breunig2000}. For example, \cite{Breunig2000} assign an anomaly score called Local outlier factor (LOF) to each data instance; this score is given by the ratio between the local density of the point and the average local density of its $k$-nearest neighbors. Local density is calculated using the radius of the smallest hyper-sphere that is centered at the data instance and contains $k$ (nearest) neighbors. \citet{Papadimitriou2002} propose a variant of the LOF called Multi granularity deviation factor (MDEF). For a given record, its MDEF is calculated as the standard deviation among its local density and the local densities of its $k$-nearest neighbors. Then they use the MDEFs to search micro clusters of anomalous records. Along the same lines, \cite{Jin2001} propose another variant of LOF that improves efficiency by avoiding unnecessary calculations. They achieve this by calculating upper and lower bounds among the micro clusters detected. Unfortunately, density-based algorithms usually are quadratic in the number of instances and thus they are not suitable for big data. Furthermore, these methods also suffer from the curse of dimensionality because of the same reasons mentioned above for the clustering methods.

On the other hand, in  supervised approaches, outlier detection can be treated as a classification problem, where a training set with class labels is used to generate a classifier that distinguishes between normal and anomalous data \citep{Gibbons1998,Aggarwal2001,Chandola2009}. Various anomaly detection algorithms have been proposed in this area, such as decision trees \citep{John1995,Arning1996} and neural networks \citep{Nairac1999,Bishop1994}. Decision trees algorithms fit the data focusing only on salient attributes, a desirable characteristic when dealing with high dimensional data. These algorithms work by modeling all points corresponding to normal classes: then points having an erroneous or unexpected classification are considered as anomalies. Similarly, neural networks are employed to model the unknown distribution of normal class points by training a feed forward network. This is achieved by adjusting the weights and thresholds while learning from the input data. Neural networks work well when training sets are representative of the unseen data. Unfortunately, this may not occur for new instances which are out of the scope of the training set. Decision trees and Neural networks are susceptible to overfitting when stopping criteria are not well determined.

\subsection{Outlier detection in astronomy}

Because synoptic sky surveys have significantly increased in the last decade \citep{Keller2007,Hodapp2004, Tyson},  astronomical anomaly detection has not been yet fully implemented in the enormous amount of data that has been gathered. As a matter of fact,  barring a few exceptions, most of the previous studies can be divided into only two different trends: clustering and subspace analysis methods.

In \citet{Rebbapragada2008}, the authors create an algorithm called Periodic curve anomaly detection (PCAD), an unsupervised outlier detection method for sets of unsynchronized periodic time series, by modifying the k-means clustering algorithm. The method samples the data and generates a set of representative light-curves centroids from which the anomaly score is calculated. In order to solve the phasing issue, during each iteration every time series is rephased to its closest centroid before recalculating the new one. The anomaly score is then calculated as the distance of the time series to its closest centroid.
Even if the anomaly detection is satisfactory on a restricted and small data set, the technique scales poorly with massive data sets. This is mainly due to the distinctive high dimensionality problem that clustering methods encounter as mentioned in the previous section. Furthermore, since the algorithm is based on the alignment of the time series periods, it is restricted to periodic light-curves, thus limiting the scope of possible astronomy applications.

Similarly, \citet{Protopapas2008} search for outliers light-curves in catalogs of periodic variable stars. To this end, they use cross-correlation as measure of similarity between two individual light-curves and then classify light-curves with lowest average similarity as outliers. Unfortunately, this method scales as $N_{LC}^2$, where $N_{LC}$ is the number of light-curves. In order to deal with this high operational cost and to apply the algorithm to large data sets they make an approximation they call \textit{universal phasing}. By using clustering they find where the signal with the highest/lowest magnitude dip occurs for each light-curve and set it to a particular phase by time-shifting the folded light-curve. Once they find an absolute phase for all the light-curves, they calculate the correlation of each one with the average of the rest of the set, reducing  the operational cost of the algorithm to $N_{LC}$. Unfortunately, this method is an approximation since it does not guarantee that the correlation between two light-curves is maximum. Furthermore this approximation also implies not taking into account the observational errors, thus losing highly valuable information. Finally, as in \citet{Rebbapragada2008}, this algorithm is also restricted to periodic light-curves.

\citet{Connolly2010} separate astronomy anomalies into two different categories: \textit{point anomalies}, which include individual anomalous objects, such as single stars or galaxies that present unique characteristics and \textit{group anomalies} (anomalous groups of objects) such as unusual clusters of the galaxies that are close together. For that end, they develop one method for each of these cases. For the former case the authors create  Mixed-error matrix factorization (MEMF), an unsupervised algorithm that explores subspaces of the data. They also assume that normal data lie in a low-dimensional subspace and that their features can be reconstructed by linear combination of a few bases features. Quite opposite, anomalies lie outside of that subspace and cannot be well reconstructed by these bases. To do so, they find a robust low-rank factorization of the data matrix and consider the low-rank approximation error to be an additive mixture of the regular Gaussian noise and the outliers that can be measured differently in the model. One limitation of MEMF is that the factorization rank $k$ has to be specified by the user and it is consequently often determined by heuristics.
For group anomalies, the authors use hierarchical probabilistic models to capture the generative mechanism of the data. In particular, they propose Dirichlet genre model (DGM), which assume that the distribution of the groups in the data set can be represented by a Dirichlet distribution. Two anomaly scores are then presented: the likelihood of the whole group and a scoring function that focus on the distribution of objects in the group. One of the main drawbacks of this method is that the inference stage considers a non convex problem and is consequently restricted to the limitation of variational approximations.

\citet{Henrion2012} propose CASOS, an algorithm to detect outliers in datasets obtained by cross-matching astronomical surveys. To do so, they compute an anomaly score for each observation in lower-dimensional subspaces of the data, where subspaces make allusion to subsets of the original data variables. In particular, any anomaly detection method that produces numerical anomaly scores can be used with this approach. The idea is to analyze the anomaly score of each observation in every possible subspace and then combine them in such way that objects with many observed variables and objects with only a few are equally likely to have high anomaly scores. Unfortunately, CASOS has the disadvantage that it will not be able to detect outliers, which are only apparent in multivariate spaces with significant numbers of variables.

Finally, \citet{Richards2012} apply a semi-supervised approach for astronomical outlier detection. Unlike the previous mentioned algorithms, in this work, the authors compute a distance metric from every candidate object to each source in a training set. To do so, they train a random forest classifier with known classes and measure the proximity value $\rho_{ij}$ for all the new instances $i$ to every $j$ object on the training set. The proximity measure $\rho_{ij}$, gives the proportion of trees in the random forest for which the feature vectors $x_i$ and $x_j$ appear in the same terminal node. Using this proximity measure they create an outlier score and evaluate each instance in the data base . A threshold on the anomaly score is then determined in order to decide whether or not an object is an outlier. This approach suffers from the same constraints as density based outlier detection methods. It is operationally expensive and slow for big data bases since every evaluated object must be compared to each instance on the training set. Furthermore, it has the problem of determining the outlier threshold, in other words what is considered as a ``far'' or ``close'' distance.

\section{Background Theory}
\label{sec:background}
Our algorithm is based on known machine learning methods, namely random forest and Bayesian networks. In this section we summarize 
the background for all these methods. Detailed explanations for each of these approaches can be found in \citet{Breiman2001}, \citet{Koller2009} and \citet{Cooper:Herskovits:1992}.
 


\subsection{Random forest}
\label{sec:RF}

Random forest developed by \citet{Breiman2001} is a very effective machine learning classification algorithm. The intuition behind this method is to train several decision trees using labeled data (training set) and then use the resulting trained decisions trees to classify new unlabeled objects in a voting system. The main principle is to follow a divide-and-conquer approach, each decision tree is trained with a random sample of the data and is consequently considered as a ``weak'' classifier. Nevertheless, the ensemble of these decision trees generates a robust or ``strong'' classifier that, based on the combinatorial power of its construction, creates an accurate and effective model.

\indent The process of training or building a RF model given some training data is as follows:

\begin{itemize}
\item Let $R$ be the number of trees in the forest (a user defined parameter) and $|F|$ be the number of features describing the data.
\item Build $R$ sets (bags) of $n$ samples taken with replacement from the training set (bootstrap samples). Note that each of the $R$ bags has the same number of elements than the training set but some of the examples are selected more than once, given that the samples are taken with replacement. 
\item For each of the $R$ sets, train a decision tree using at each node a feature picked from a  random sample of $|F\rq|$ features ($|F\rq|$ is a model parameter where $|F\rq| \ll |F|$) that optimizes the split.
\end{itemize}

Each decision tree is created independently and randomly using two principles. First, training each individual tree on different samples of the training set. Growing trees from different samples of the training set, creates the expected diversity among the individual classifiers. The second principle is the random feature selection, which means that for each tree the splitting (decisive) feature in every node is chosen from a random subset of the features. This contributes to the reduction of the dimensionality and has been shown to significantly improve the RF accuracy \citep{Bernard2008,Geurts2006}. Furthermore, each tree is grown to the maximum possible subject to the minimum size chosen for the terminal nodes (model parameter). For this paper we set the terminal nodes minimum size to be one, so the trees can be as large as the model allows it.

When classifying a new instance, each tree gives a classification or ``votes'' by following the decision rules in every node of the tree until reaching a terminal node.  Since RF is a composition of many trees the output corresponds to the votes of all the trees. 
The  class probability,   $P(class \: y_j/features(i))$ is  the proportion of trees that voted for the class $j$ ($j  \in \{ 1,\ldots, k \}$, where $k$ is the total number of classes).

\subsection{Bayesian Networks} \label{sec:BNs}

A Bayesian network is a directed acyclic graph (DAG), a particular probabilistic graphical model that encodes local statistical dependencies among random variables. A BN is defined by a set of nodes representing random variables $V=\{ v_1, \ldots, v_k \}$ and a set of edges $\varepsilon = \{ \varepsilon_1, \ldots, \varepsilon_b \}$ connecting the variables. One of the applications of BN is to estimate joint probability  density functions (PDF). This is done by assuming that the variables in the PDF are the nodes in the BN and that the connections between the nodes determine certain dependence relationships that simplify the joint distribution. More formally, if we want to estimate the joint probability distribution $P(v_1, \ldots, v_k)$ and we have a BN describing connections between these variables, we can simplify it as:\\ 

\begin{equation}
P(v_1, \ldots, v_k) = \prod_{i= 1}^{k} P(v_i | Pa_{BN}(v_i) )
\end{equation}

\noindent where $Pa_{BN}(v_i)$ corresponds to the parents of the node $v_i$ in the BN. Note that the PDF has been decomposed in a product of smaller factors (conditional probabilities).


 The main challenges of learning a BN that models a PDF over a set of variables are a) to learn the set of edges $\varepsilon$, or in other words the BN structure,  and b) to learn the conditional probabilities $P(v_i | Pa_{BN}(v_i) )$.  

 
 \subsection{Learning the edges of the BN} \label{sec:BNLearnStruct}
 
 A BN is a directed acyclic graph where each node represents a random variable. In our case the random variables we are modelling are the RF outputs, in other words, a probability vector $[v_1,\ldots,v_k ] \; v_j \in [0 ,1]$, representing the probabilities of belonging to each of the possible classes $c_j$ ($j \in [1 \ldots k]$, with $k$ being the number of known classes). Given that the amount of possible network structures is exponential in the number of variables, it is necessary to use heuristics to find the optimal  network. 
In our work we use a greedy algorithm proposed by \citet{Cooper:Herskovits:1992}. They define a score to evaluate each possible network structure and greedily search for the structure with the maximum score. First, they decide an order of the variables (topological order) from where possible structures will be explored.  A topological order $\{ 1, \ldots, k \}$ is such that if $i$ is smaller than $j$ in the order, then $v_i$ is an ancestor of $v_j$ in the network structure. After deciding on a particular order, the algorithm proceeds by finding the best set of parents per each node, greedily adding a new candidate parent and checking if the new addition creates a better network score or not. In case the edge addition improves the network score, the edge remains in the actual network. Note that the maximum number of parents per node is an input parameter of the algorithm. 
  
 Finally, to calculate the network score, they evaluate the probability of the structure given the data, which corresponds to apply the same factorization imposed by the structure to the data and use multinomial distributions over each factor.
How exactly   score is assigned to a given structure is well described in the original work \citep{Cooper:Herskovits:1992,Pichara2013}.

 \subsection{Learning the parameters of the conditional distributions}\label{sec:BNLearnParams}
In order to model the conditional probabilities, we may assume that all variables (votes) are continuous and normally distributed. Since $V$ comes from the RF votes, its distribution is multimodal and consequently a single Gaussian would not describe the data.  A better solution is to discretise the continuous data \citep{Montit}, so as to use multinomial distributions. Even if this process only gets rough characteristics of the distribution of the continuous variables, it better describes the data by capturing its multimodality. To do the discretisation, the data is divided into a set of bins, thus every data value which falls in a given interval, is replaced by a representative value  of that interval.

Given that our data are now discrete, we use multinomial distributions to model each conditional probability $P(v_{j} |  Pa_{BN}(v_j) )$. The number of parameters to be estimated depends on the number of values that variables $v_j$ and $Pa_{BN}(v_j) $ can take. For example, suppose that the parents of variable $v_j$ are $\{v_a,v_b \}$, where each of the three variables $\{v_j,v_a,v_b \}$ can take two different values (for simplicity say 1 and 2). The probability distribution $P(v_j | v_a,v_b)$ is then completely determined by Table~\ref{Fig:ExProbs}.


\begin{table}[h!]
\caption{\label{Fig:ExProbs} Probability of $v_j$ given the different values of the parents, $P((v_j  | v_a , v_b))$. There is one multinomial distribution per each combination of the values of the parents.
The number of outcomes of each distribution corresponds to the number of values of variable $v_j$.}

\centering    
\begin{tabular}{|c|c|c|}
\hline
  & $v_j = 1$ & $v_j = 2$ \\  
\hline
$v_a = 1, v_b = 1$ & $\theta_1$ & $1-\theta_1$ \\
$v_a = 1, v_b = 2$ & $\theta_2$ & $1-\theta_2$ \\
$v_a = 2, v_b = 1$ & $\theta_3$ & $1-\theta_3$ \\
$v_a = 2, v_b = 2$ & $\theta_4$ & $1-\theta_4$ \\
\hline
\end{tabular}
\end{table}

The number of parameters for each variable is consequently given by the following expression: 

\begin{equation}
(N_{bins}-1) \times (N_{bins})^{N_{parents}}
\end{equation}

where $N_{bins}$ is the number of bins chosen for the discretization and $N_{parents}$ corresponds to the number of parents of the variable. In the example given above, the number of parameters we have to estimate is ${(2-1)\times 2^2=4}$. To estimate the parameters, we use the {\em maximum a posteriori}  (MAP) approach, where we select the value for the unknown parameter as the value with maximum probability under the posterior distribution of the parameter. 
The posterior distribution of, lets say $\theta_1$, is calculated as:

\begin{equation}
 P(\theta_1 | data) = \frac{P(data | \theta_1) \times P(\theta_1)}{ \sum_{\theta_1} P(data | \theta_1)\times P(\theta_1)}
 \end{equation}
 
 Where $P(data | \theta_1)$ is the likelihood of the model and $P(\theta_1)$ is the prior of the parameter $\theta_1$. The likelihood is calculated as: 

\begin{equation}
 P(data | \theta_1) = \theta_1^{N_1} \times (1- \theta_1 )^{N_2}
 \end{equation}
 
 Where $N_1$ is the number of cases in the data where $v_j$'s  take a particular value. Following the example above, $N_1$ is the number of cases where $v_j = 1$ and $v_a = 1$, $v_b = 1$ and $N_2$ is the number of cases where $v_j = 2$ and $v_a = 1$, $v_b = 1$.\\ 
 
The main purpose of the priors is to avoid overfitting. In other words, in cases where we have just few cases in the data with a given combination of values, the estimation of the parameters  should tend to stay in a predefined value until the data cases increase. Priors are a manner to simulate previously seen \lq\lq imaginary data'' in order to compensate situations of few cases. We choose conjugate priors to simplify the calculations of the posteriors.  In our case, given that the likelihood is a multinomial, the chosen prior for $P(\theta)$ is a Dirichlet distribution, which is the conjugate distribution for the multinomial. Using a Dirichlet
 prior the obtained posterior is:
 
 \begin{equation}
 P(\theta_1 | data) \propto  \theta_1^{N_1 + \alpha_1} \times (1- \theta_1 )^{N_2 + \alpha_2 }
 \end{equation}
 
 where $\{ \alpha_1,  \alpha_2 \}$ are the Dirichlet distribution parameters. The values of $\{ \alpha_1,  \alpha_2 \}$ act as the \lq\lq imaginary data'' that we count, and we just assume that 
  all combinations of values have the same number of previously seen cases. 
  Analogously, we can find the value of every parameter $\theta_j$ for variables with any number of different values.

\section{Methodology}
\label{sec:methodology}

\begin{figure*}
\centering
\includegraphics[width=15cm]{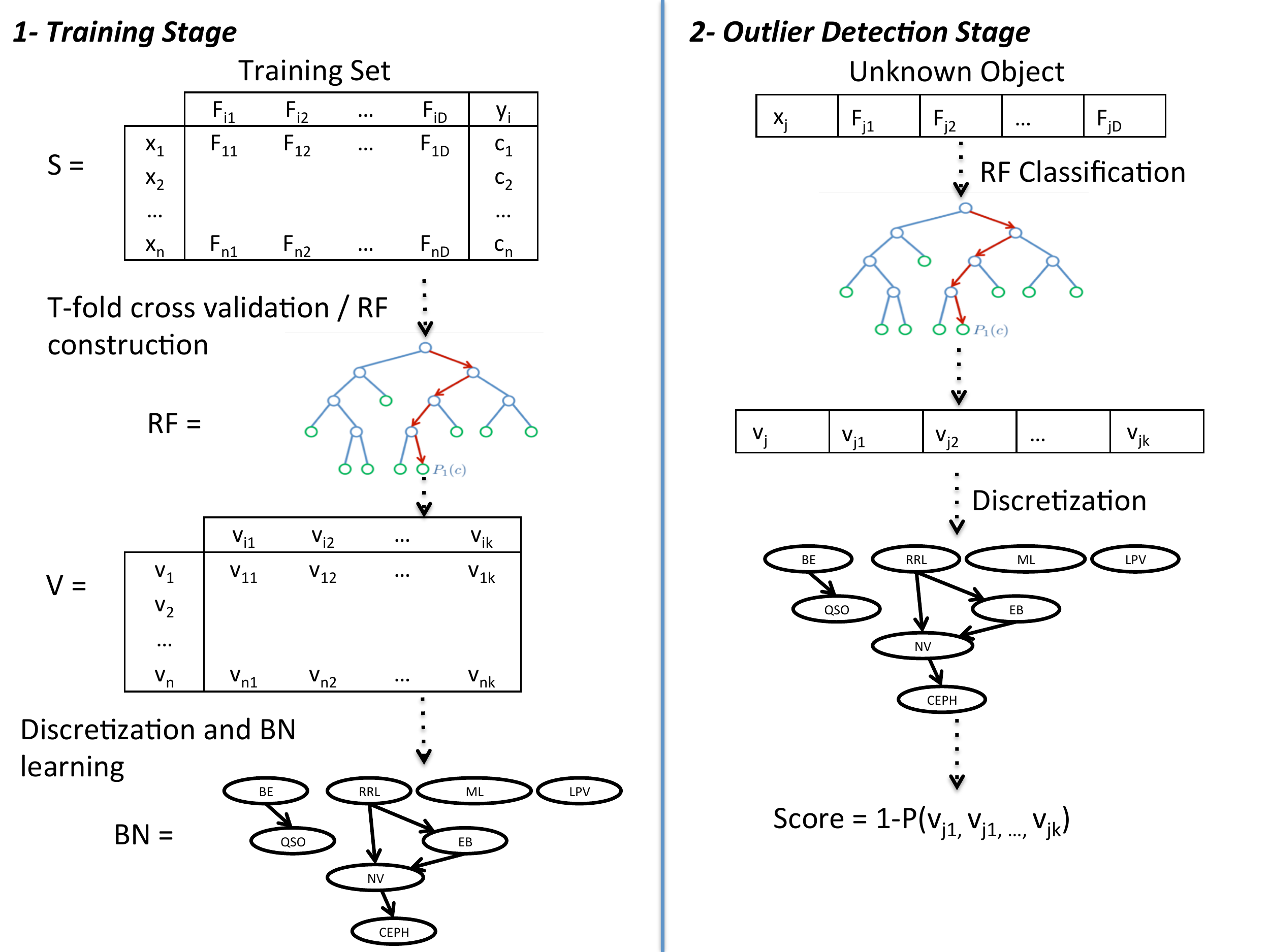}
\caption{Pictorial representation of the algorithm and its two main stages: the training stage (left panel) and the outlier detection stage (right panel).}
\label{fig:chart}
\end{figure*}

In the next section we detail our work and methodology. For illustration, we present in Figure~\ref{fig:chart} a pictorial representation of our algorithm and its two main stages: the training stage (left panel) and the outlier detection stage (right panel). In the training stage, we start with a training set followed by the training of the RF, discretization of the probabilities and finally the construction of the BN. In the outlier discovery stage, every new instance is passed through the already learned RF and BN resulting in a score for being outlier.


As we previously mentioned, the idea behind our method is to train a classifier with known classes and learn its decision mechanism with a model. In this manner, when an outlier is being analyzed, the classifier will present an abnormal voting confusion that will be immediately flagged by our model. 

  Our method starts with a set of $n$ labeled instances (training set) $S = \{ (x_1,y_1),\ldots, (x_n,y_n) \}$, where each $x_i = \{ x_{i1},\ldots, x_{iD} \}$ is a vector in a $D-$dimensional space - the statistical descriptors or features that represent each light-curve -  and $y_i$ corresponding to the label of $x_i$ ($y_i \in \{ c_1,\ldots, c_k \}$, are all the known classes in the training set). In Section \ref{sec:data}, we give details about the classes and statistical descriptors we used. 
  
We train a RF classifier and obtain voted labels for each element using the set $S$. Since trees are constructed from different bootstrapped samples of the original data (as explained in section 3.1), about one-third of the cases are left out of the \textquotedblleft bag\textquotedblright  and not used in the  construction of each tree.  By putting  these out-of-bag (oob) observations down the trees that were not trained with oob data,  we end up with unbiased predicted labels for $S$. Each prediction obtained from the RF comes as a vector $\{v_{i1}, \ldots,v_{ik} \}$ where each $v_{ij} \in [0,1],  \;  j \in [1 \ldots k]$, tell us the probability that the element $x_i$ belongs to the class $y_j$, $\sum_{j=1}^{k} v_{ij} =  1, \; \forall i \in [1 \ldots n]$ or as we explained in section \ref{sec:RF}, $v_{ij}$\\ corresponds to $P(class \: y_j/features(i))$. 

In our experiments we use 20 bins for the discretization and a maximum of two parents.\\

 After this step is performed, we end with  a new dataset $V = \{ v_{1}, \ldots, v_{n} \}$ where each $v_i = \{v_{i1}, \ldots,v_{ik}\}$.  This dataset gives us information about how the RF votes among objects that belong to each of the known variability classes. We want to use this dataset to decide if an unlabeled object belongs to an unknown variability class or not, simply by comparing the RF votes of this unlabelled object with the \lq\lq usual'' votes of the RF obtained from the dataset $V$. If the voting vector for the unlabeled object is too different from the voting vectors stored in the dataset $V$, we flag it as an outlier. To do this comparison, we learn the joint probability distribution over the dataset $V$  using a BN. Recall that BNs estimate joint probability distributions as a product of smaller factors. These factors are conditional probability distributions and in our case, the joint probability we aim to model is the joint probability of the various votes, $P(v_{1}, \ldots,v_{k})$. \\
 
When analyzing an unlabelled object $i$, we first obtain its RF votes $\{v_{i1}, \ldots,v_{ik} \}$ using the already trained RF and next we calculate the joint probability associated to this vector $P(v_{i1}, \ldots,v_{ik})$ using the already learned BN.  Our outlier score is calculated as one minus the joint probability; the lower the joint probability is, the higher the score is and therefore the more outlying the corresponding object is.\\
  
In section \ref{sec:BNLearnParams} we mentioned the necessity of a prior in order to include all the possible cases in our model. We assume the same value for all the needed priors. To chose the value of $\alpha$, we calculate the number of instances one would hope to see if the data were uniformly distributed. Three parameters are considered for this estimation: the size of our data (5646), the number of bins in the discretization process (20 bins) and the maximum number of parents a node can have on the BN. Given that the minimum number of parents is zero and the maximum is three, a reasonable number for $\alpha$ is four. We also empirically tested with different values of $\alpha$ and found that the results are not sensitive to the choice of $\alpha$.

\section{Data}
\label{sec:data}

\subsection{MACHO catalog}

MACHO (Massive Compact Halo Object) is a survey which observed the sky, starting in July 1992 and ending in 1999, to detect microlensing events produced by Milky Way halo objects. Several tens of millions of stars where observed in the Large Magellanic Cloud (LMC), Small Magellanic Cloud (SMC) and Galactic bulge. The average number of observations per object is several hundreds, with the center of the LMC being observed more frequently than the periphery. The reader can find detailed MACHO description in \citet{Alcock1997ApJ...479..119A}.

Every light-curve is described by 13 features corresponding to the blue non standard pass with a bandpass of 440-590nm (see \citet{Pichara2012} for more details).

\subsection{Training set} 
\label{sec:training_set}

\begin{figure*}
\centering
\begin{minipage}{\textwidth}
  \centering
  \includegraphics[width=.4\linewidth]{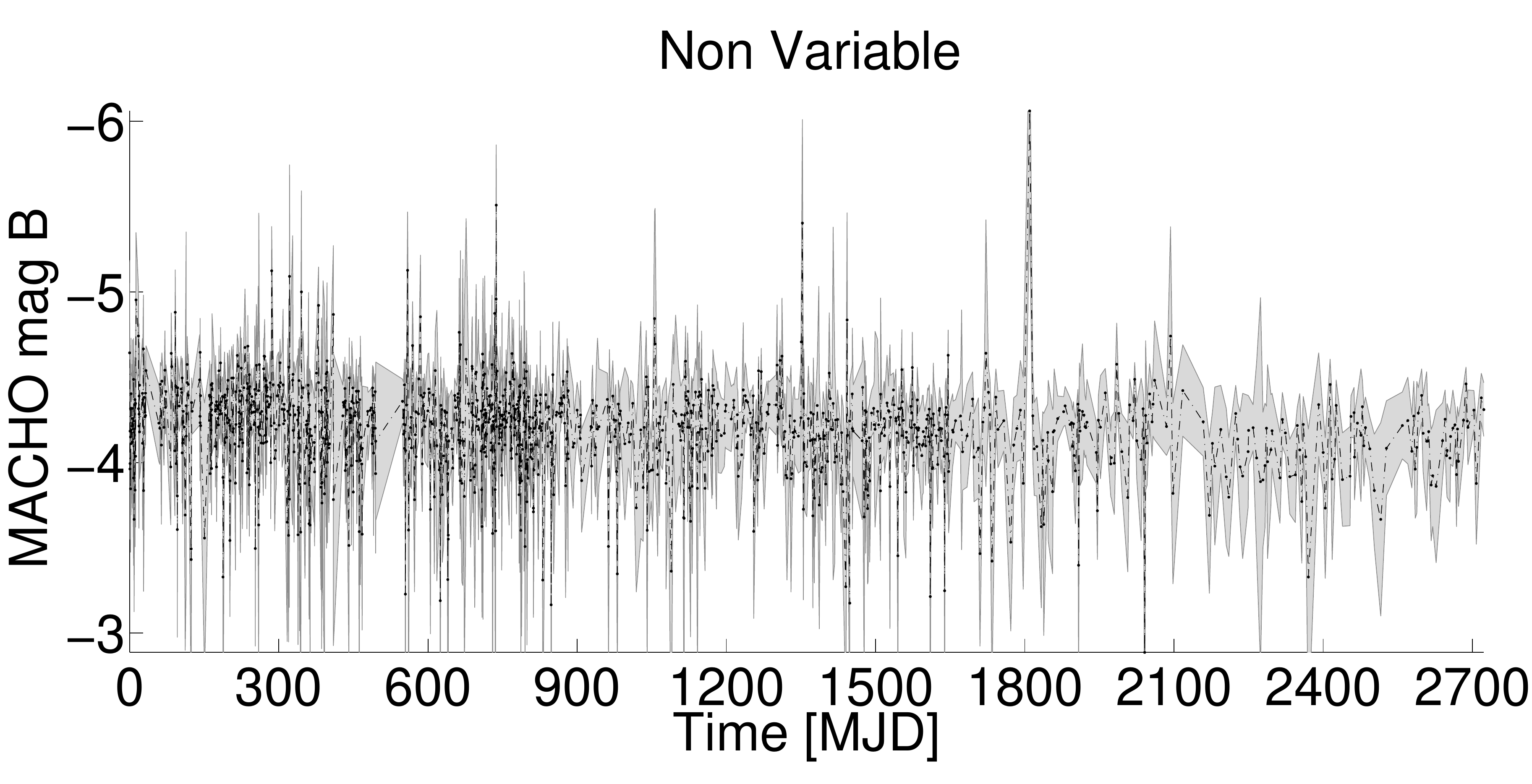}
  \includegraphics[width=.4\linewidth]{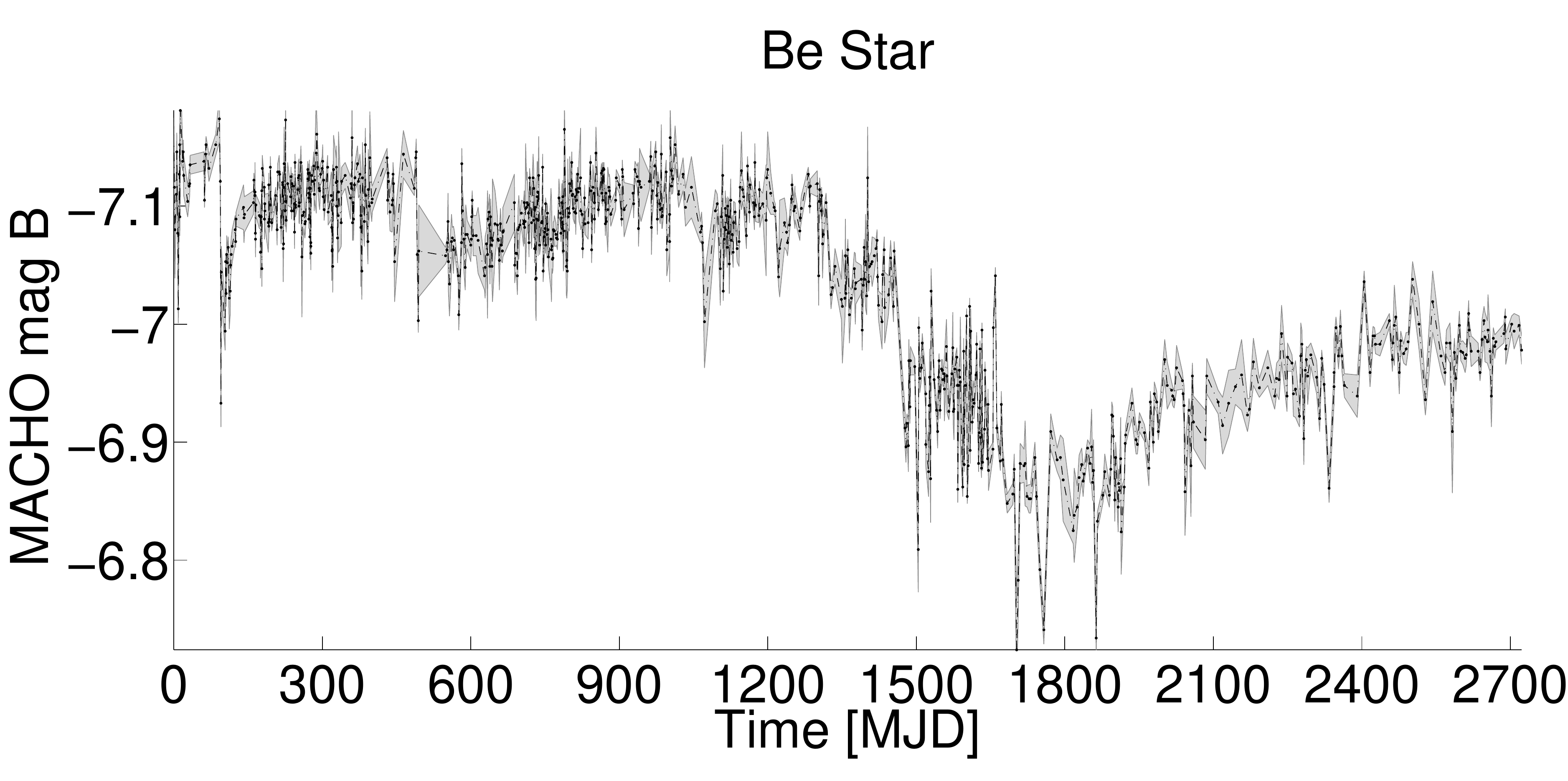}
  \includegraphics[width=.4\linewidth]{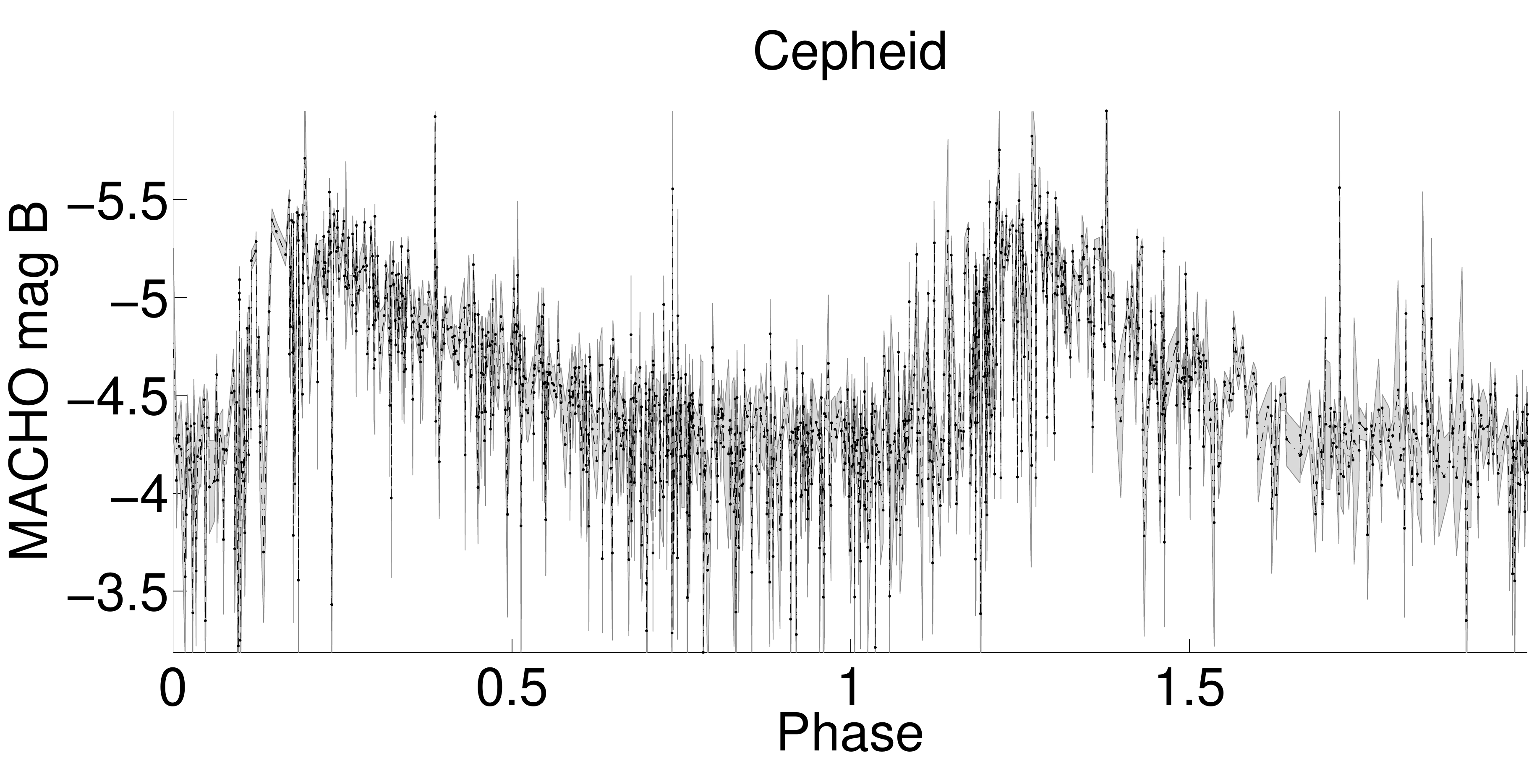}
  \includegraphics[width=.4\linewidth]{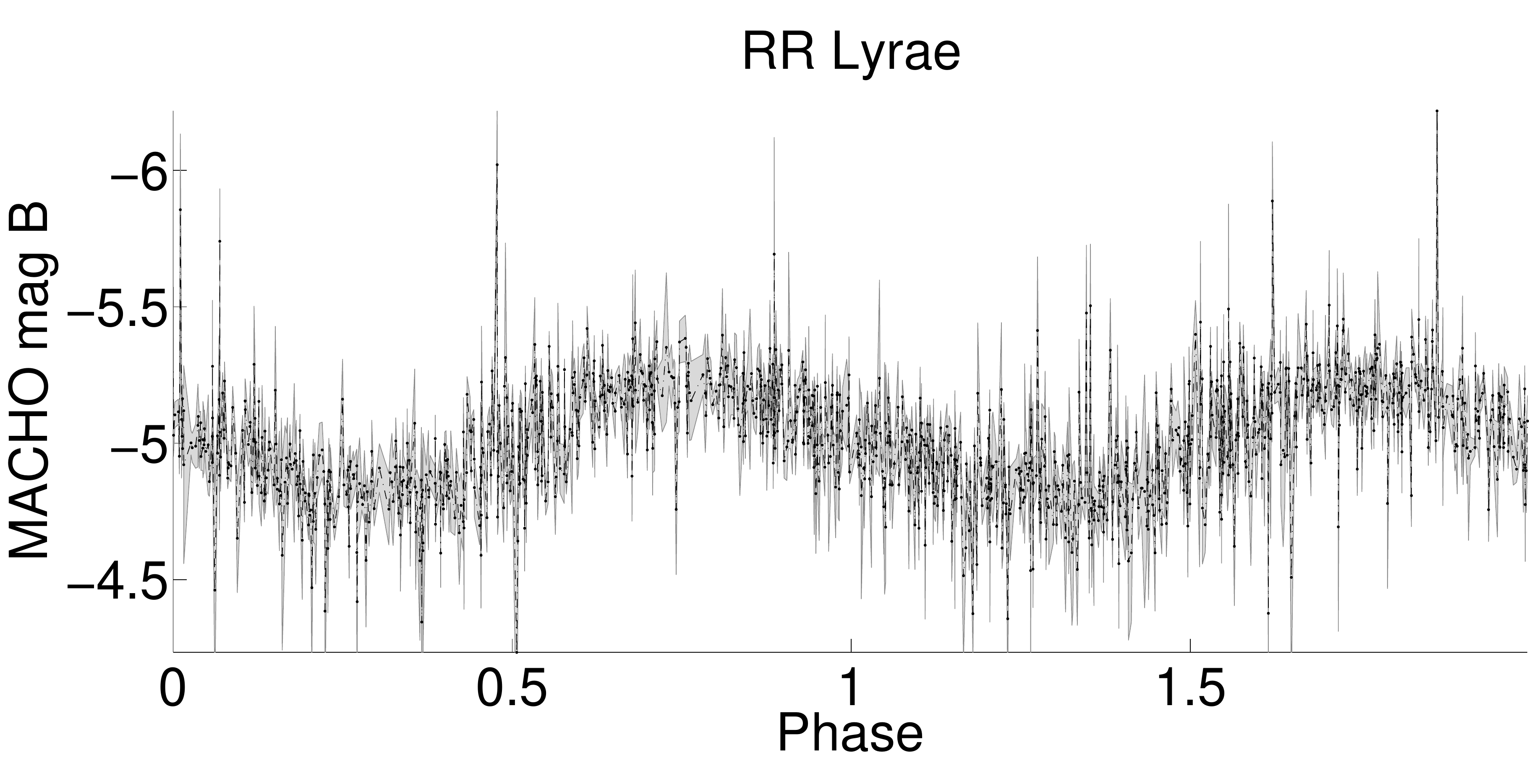}
    \includegraphics[width=.4\linewidth]{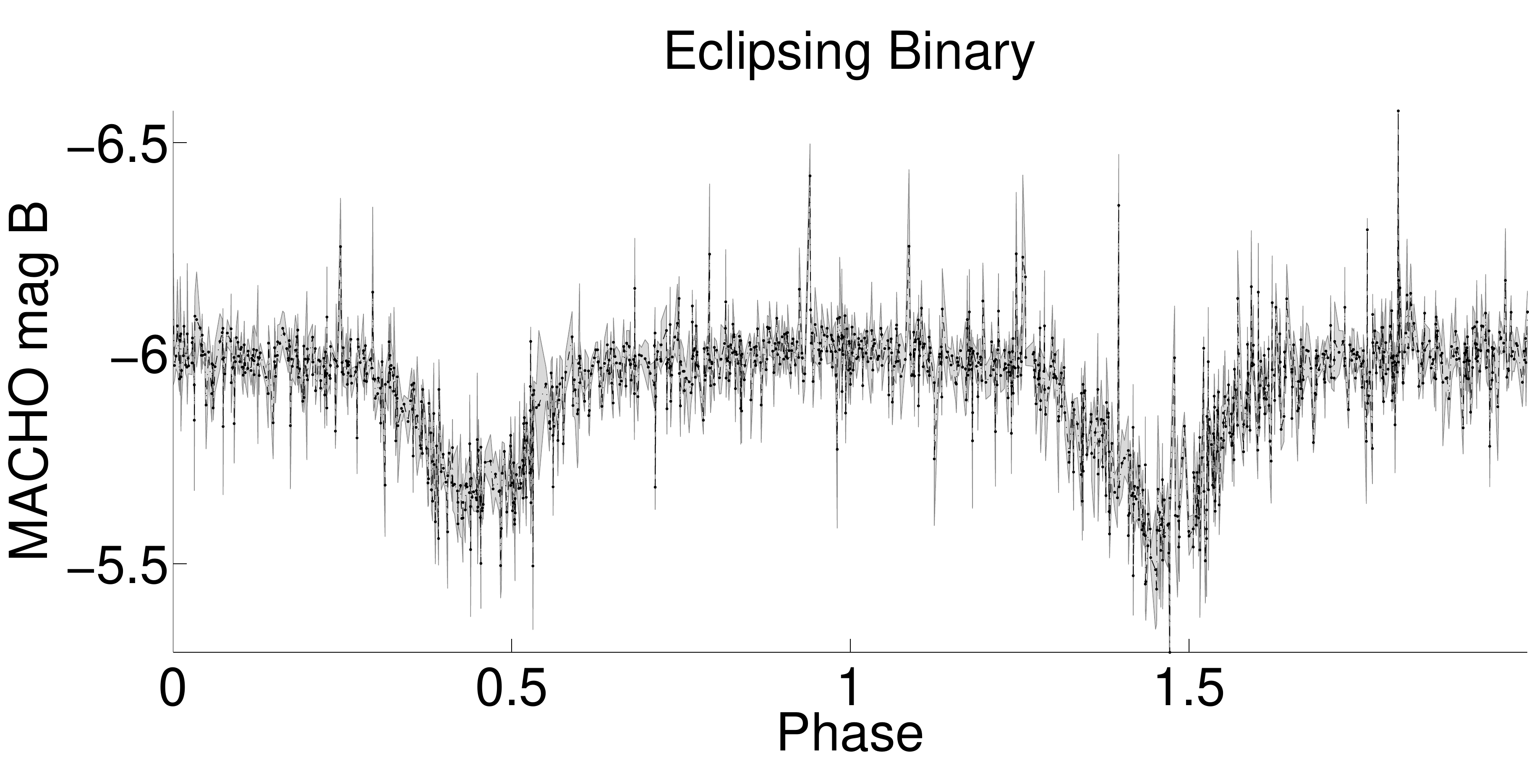}
      \includegraphics[width=.4\linewidth]{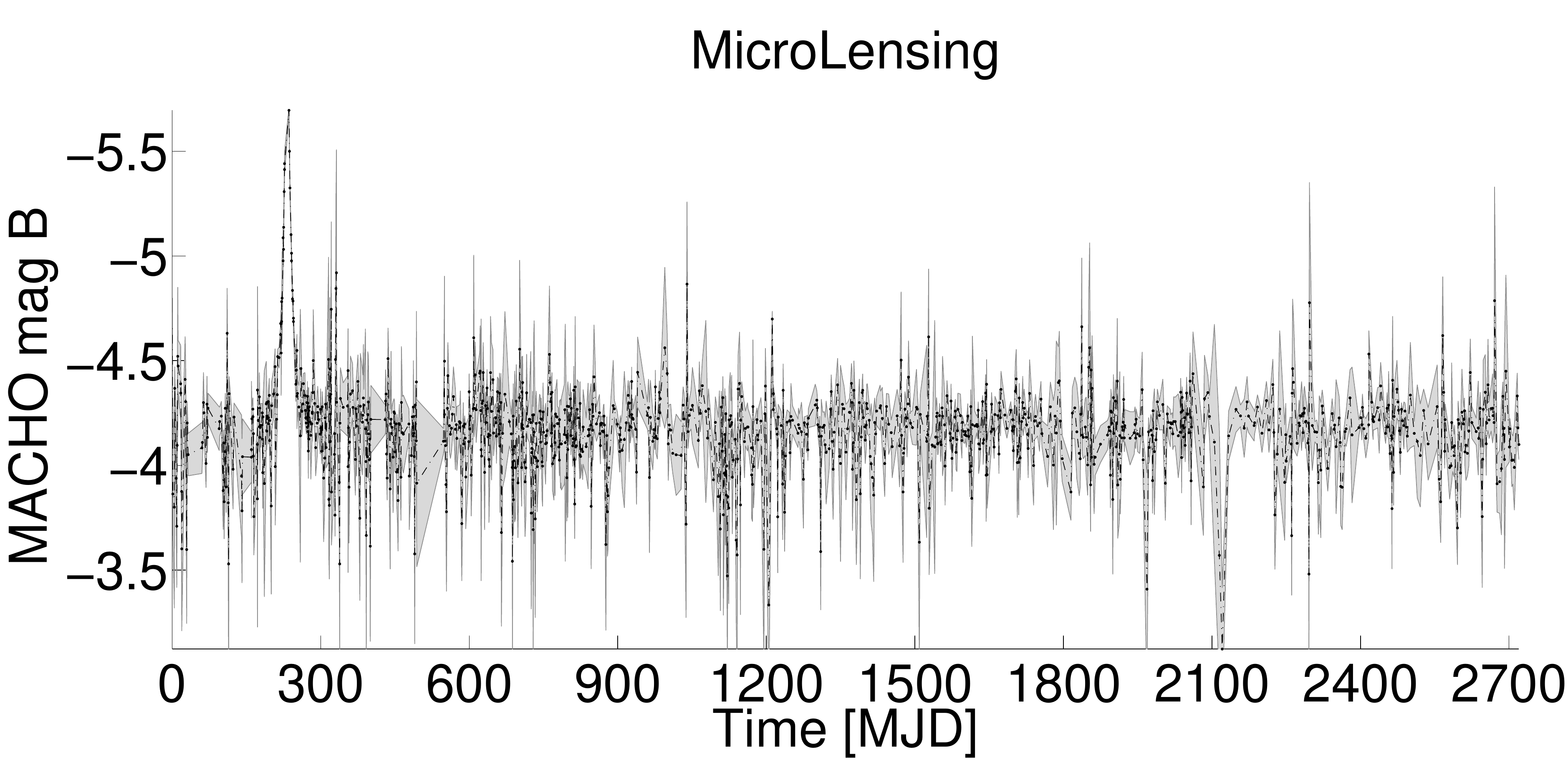}
        \includegraphics[width=.4\linewidth]{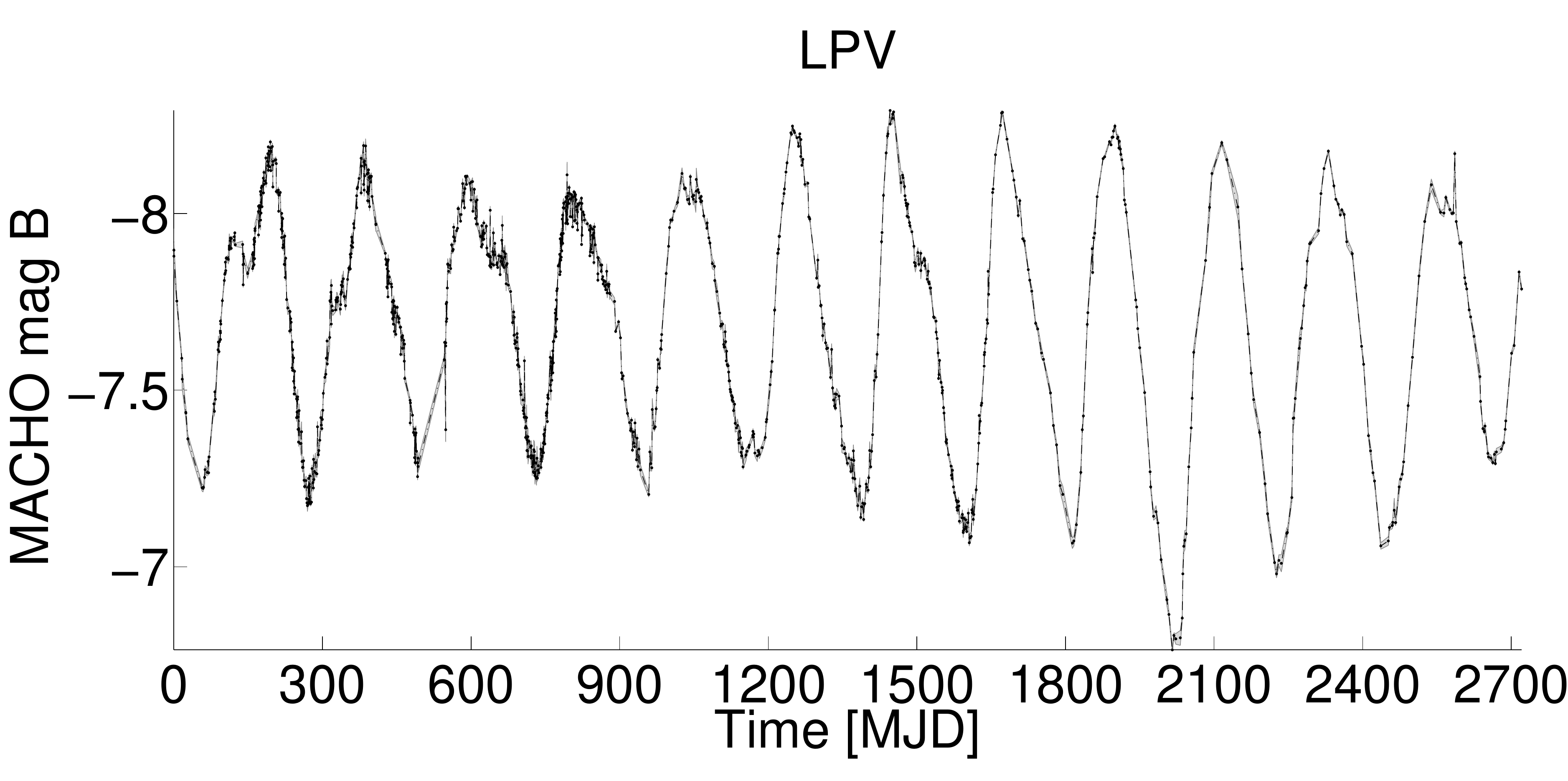}
          \includegraphics[width=.4\linewidth]{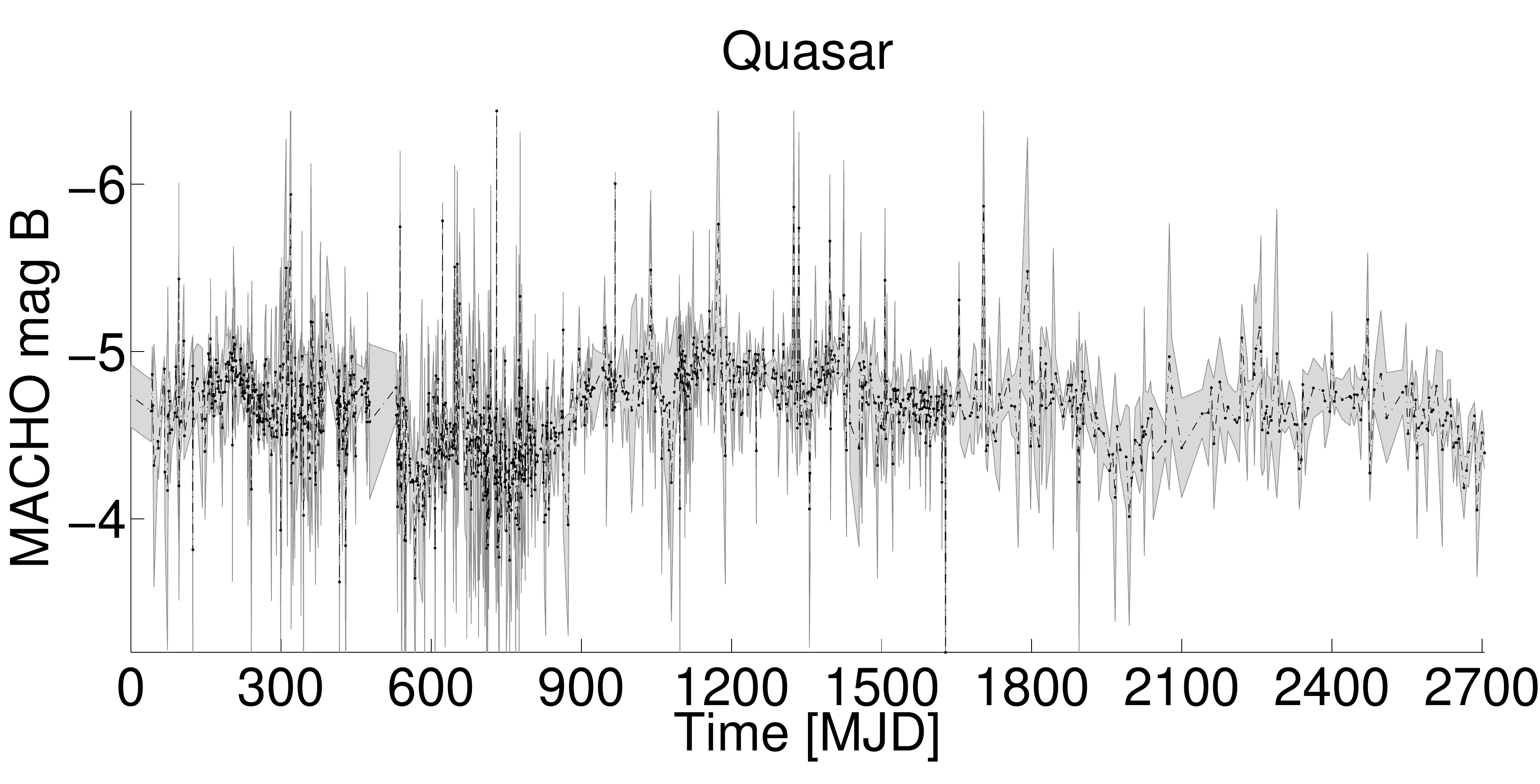}
\caption{Example light-curves of each class in the MACHO training set. The \textit{x}-axis is the modified Julian Date (MJD), and the \textit{y}-axis is the MACHO B-magnitude. Note that Cepheid, RR lyrae and Eclipsing Binary light-curves are folded since they are periodic.}
\label{fig:curvas_ejemplos}
\end{minipage}%
\end{figure*}

The training set is composed of a subset of 5646 labeled observations from the MACHO catalog \citep{Kim2011}\footnote{We collected these variables from the MACHO variable catalog found at: http://vizier.u-strasbg.fr/viz-bin/VizieR?-source=II/247}.
The constitution of the training set is presented in Table 1 and a representative example of each class light-curve is shown in Figure~\ref{fig:curvas_ejemplos}. 

The catalog  comprises several sources from  MACHO variable studies 
\citep{Alcock1996AJ, Alcock1997ApJ...482...89A, Alcock1997AJ, Alcock1999AJ....117..920A},
the MACHO microlensing studies
\citep{Alcock1997ApJ...491L..11A,  Alcock1997ApJ...479..119A, Alcock1997ApJ...486..697A, Thomas2005ApJ...631..906T},
and the LMC long-period variable study \citep{Wood2000PASA...17...18W}.
Quasars in the training set were collected from
\citet{Blanco1986PASP...98..635B, Schmidtke1999AJ....117..927S, Dobrzycki2002ApJ...569L..15D, Geha2003AJ....125....1G}.
Be stars were obtained from private communication with Geha, M. 
The non-variables were randomly chosen from the MACHO LMC database, 
and any previously known MACHO variables were removed from the non-variable set.

\begin{table}[H]
\caption{\label{Table:Training}Training Set Composition}
\centering
\begin{tabular}{ c | c | c |}  
\cline{2-3}
{} & Class & Number of objects \\
\hline
\multicolumn{1}{ |c| }{1} & Non variable & $3969$ \\
\multicolumn{1}{ |c| }{2} & Quasars & $58$ \\
\multicolumn{1}{ |c| }{3} & Be Stars & $127$ \\
\multicolumn{1}{ |c| }{4} & Cepheid & $78$ \\
\multicolumn{1}{ |c| }{5} & RR Lyrae & $288$ \\
\multicolumn{1}{ |c| }{6} & Eclipsing Binaries & $193$ \\
\multicolumn{1}{ |c| }{7} & MicroLensing & $574$ \\
\multicolumn{1}{ |c| }{8} & Long Period Variable & $359$ \\
\hline
\end{tabular}
\end{table}

\section{Results}
\label{sec:Results}

In this section, we show how we applied the above methods to the MACHO catalog. 

\subsection{Performance Test} 
To prove the performance of our algorithm, we created a test set leaving one class out of the MACHO training set; we trained our algorithm with the remaining classes and considered the excluded class as unknown objects that we want to discover. In other words, we expected these light-curves to have the highest outlierness score as they have never been seen by the model. 



We performed three different tests, each time leaving out of the training set one of the classes: quasars, eclipsing binaries and Be stars. The RF considered 500 trees, with  $|F'|=\lfloor \sqrt{|F|} \rfloor$ features in every node.  

\begin{figure}[h]
\centering
\includegraphics[width=8.5cm]{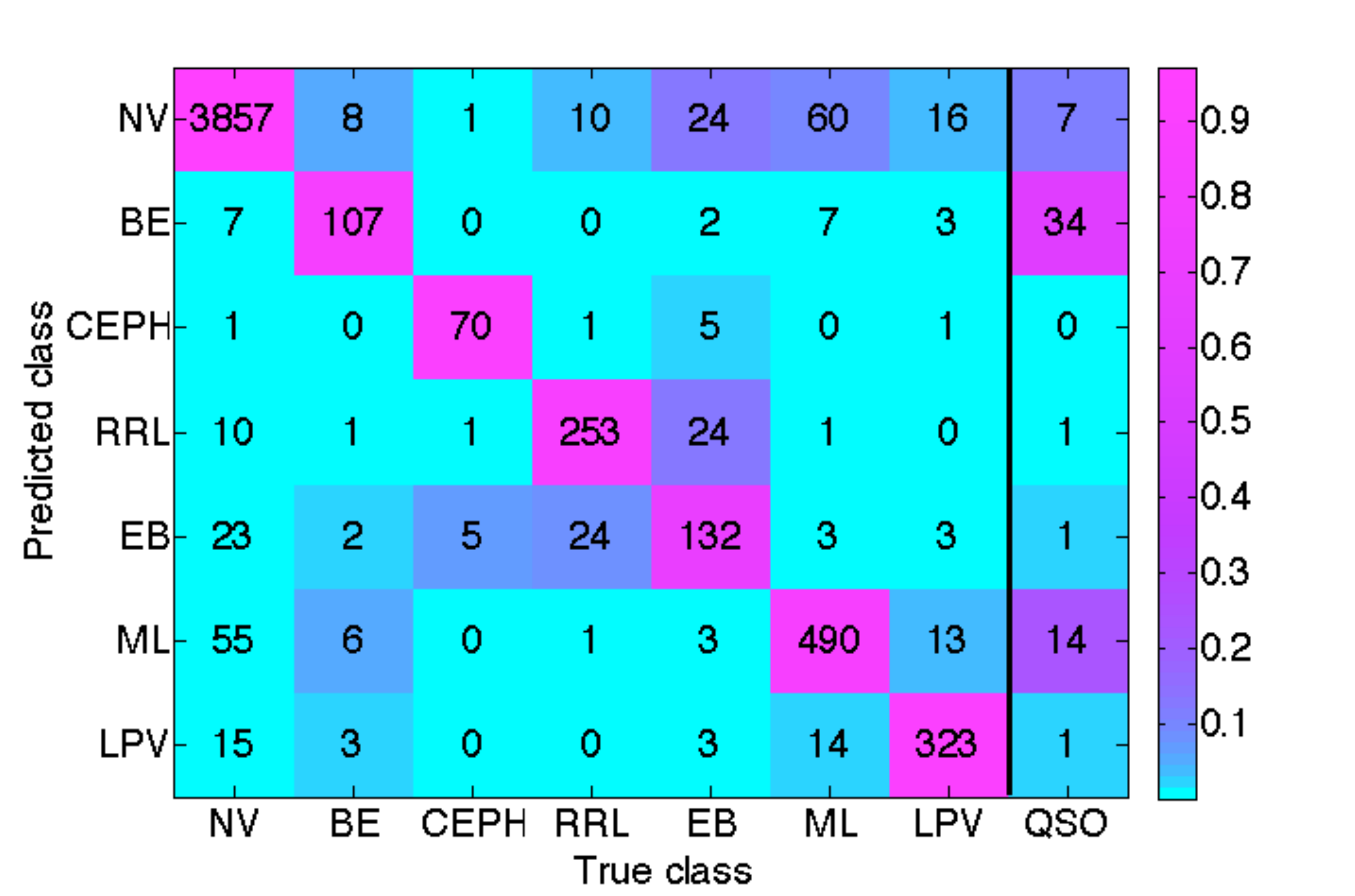}
\caption{RF votes distribution (NV: Non variable, BE: BE stars, CEPH: Cepheid, RRL: RR lyrae, EB: Eclipsing Binaries, ML: Microlensing, LPV: Long Period Variable, QSO: Quasars).
}
\label{fig:test_RF}
\end{figure}

\begin{figure}[h]
\centering
 \includegraphics[width=8cm]{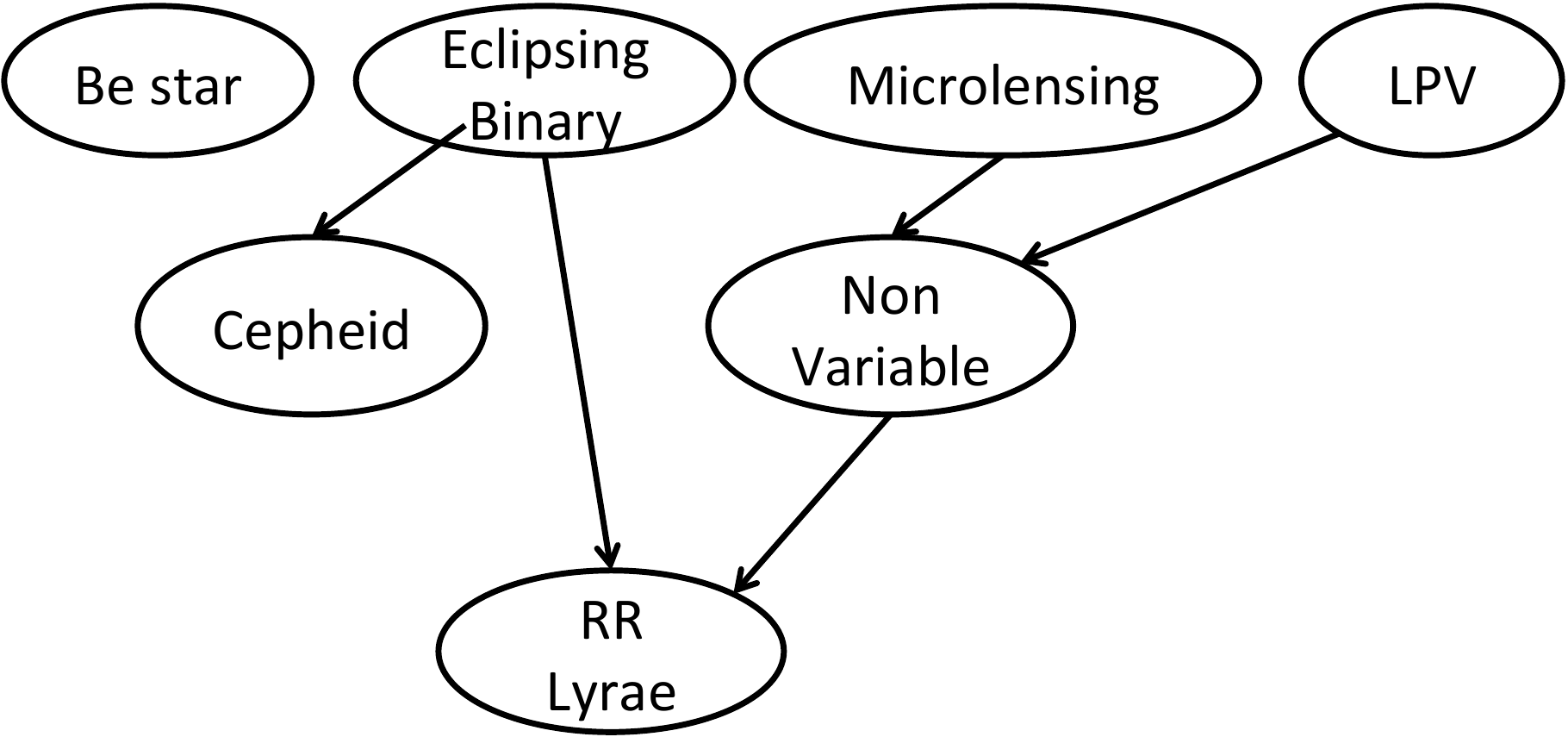}
\caption{BN structure for the performance test.}
\label{fig:quasardag}
\end{figure}

Next, we present the results for the test leaving the quasars out of the training set. In order to visualize the voting database $V$, we present the average number of objects voted by the RF for each class in Figure~\ref{fig:test_RF}. By using a color scale, we also show the average distribution of the votes among the different classes during the training phase and for the test class (quasars) during the testing phase (right vertical line). For example, when the RF is classifying a RR Lyrae it doubts mainly between non variables, eclipsing binaries and the true class, RR Lyrae. 
This is shown in the colors along the vertical line labeled RRL. 
This hesitation is learned by the BN and the relationship between classes is represented on a graph as shown in Figure~\ref{fig:quasardag}. RR Lyrae node is a child node of Cepheid and non variable nodes meaning that, when the light-curve to be classified is from RR Lyrae class, the voting vector will present high values in these other two classes. On the other hand, Be stars node is independent of the other classes, as expected.

\begin{figure}[h]
 \includegraphics[width=8cm]{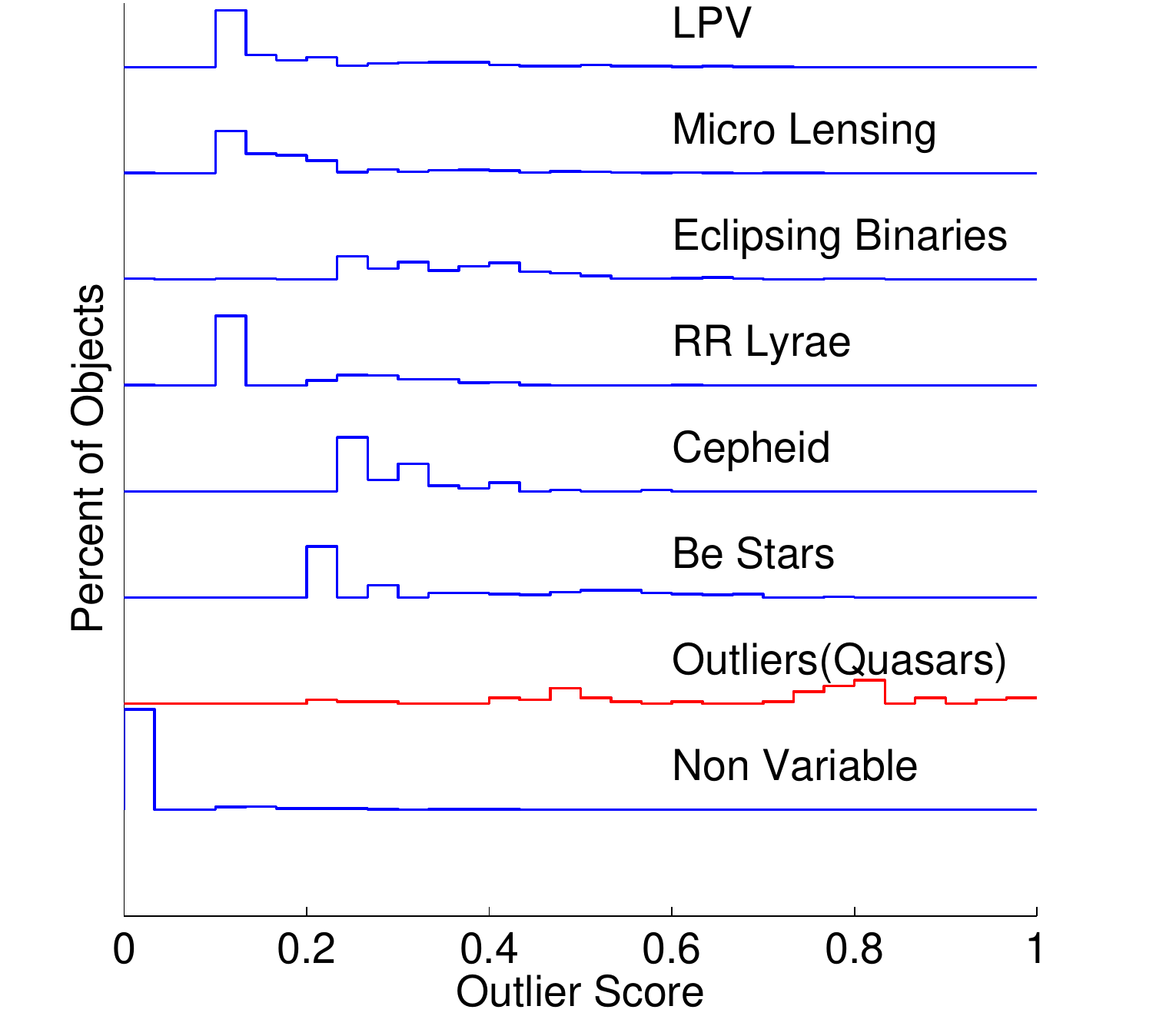}
\caption{Stacked plot of the outlier score distribution for the performed test. Each layer represents the outlier score distribution of the objects of a class (blue lines show the training set classes and the red line the outlier class). The y-axis scale for each layer goes from zero to one but it was removed for visual clearness.}
%
\label{fig:joint}
\end{figure}

\begin{figure}[h]
\centering
 \includegraphics[width=8cm]{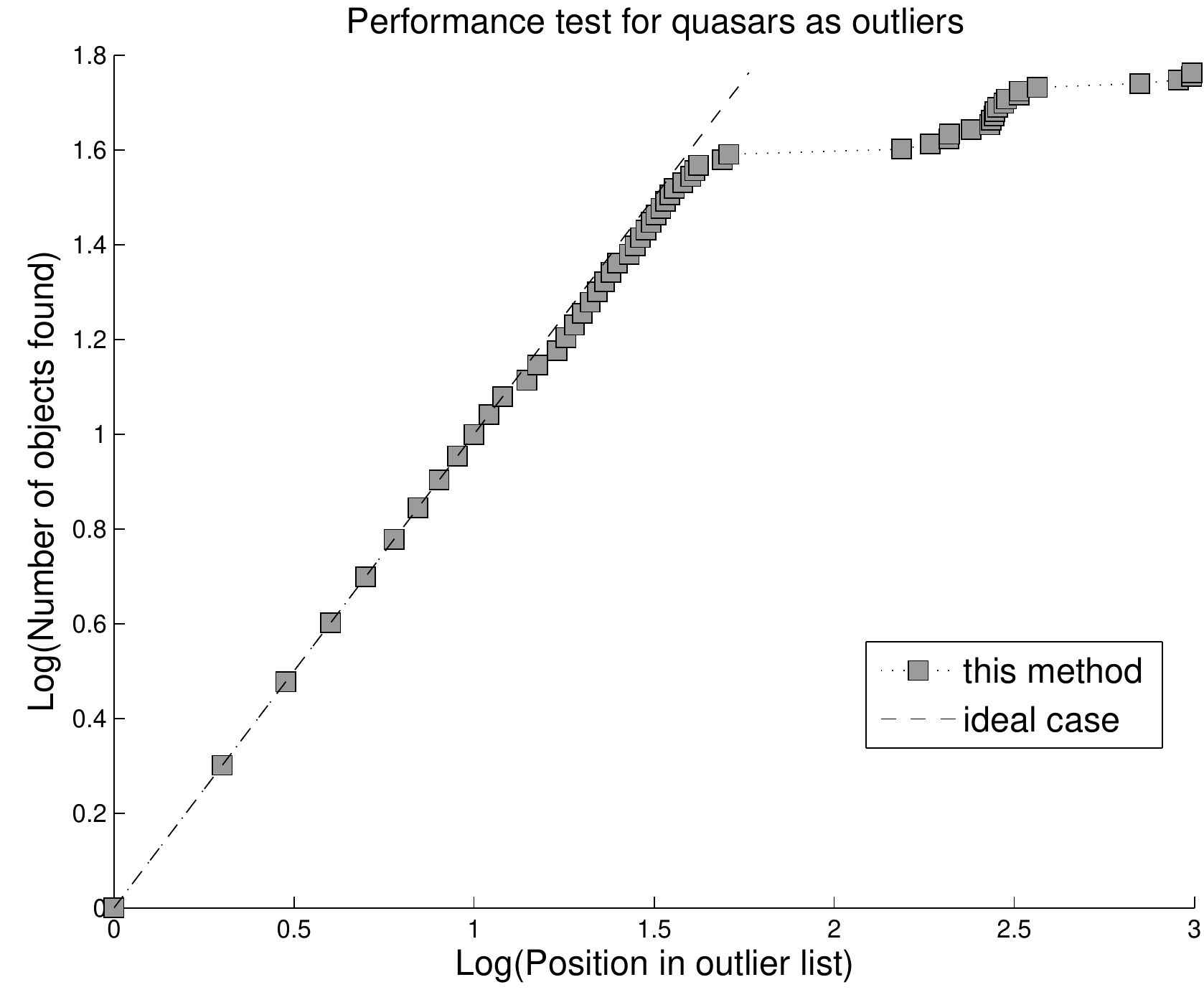}
  \includegraphics[width=8cm]{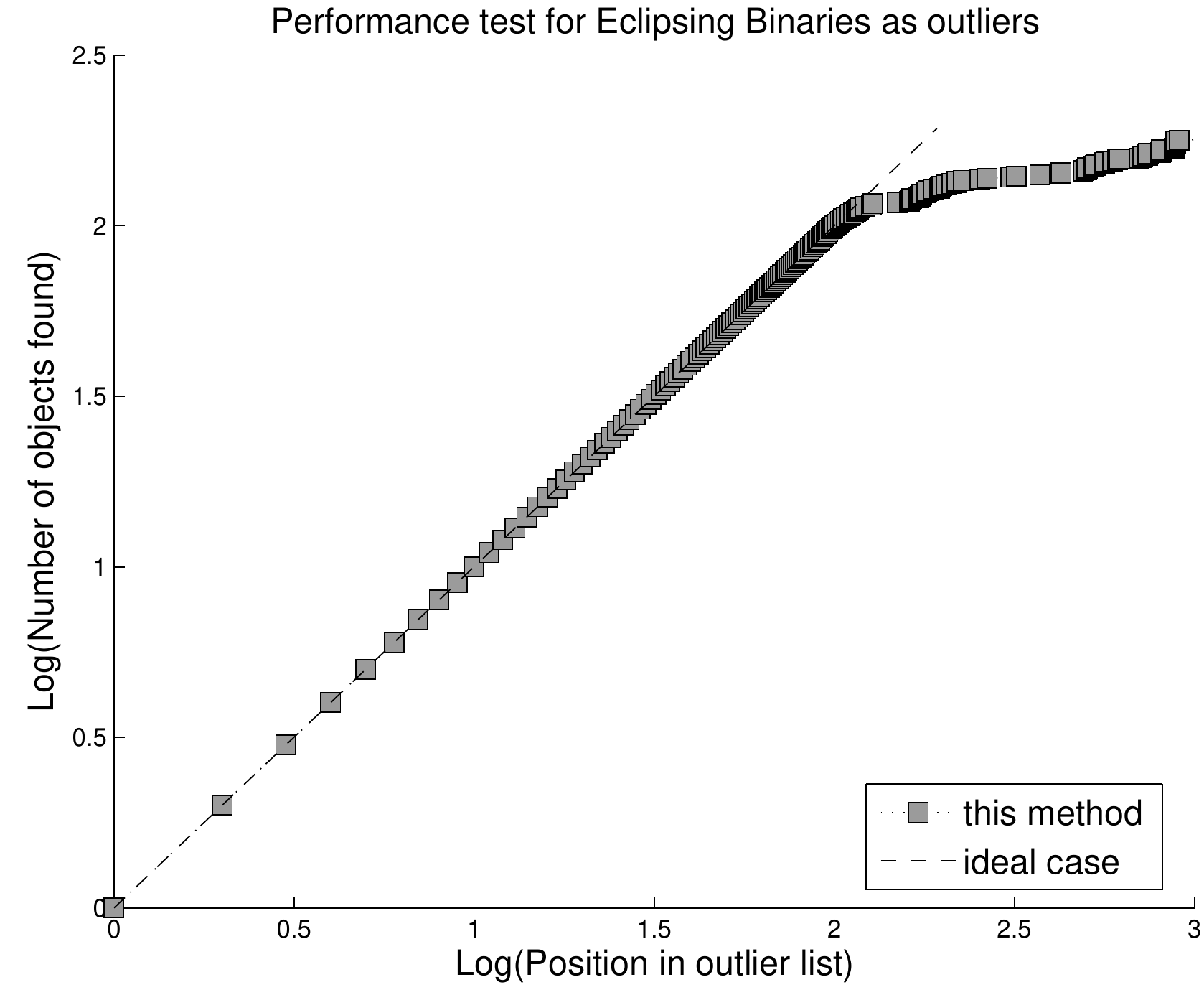}
    \includegraphics[width=8cm]{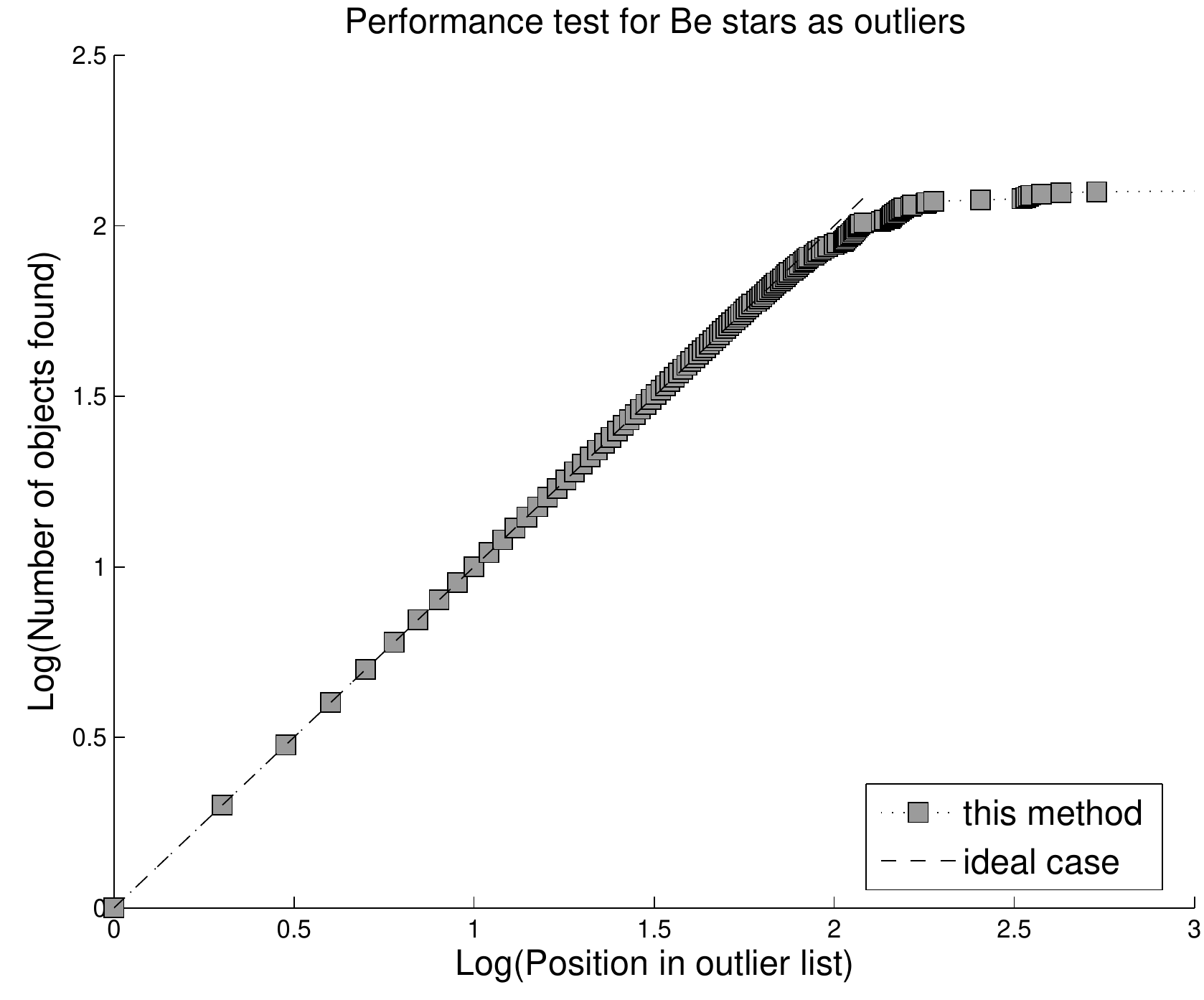}
\caption{Performance test results for Quasars, Eclipsing binaries and Be stars as outliers. The dashed line represents the ideal result, where the class left out use the top positions in the outlier list. Grey squares shows the actual obtained positions.}
\label{fig:quasartest}
\end{figure}

`After the algorithm training stage was completed,  the outlier detection stage was performed.  We first obtained from the trained RF a vector $v_i$  for every object in the training set, quasars included. We then determined the joint probability and thus the outlier score of each $v_i$  by using the already learned BN. We were expecting the quasars to have the highest outlier scores and thus to find them on the top of the resulting outlier list. Figure~\ref{fig:joint} shows how objects of ``known'' classes present high joint probabilities while outliers objects (quasars) have the lowest values. Finally, the top left panel of Figure~\ref{fig:quasartest} represents our algorithm performance, comparing the imputed outliers (quasars) positions in the top outliers list with the ideal case result - the 58 quasars will be using the 58 first places in the outlier list. It can be seen that the top 40-60 outliers are quasars and all imputed outliers (quasars) are in the top 200 list. The same behavior is observed when we choose other classes as the outlier class, as shown in the top right and bottom panel in Figure~\ref{fig:quasartest}.




%

\subsection{Running on the whole dataset}


Once we tested the accuracy of our method, we trained a RF (F-score= 0.9080)  with the complete training set and learned a new BN. The same parameters of the performance test were used in this stage.


 We ran our model on the whole MACHO data set (about 20 million of light-curves) to obtain a list of outlier candidates. 
Fortunately the main computational cost of the algorithm occurs during the training phase, for which the model needs to run the build the RF and learn the BN structure and parameters. After training the model, performing the inference for a light-curve takes a fraction of second and it is easily parallelizable.

%


\begin{figure*}
\centering
\begin{minipage}{\textwidth}
  \centering
  \includegraphics[width=.4\linewidth]{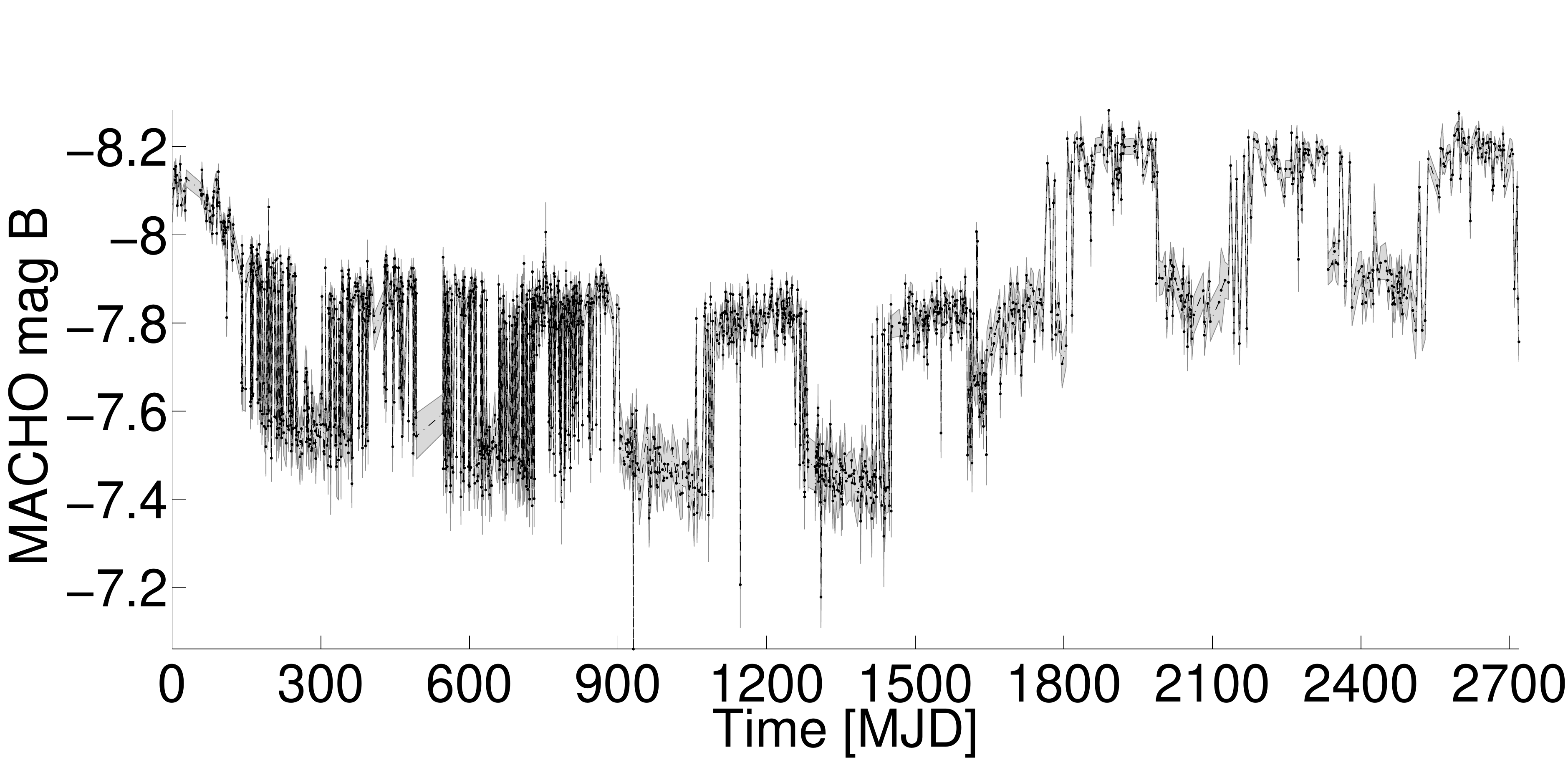}
  \includegraphics[width=.4\linewidth]{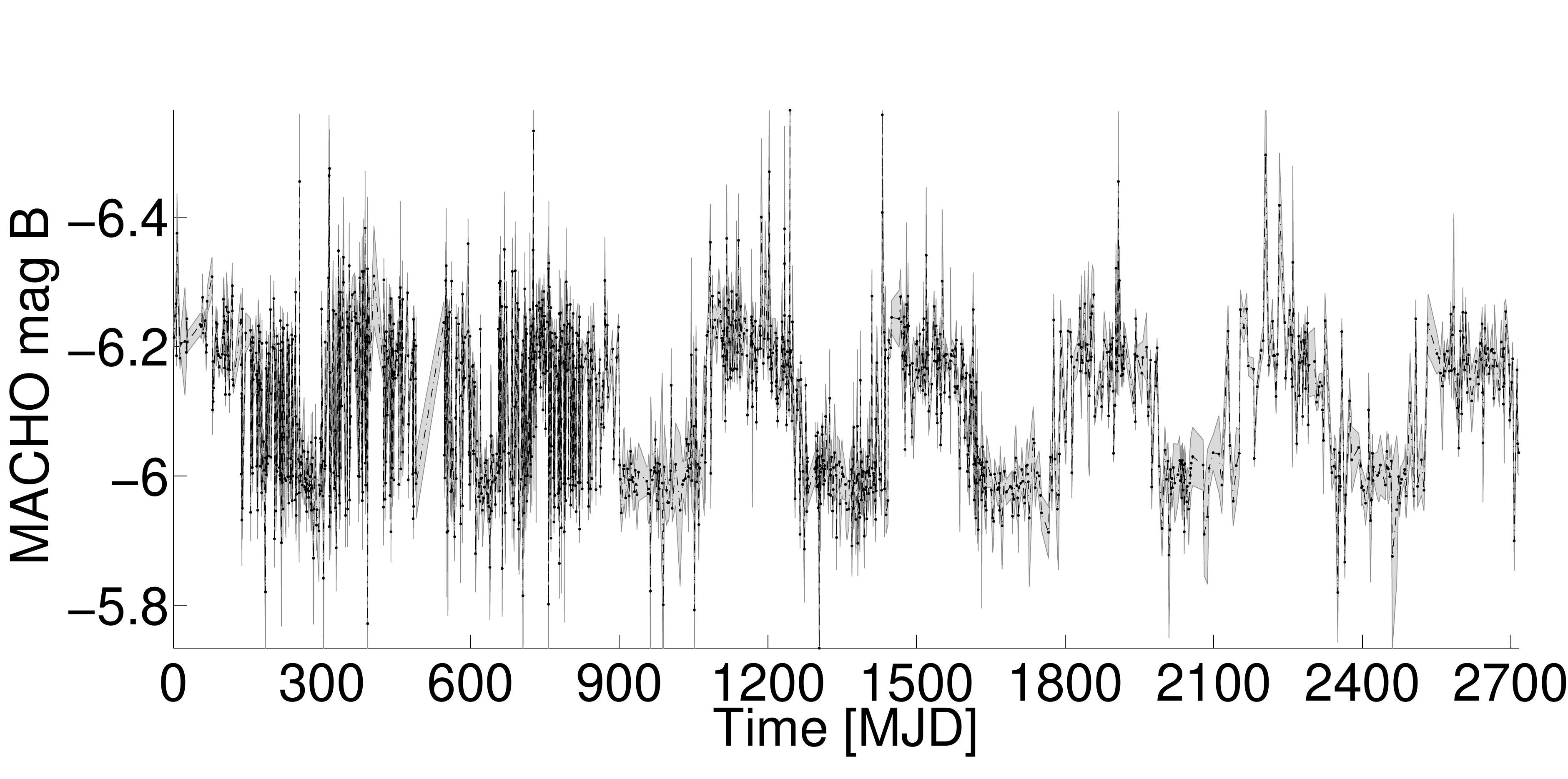}
  \includegraphics[width=.4\linewidth]{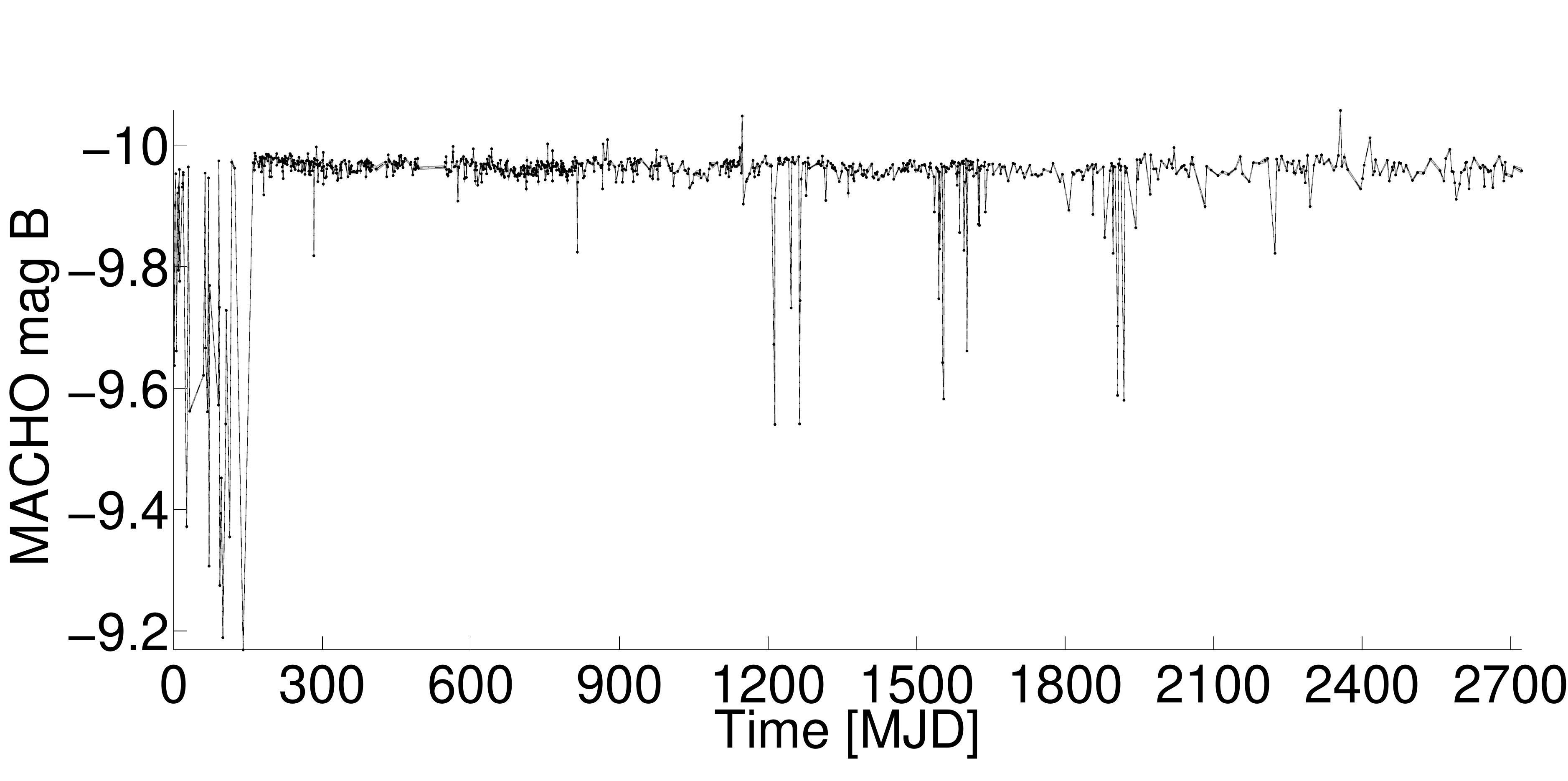}
  \includegraphics[width=.4\linewidth]{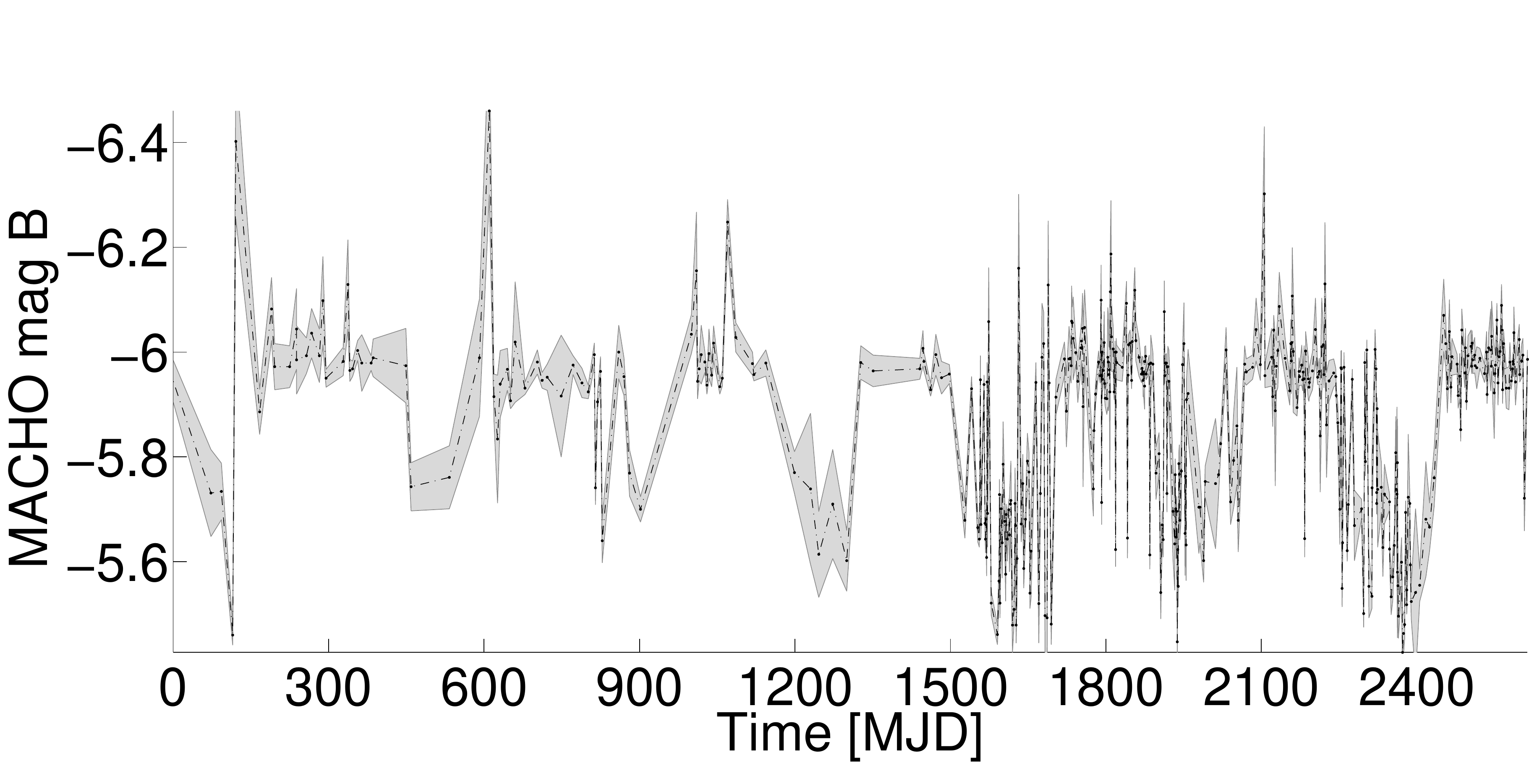}
  \caption{\label{fig:artifacts} Top left panel one day period artifact MACHO\_77.7187.271, top right panel one day period artifact MACHO\_79.4780.358, bottom left panel sampling artifact MACHO\_5.5010.986 and bottom right panel 370 days period MACHO\_49.5899.715.}
\end{minipage}%
\end{figure*}




\subsubsection{Removal of spurious outliers}
Figure~\ref{fig:artifacts} shows some of the outliers we obtained from this first iteration. The top left and right outliers in Figure~\ref{fig:artifacts} are characterized by having one day period, while  the bottom right has a period of approximately a  year. This is probably caused by MACHO\rq s nightly and seasonal observational pattern and not by an intrinsic anomalous behavior. 
We also faced other kind of artifacts like the  outlier  in Figure~\ref{fig:artifacts} bottom left panel, which is obviously due to some instrumentation problems - this behavior at the beginning of the light-curve appeared in many light-curves.

In order to remove the spurious outliers we do the following steps:
\begin{enumerate}
\item Filter all outlier candidates that have periods very close to sidereal day or a year. There is no doubt that those light-curves exhibit strange behavior due to variable seeing conditions during the night or seasonal aliases. 
\item We run the whole analysis in the MACHO red non standard bandpass. MACHO was observed in two bandpasses simultaneously and therefore there are corresponding red-band light-curves for each object. For every outlier candidate that is not in the top 20,000 list of the equivalent list in the red candidate list, we consider it as an artifact/spurious and therefore it is removed from the candidate list. 
\item We visually inspect all candidates and group those that are obviously spurious, like the  examples in Figure~\ref{fig:artifacts}, into groups of similar shapes and behaviors. We add these new classes to the training set,  re-train and then we predict outliers again as explained above. 

\item Repeat previous steps until finding no artifacts on the top outlier list. 
\end{enumerate}


We expect that once we filter the artifacts, the `true' outliers will be the only ones remaining.

\section{Post analysis}
\label{sec:post}

As a first step, we visually inspected all the candidates starting from the top of the list (``strongest'' outliers) and moving our way to the ``weakest'' outliers.   We determined that about 4,000 candidates  was a good number of candidates to start, since  candidates beyond this point either were  not showing any significant variation or had low signal-to-noise ratio (SNR), and therefore not interesting. 
  
 As a second step, we cross-matched our candidates with other astronomical catalogs of known types, or catalogs with additional contextual information.  Some of these catalogs are  collections of known types; for example, LMC Long Period Variables \citep{Fraser2008}, is a collection of long period variables from LMC. On the other hand, catalogs like XMM-Newton  \citet{Watson2009}  contain X-ray information, which can be useful to further understand the nature of the candidates. Having additional information for some of the outlier candidates could be helpful to identify the nature  of these objects. Table~\ref{table:xmatchcat} summarizes all the catalogs used in the analysis and the resulting  cross-matched numbers ($N_{{\rm x-matched}}$).

The fact that some of the outlier candidates appear in catalogs of known objects, as it is shown in table  ~\ref{table:xmatchcat}, could be explained by the following reasons:


 \begin{enumerate} 
 \item Known classes with a small number of objects were not included in our original training set (i.e. Cataclysmic variables, R Coronae Borealis, etc.) . Since these are rare classes we were expecting to find more objects of their kind.
 
 
 
  
  \item The objects in these catalogs were mislabeled or incorrectly classified. Many of these catalogs are guided by algorithms or done automatically, so unavoidably they contain errors. Even when humans are involved in the classification, biases  are always  present.  These \textquotedblleft errors" should present themselves as outliers in our final analysis.  
  Indeed, 45 of our outliers that were labeled as eclipsing binaries, Cepheids RR Lyrae in other catalogs, do not have  the characteristics or the light curve shapes of these classes and therefore were flagged as outliers. 
    
 \item The features considered in this work and the features used by the other catalogs are not consistent.  
 For example,  the period of \verb+MACHO_77.7428.190+ is 906.3559 days, while in  \cite{Soszynski2008}  is  0.2843359 days. Because of this,  this light-curve does not appear to be an RRL in our model  and therefore it is identified as an outlier.  It is  known that uncertainties in features could result in low quality classification and consequently erronous outlier predictions. Dealing with feature uncertainties  is a topic of future work.  
  
 \item The SNR of the light-curves is survey dependent and therefore features that  depend on the actual amplitude of the variability vary from catalog to catalog. For example, if a catalog is compiled using a survey that is more sensitive than  our survey,  the fainter objects are indistinguishable from the non-variables in our database even if it is a true known variable. Moreover, as described above, low signal-to-noise ratio light-curves  have uncertain features and therefore higher probability of being false-positive. 
 
 \end{enumerate}
  
Most  of these reasons can be attributed to the lack of a perfect training set and high quality features.  
 Because our method is based on a supervised classification,  the results heavily depend on the choice of these representative objects. In the ideal scenario, one would compile a training set that contains every possible known objects with high quality features. In our case, we started with a trustworthy training set that was missing some of the known but rare types. This served as a blind test, since some of these types were never presented to the method, never trained with them and therefore they should have been discovered by our method. As expected, we recovered most of these objects in the candidate list.


\begin{table*}[]

\caption{\label{table:xmatchcat}Catalogs used  for post-analysis.}  
\begin{minipage}{6cm}
\begin{tabular}{|c|c|c|c|}
\hline
  Catalog & Reference  & Number of objects in catalog& $N_{{\rm x-matched}}$  \\  
  \hline 
  LMC LPVs from MACHO & \citet{Fraser2008} & 56,453 & 52\footnote{52 were types 0, 9 or no types in paper} \\
  XMM-Newton  & \citet{Watson2009} & 262,902 & 13\\
  ROSAT All-Sky Bright Source Catalogue (1RXS) & \citet{Voges1999} & 18,806 & 2 \\
LMC Blue variable stars from MACHO & \citet{Keller2002} &1280 &  91 \\
OGLE eclipsing binaries in LMC &  \citet{Observatory2003} &2720  & 29 \\
OGLE RR Lyrae in LMC &  \citet{Alamos2003} &7661 &  8 \\
LMC Cepheids in OGLE and MACHO data & \citet{Poleski2009} &2946 & 8 \\
OGLE+2MASS+DENIS LPV in Magellanic Clouds & \citet{Groenewegen2004} & 2919  & 9 \\ 
Variable Stars in the Large Magellanic Clouds  & \citet{Alcock2004}  &21474 & 334 \\
Machine-learned ASAS Classification Cat. (MACC) &  \citet{Richards2012} &50124  &5 \\
QSO Candidates in the MACHO LMC database & \citet{Kim2012} &2566 &51 \\
EROS Periodic Variable Candidates &  \cite{Kim2014} & 150,115 & 432  \\
Type II and anomalous Cepheids in LMC &\citet{Soszynski2008} &286& 19 \\
OGLE Variables in Magellanic Clouds & \citet{Ita2004} &8852 & 134 \\
GCVS, Vol. V.: Extragalactic Variable Stars & \cite{artyukhina1996gcvs} &10979 & 74 \\
High proper-motion stars from MACHO astrometry & \cite{Alcock2001} & 154 &0\\
\hline
\hline
\end{tabular}
\end{minipage}
\end{table*}

\begin{figure*}[]
        \centering
                     \includegraphics[width=\textwidth]{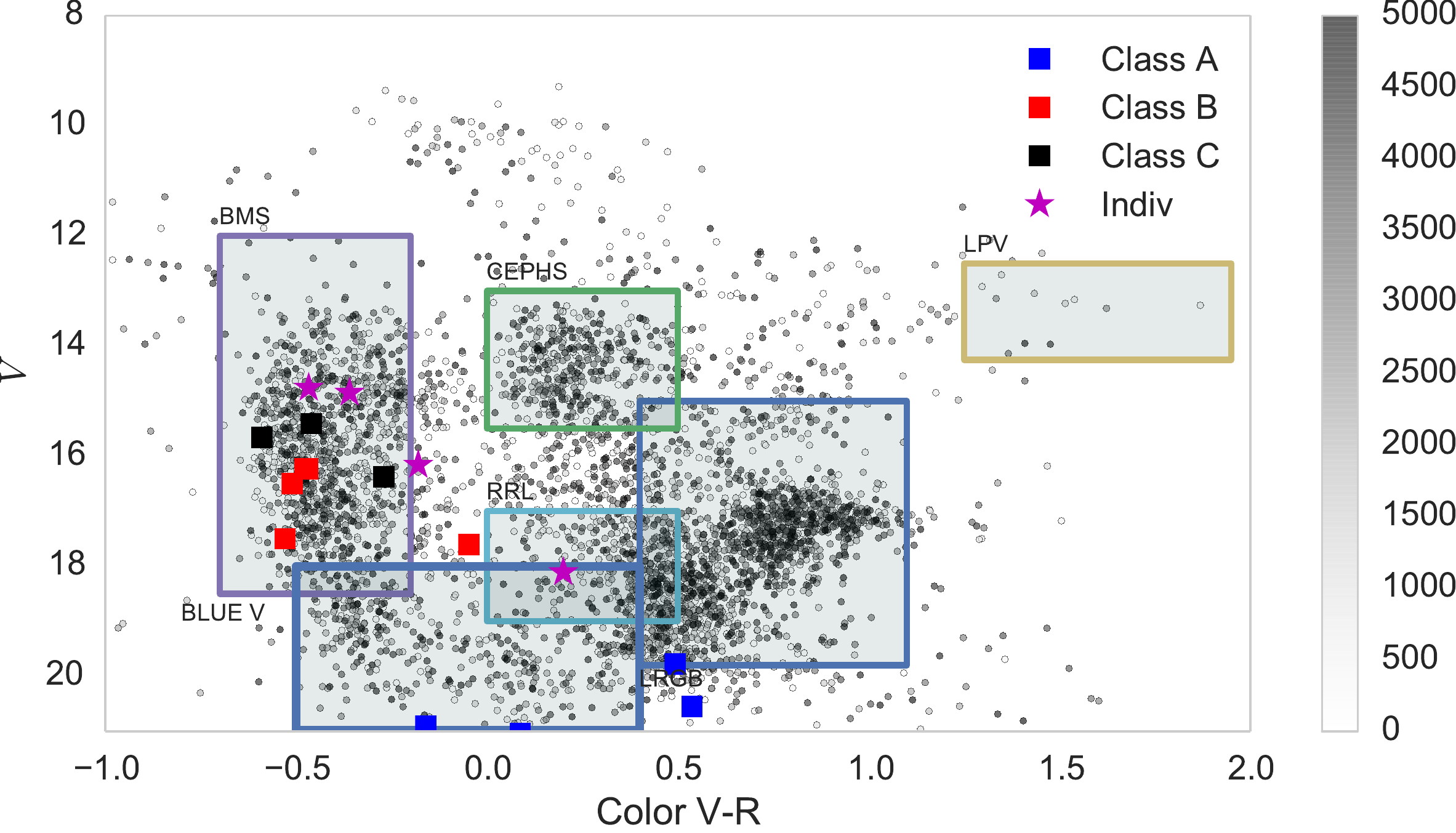}
                \caption{      \label{fig:CMD} Color magnitude diagram of all the outliers. The outlier rank is indicated by the color of each data point. The bluer, the higher the outlier score. Black boxes mark the location of  blue main sequence (BMS),   lower red giant branch (LRGB), long period variables (LPV), RR Lyrae (LLR) and Cepheid (CEPH).}
\end{figure*}

As a third step, we examined the color magnitude diagram (CMD) of the candidate list and identified regions where objects were most likely from a known type. One of the advantages of the LMC, is that all stellar populations are at essentially the same distance and thus we can use CMDs as an additional way to separate and identify the sources. Figure~\ref{fig:CMD} shows the CMD for the outliers.

As a fourth step, we grouped the outliers into sets  based on the morphology of the light-curves. Here we present the most interesting subgroups, some of which are known but rare classes, while others do not obviously belong to any known class of objects.


\begin{enumerate}
\item \underline{Eclipsing Cepheid:}
Eclipsing Cepheids have  been discussed  in  papers of  the MACHO, OGLE and EROS-2 surveys \citep{Alcock2002,Marconi2013,Cassisi2011}. These objects are Cepheids in binary systems where there are flux drops during the pulsating cycle caused by the transit of a companion star.  Although it is known that 50\% of Galactic Cepheids are in binary systems, only about 20 such Cepheids are known in the LMC, which is mainly due to their faint magnitudes caused by the distance to the LMC. Recently, \cite{Pietrzynski2010} have used such a system to  limit the distance uncertainty  to the LMC, so finding such systems is very valuable for precision cosmology. By simply looking through our catalog of outliers, we found few objects of this kind. Figure~\ref{fig:EBCeph} shows one of these examples. 

\begin{figure}[h]
        \centering
                     \includegraphics[width=0.45\textwidth]{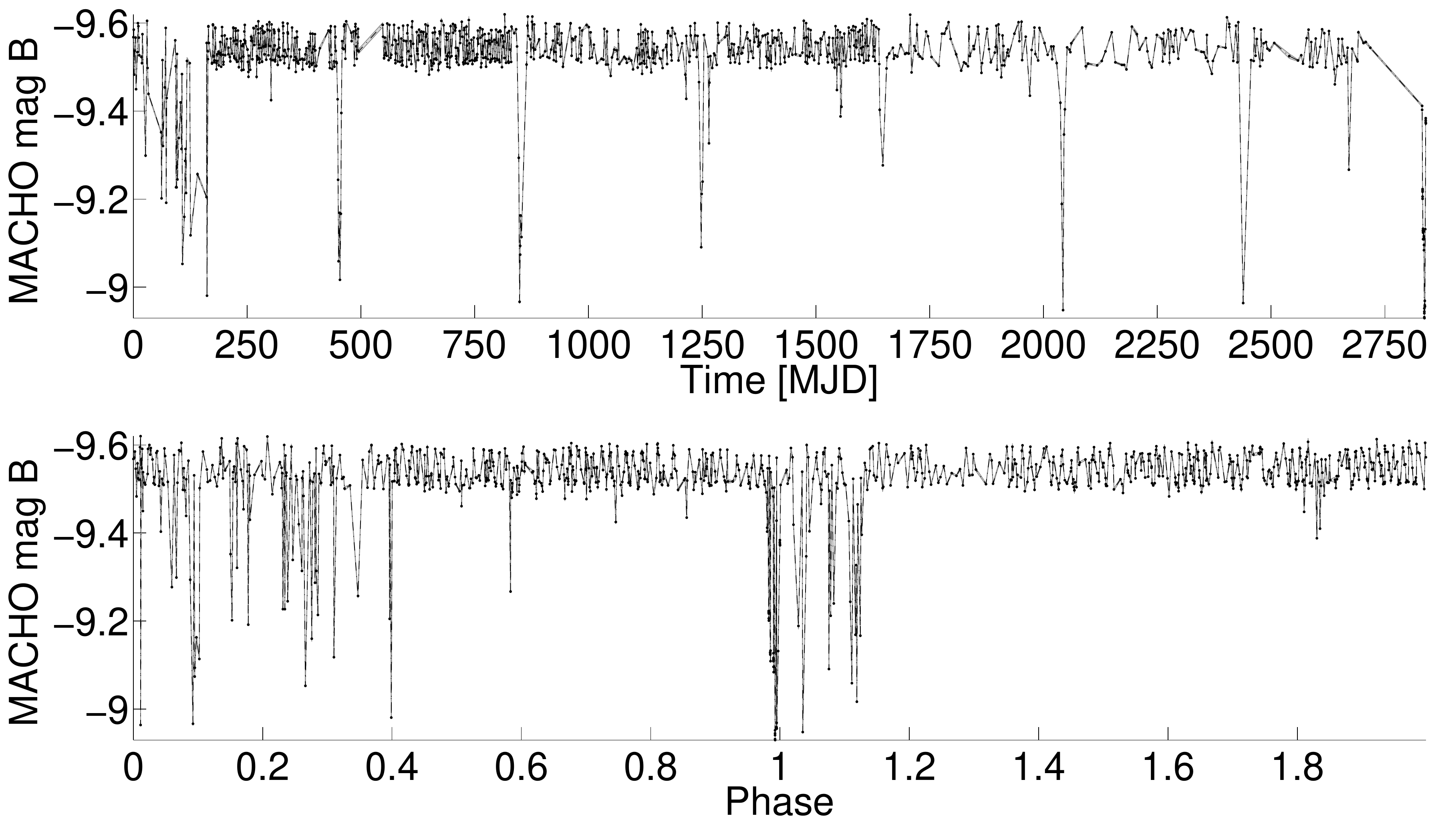}
                \caption{\label{fig:EBCeph} Top panel Eclipsing Cepheid MACHO\_6.6454.5 and bottom panel its folded light-curve.}
\end{figure}

\item \underline {Cataclysmic Variables (CV):} 
Another interesting group of outliers are CVs or novae or novae-like looking objects. Because there are no unified variability characteristics, this group was not included in the training set and therefore few CVs are in our candidate list. These objects  can increase more than 20 magnitudes, becoming approximately $10^8$ times brighter. Novae and Recurrent Novae are close binary systems that are variable due to explosions on their surfaces. The eruptions can last from a few days to almost a year, and can be quasi-periodic as the recurrent Novae \citep{Schaefer2010,Knigge2011}. This a subject of an extensive research,  and recently  the interests focused on  superluminous SNe \citep{Quimby2011}.
 Figure~\ref{fig:nova} shows \verb+MACHO_77.7546.2744+, one of this class example, where the change in magnitude is 2.5 and the relaxation time is about a year.   
Our candidate list contains a few dozen of these objects, nevertheless, some of them are already known, such as those presented in \cite{Shafter2013}.

\begin{figure}[h]
        \centering
                      \includegraphics[width=0.45\textwidth]{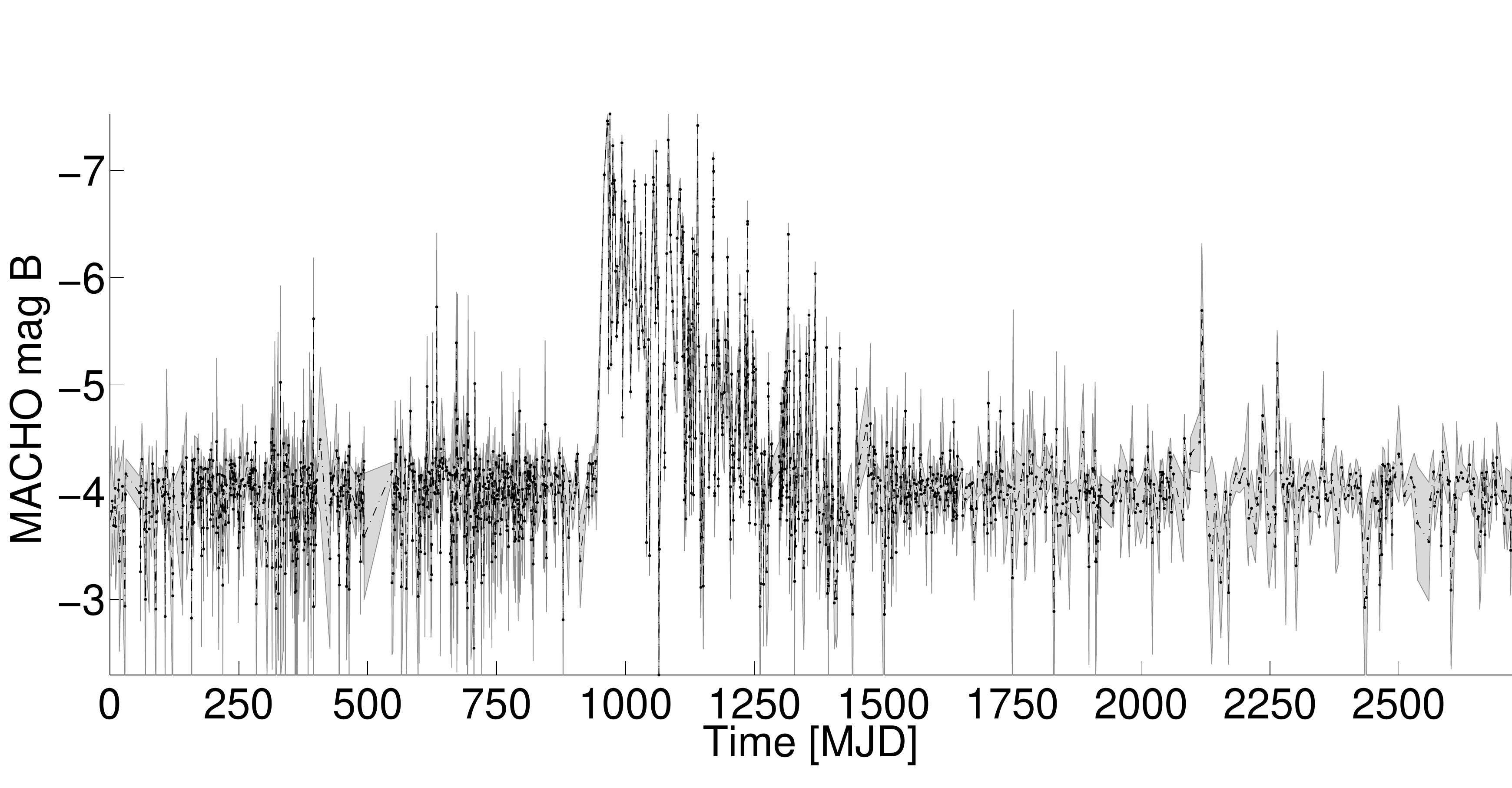}
                \caption{      \label{fig:nova} Nova like variable MACHO\_77.7546.2744.}
\end{figure}

\item \underline {Blue Variables:} 
The class coined Blue Variables is a generic class without a unified light-curve morphology or features. Because of this, we did not include such a class in the training set. \cite{Keller2002a}  proposed that the variability of these stars is the result of processes related to the establishment, maintenance and dissipation of the Be disk. The emission that characterizes Be stars originates in a gaseous circumstellar quasi-Keplerian disk.  
These objects appear to be blue and are simply variable. Sixty-eight of our candidates fall into this category. An example of such light-curve is shown in Figure~\ref{fig:BlueVariable} and the locations of all the members in the CMD are shown in Figure~\ref{fig:CMD}.  

\begin{figure}[h!]
        \centering
                      \includegraphics[width=0.45\textwidth]{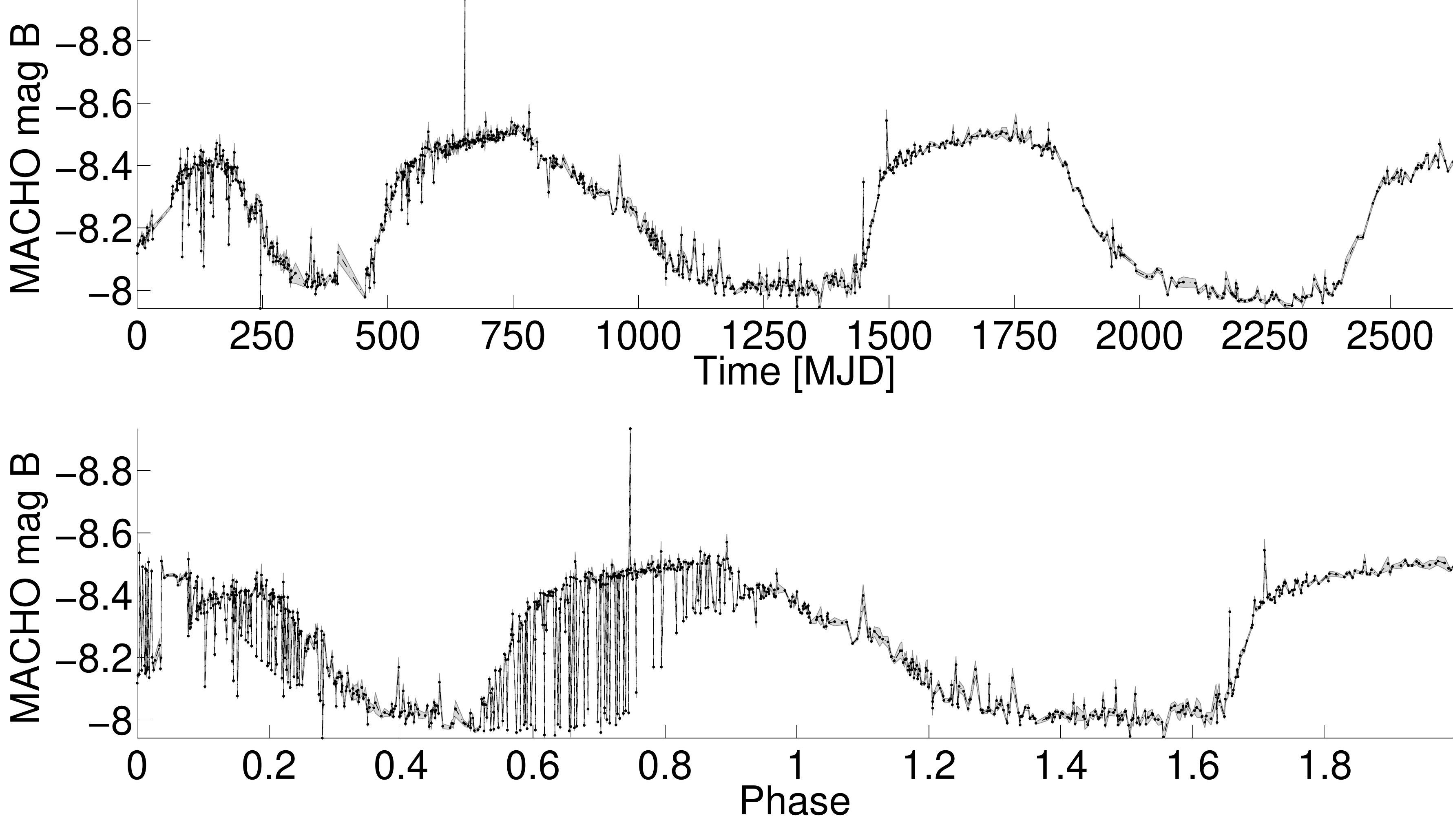}
                \caption{      \label{fig:BlueVariable} Blue variable MACHO\_81.9727.662.}
\end{figure}
\item \underline{ X-ray Sources:}
There are two sources cross-matched with the ROSAT all-sky survey bright source catalog
\citep{Voges1999AA...349..389V} 
 and 13 with the second XMM-Newton serendipitous source catalog 
 \citep{Watson2009AA...493..339W}.  Among these X-ray sources, 
 \verb+MACHO_61.9045.32+ is a confirmed high-mass X-ray binary \citep{Liu2005AA...442.1135L}
 hosting a radio pulsar \citep{Ridley2013MNRAS.433..138R}
 but the other 14 counterparts are not carefully studied for their X-ray origins. 
 These remaining objects are interesting sources 
 since they show strong optical variability, either periodic or non-periodic 
 and X-ray emission simultaneously. They could be either W UMa-type contact binaries, 
 X-ray binaries, or other types of X-ray emitters (e.g. see 
 \citealt{Ness2002AA...387.1032N, Chen2006AJ....131..990C, Liu2007AA...469..807L}
 and references therein). Particularly, X-ray binaries are most interesting sources 
 since they are known to host either neutron stars or black holes 
 (i.e. accretor) together with a companion star. Their X-ray emission is caused by 
 accreting material falling from the companion star into the accretor 
 \citep{Heuvel1992AA...262...97V, Done2007AARv..15....1D}.
 Thus studying X-ray binaries help us to understand the process of accretion and 
 the fundamental physics of the binaries such as mass, radius, orbit, jets, etc (e.g. see 
 \citealt{Klis2000ARAA..38..717V, Fender2004MNRAS.355.1105F}). Figure~\ref{fig:XMM} shows one representative example.
 \begin{figure}[h]
        \centering
                      \includegraphics[width=0.45\textwidth]{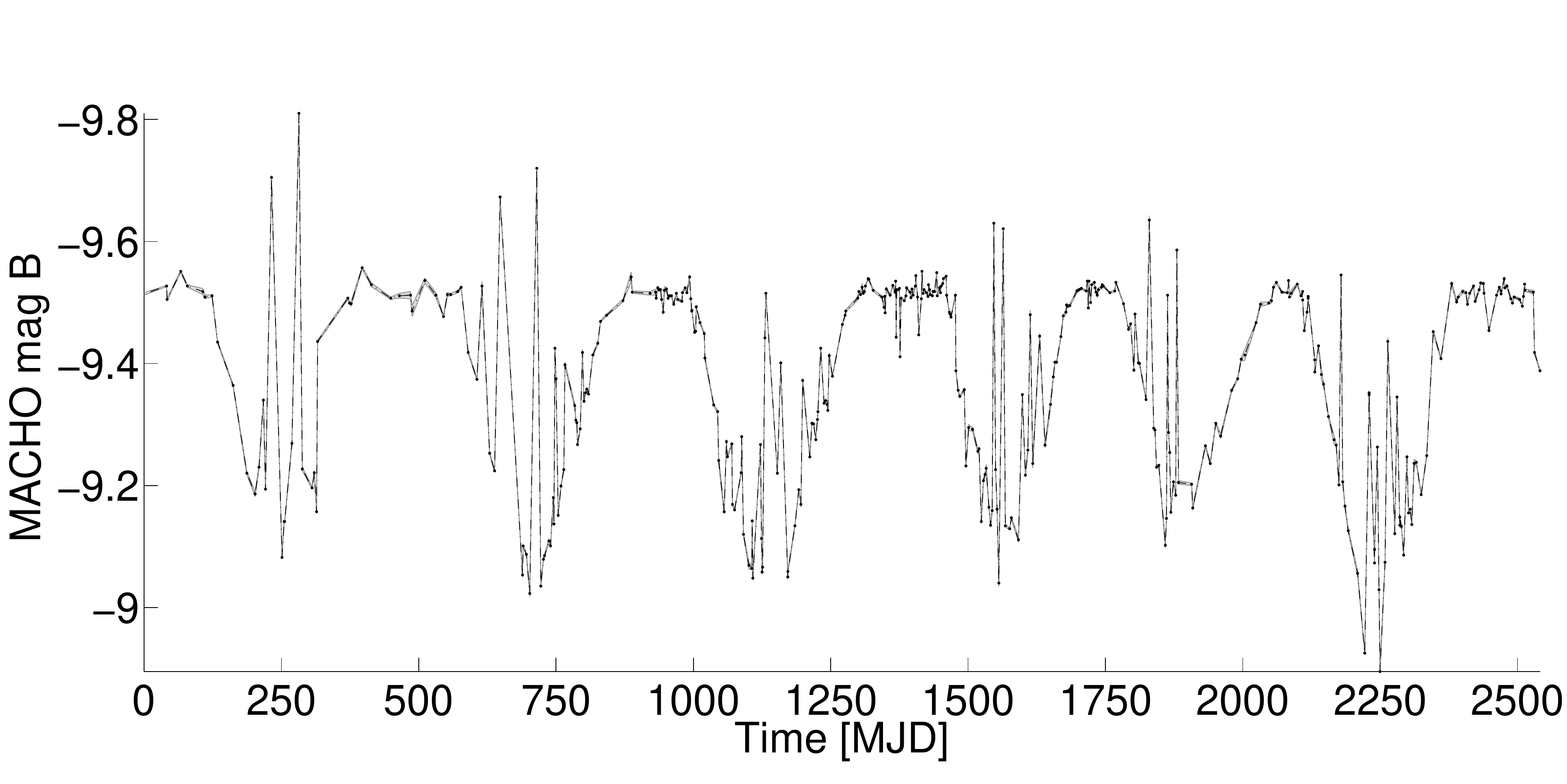}
                \caption{      \label{fig:XMM} X-Ray binary MACHO\_61.9045.32.}
\end{figure}
  
 \item \underline{ R Coronae Borealis:}
  Within our outliers we identified one object belonging to one of the most rare and interesting classes among the variable stars. \verb+MACHO_6.6696.60+ is a R Coronae Borealis (RCB) star. These kinds of objects are yellow supergiant stars whose atmospheres are carbon rich and extremely hydrogen deficient. This causes irregular intervals of dust-formation episodes that result in a drop in brightness of up to 8 magnitudes in a short period\citep{Clayton1996}. An example of this type of light-curve is shown in Figure~\ref{fig:RCB}.
 
  \begin{figure}[h]
        \centering
                      \includegraphics[width=0.45\textwidth]{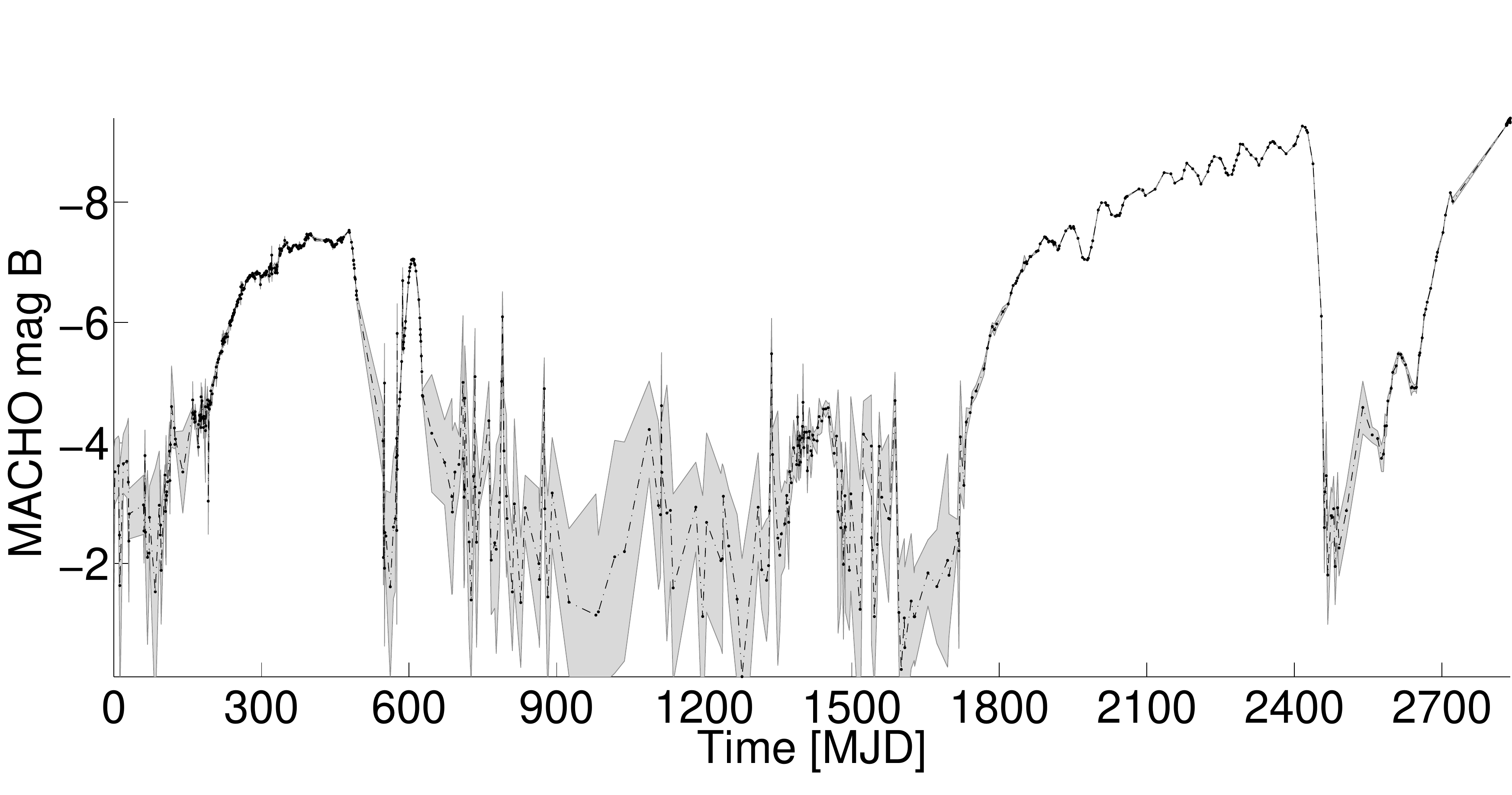}
                \caption{      \label{fig:RCB} R Coronae Borealis MACHO\_6.6696.60.}
\end{figure}
 

\item \underline{ {\bf The OTHERS:}}
Undoubtedly, there are  many variable classes and it is out of the scope of this work to analyze and comment on every outlier from our list. Our goal was to find novel objects that have not been identified before. For this end, we first ran a clustering algorithm on all the candidates and then visually inspected all the light-curves that are not in the categories mentioned above, identified a few classes of objects and a few individual objects that could not be assigned to known classes. 
We show three classes and four individual outliers in Table~\ref{table:theothers} ,  Figures~\ref{fig:test1}, \ref{fig:test2}, \ref{fig:test3}, \ref{fig:test4}, and also in the CMD in Figure~\ref{fig:CMD}.

Nevertheless, we had to perform a more specific analysis for outliers in Class A. We noticed that the objects belonging to this class are neighbors (they are located in the same field, number 82), and therefore it is  likely that the perturbation on the light-curves was caused by a high proper motion star moving close to these sources.
In order to confirm or reject this hypothesis, we calculated the distance between these objects and the time differences of the peaks of the variation. The time difference of the variation was on average of 400 days, but the objects were $\sim$100 arcsec  apart. Since typical proper motions are less than few arsec/year, the hypothesis was  rejected. Objects of Class A are consequently good candidates to conform a new variability class.

%
%

\end{enumerate}


\begin{table*}[]
\centering
\label{table:theothers}
\caption{The others: examples of new variability classes and individual outliers.}
\begin{tabular}{|c|c|c|c|c|c|c|c|c|}

\hline
\hline
 Class & MACHO id &  RA  & Dec & Period [days] & V &  R & Color & SNR  \\  
 \hline 
 Class A&82.8887.471 &5.59031 &-69.2956 &657.19& 19.78 &19.28 &0.49 &1.53 \\
 Class A&82.9009.834 &5.59633 &-69.2722 &525.75&20.25 &19.71&0.536&2.38 \\
 Class A&82.9009.1850 &5.59655 &-69.2762 &525.75&21.04 &20.96 &0.08 &1.66  \\
Class A&82.8887.2395 &5.59106 &-69.2954 &876.25&21.04 &21.20 &-0.16&1.59  \\
\hline
Class B&56.5178.29&5.19911&-66.5471&363.00&16.50&17.02&-0.51&4.44\\
Class B&44.1616.257&4.84559&-70.0673&871.32&16.23&16.704&-0.47&3.84\\
Class B&35.7272.13&5.42992&-72.127&374.98&16.23&16.70&-0.47&5.63\\
Class B&48.2864.67&4.96026&-67.5326&872.94&17.00&17.53&-0.52&2.98\\
Class B&82.8284.126&5.51805&-69.202&438.12&17.56&17.61&-0.04&2.34\\
\hline
Class C&17.2711.26&	4.9556&	-69.6723&	680.70&	15.41&	15.87&	-0.46&9.14\\
Class C&82.8283.41&	5.5218&	-69.2594&	525.75&	15.07&	15.66&	-0.59&8.53\\
Class C&62.7361.30&	5.4249&	-66.2181&	848.30&	16.38&	16.65&	-0.27&5.44\\
\hline
Individual Outlier&13.5835.11	&5.2742&	-71.0974&	296.98&	14.85&	15.21&	-0.36&	51.52\\
Individual Outlier&18.2478.9&	4.9342&	-69.0323&	226.90&	14.76&	15.23&	-0.46&	36.69\\
Individual Outlier &78.6462.561&	5.3366&	-69.6743&	678.95&	18.11&	17.91&	0.20	&7.02\\
Individual Outlier&62.7241.19&	5.4114&	-66.1581&	636.23&	16.16&	16.34&	-0.18&	52.51\\
\hline
\end{tabular}
\end{table*}


\begin{figure*}
\centering
\begin{minipage}{\textwidth}
  \centering
 \includegraphics[width=.4\linewidth]{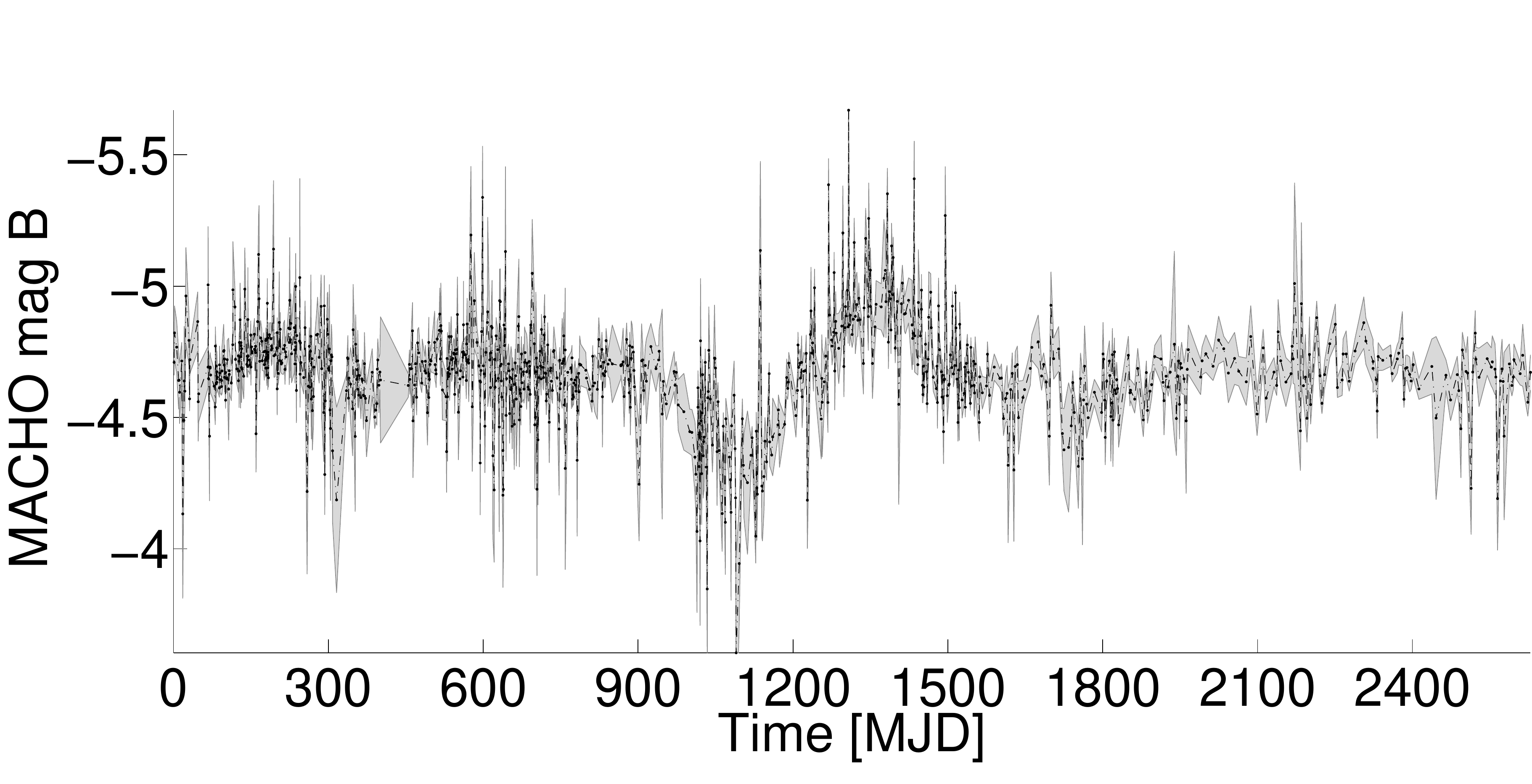}
  \includegraphics[width=.4\linewidth]{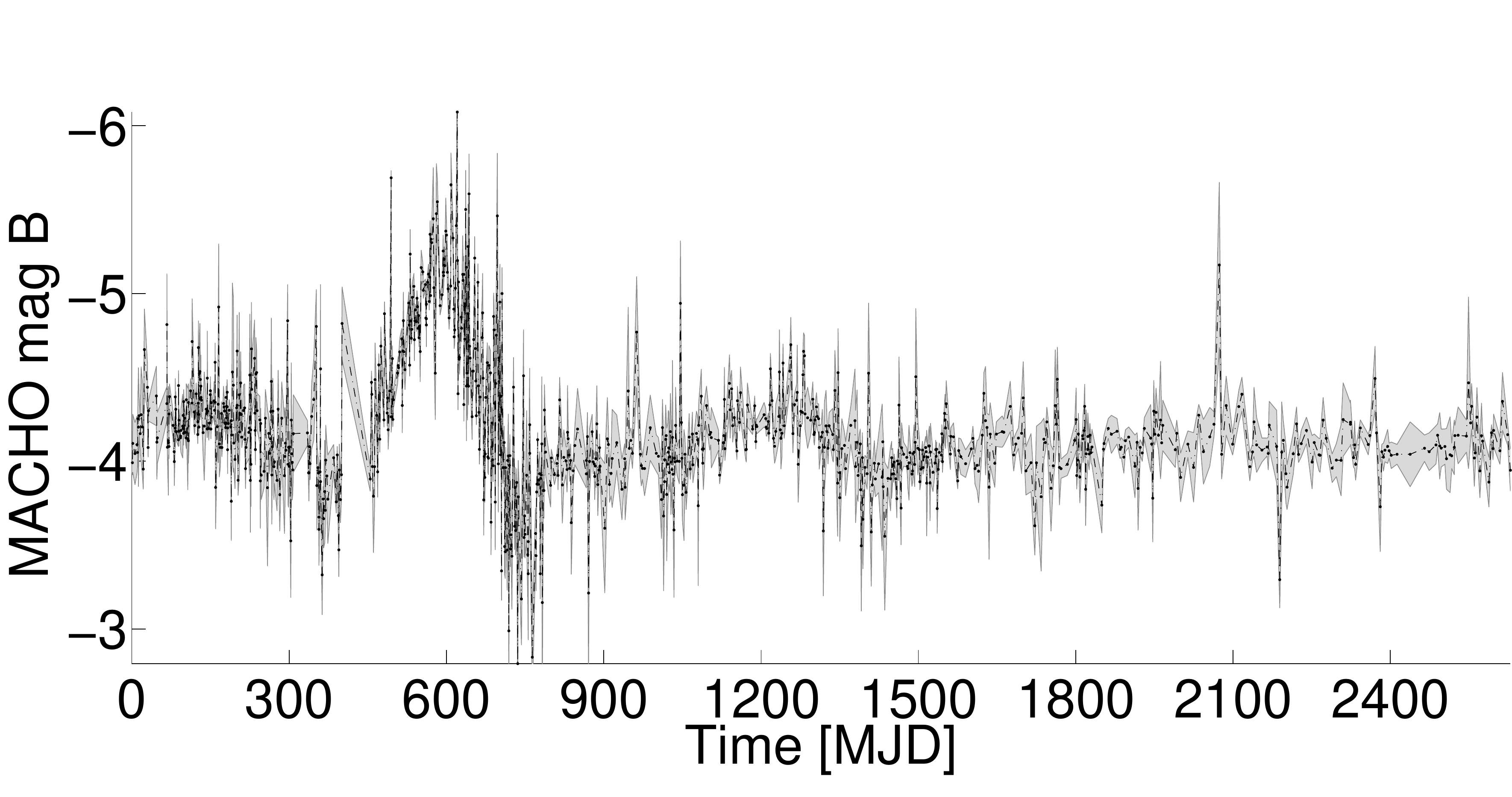}
  \includegraphics[width=.4\linewidth]{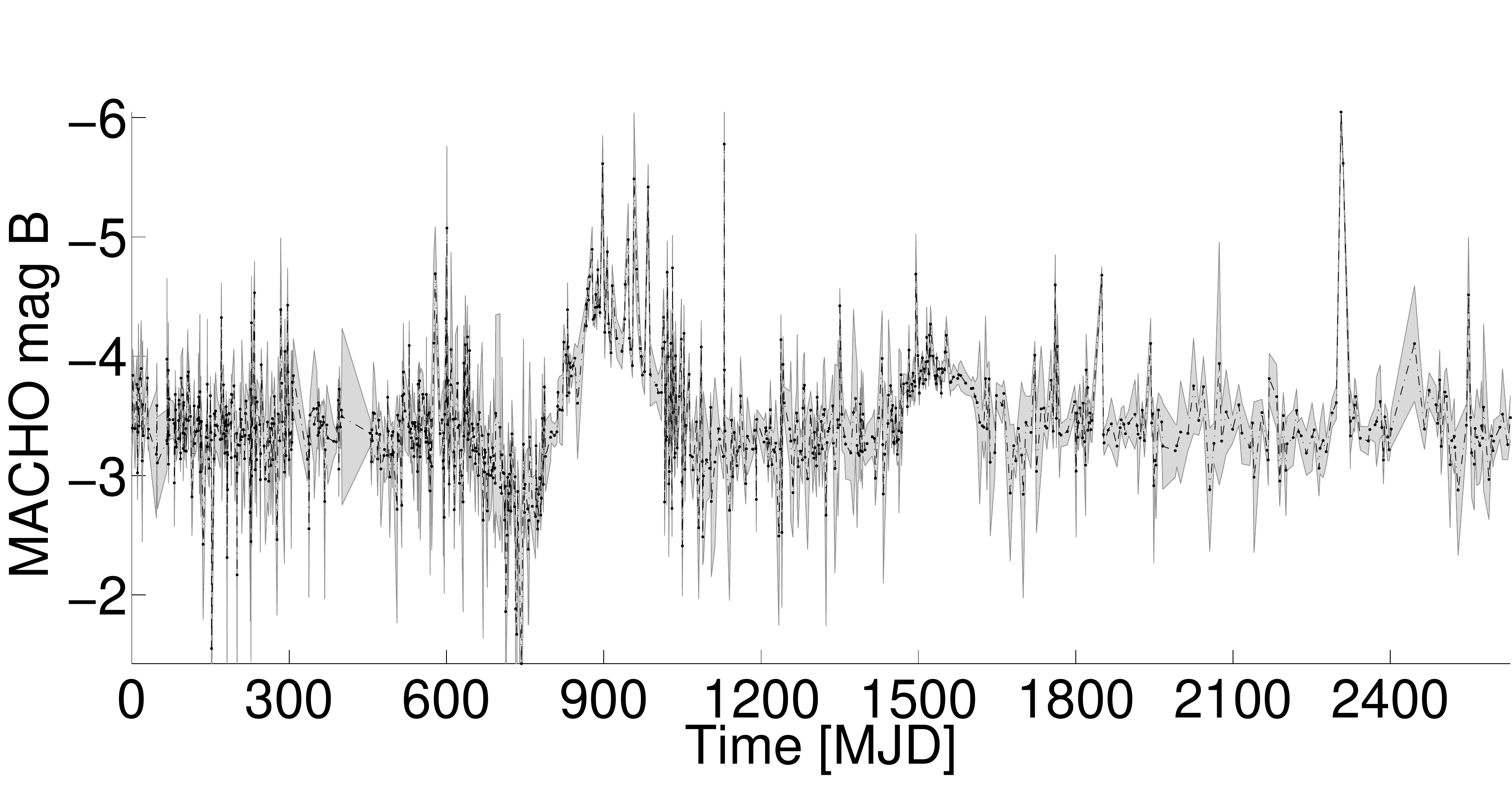}
  \includegraphics[width=.4\linewidth]{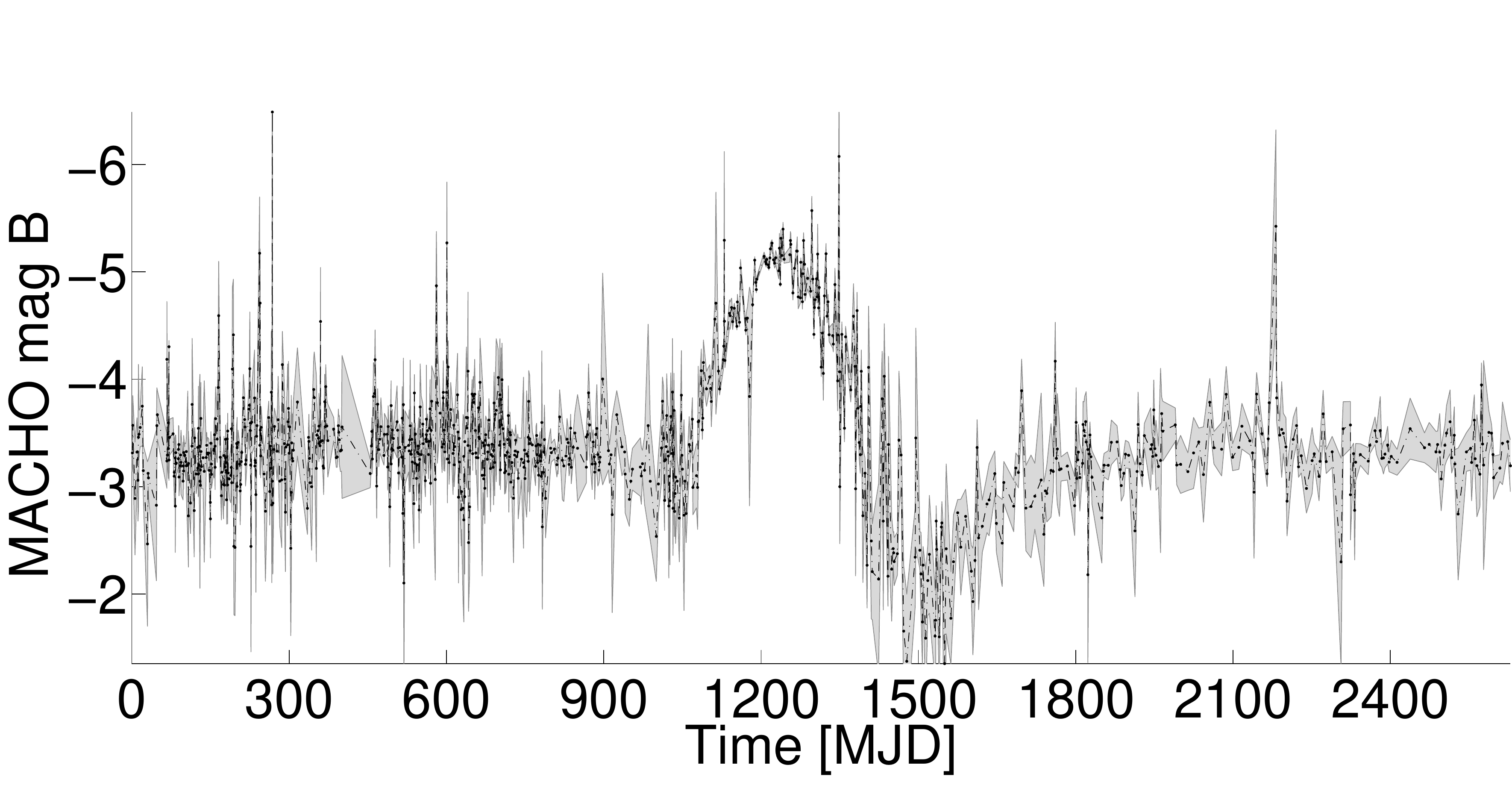}
  \caption{\label{fig:test1} Top left panel Class\_A MACHO\_82.8887.471, top right panel Class\_A MACHO\_82.9009.834, bottom left panel Class\_A MACHO\_82.9009.1850 and bottom right panel Class\_A MACHO\_82.8887.2395.}
\end{minipage}%
 \vspace{-.3cm}
\begin{minipage}{\textwidth}
  \centering

  \label{fig:test2}
\end{minipage}
\begin{minipage}{\textwidth}
  \centering

  \includegraphics[width=.4\linewidth]{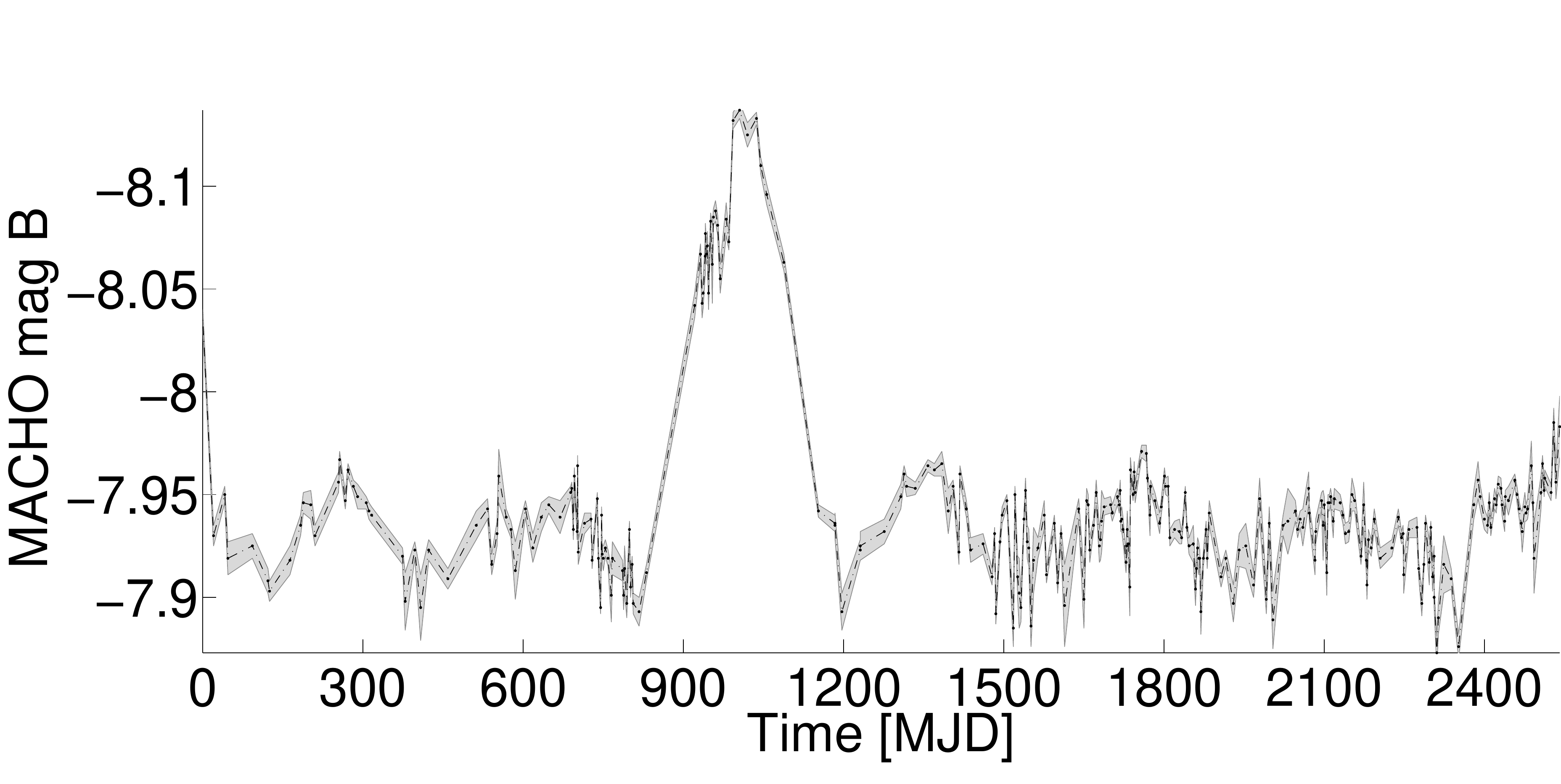}
  \includegraphics[width=.4\linewidth]{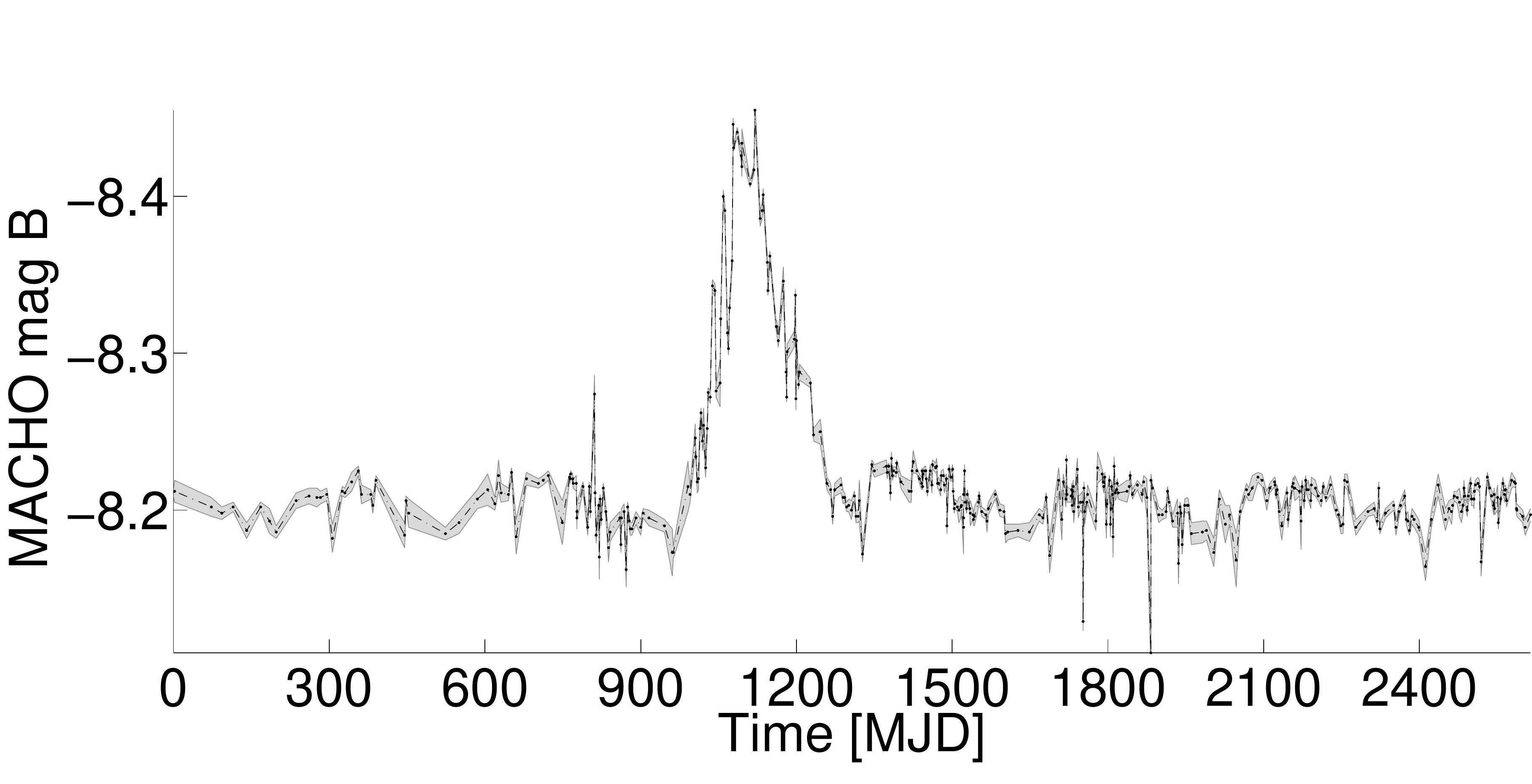}
  \caption{\label{fig:test2} Left panel Class\_B MACHO\_56.5178.29 and right panel Class\_B MACHO\_44.1616.257. }
\end{minipage}
 \vspace{-.3cm}
\begin{minipage}{\textwidth}
  \centering
  \includegraphics[width=.4\linewidth]{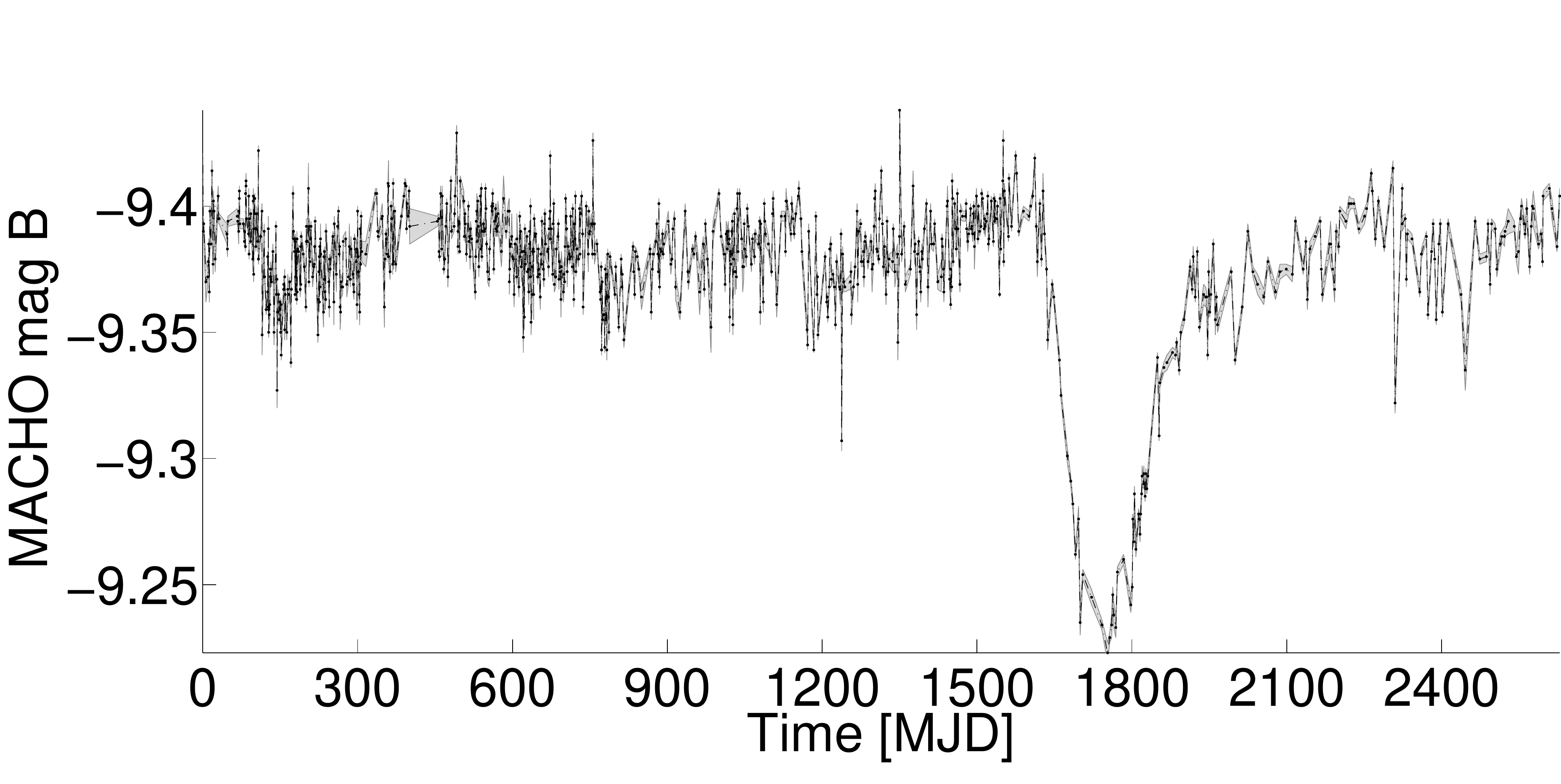}
  \includegraphics[width=.4\linewidth]{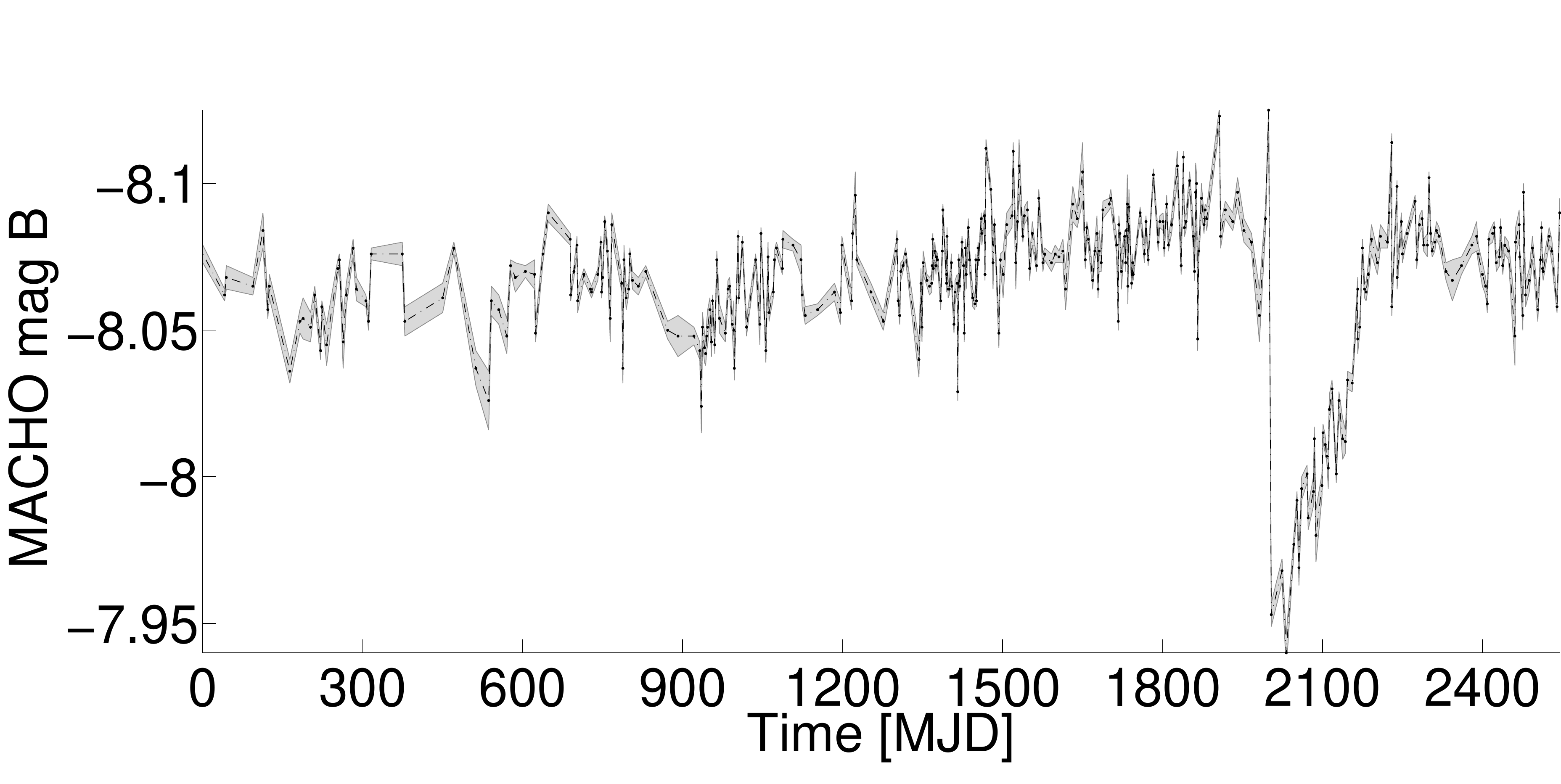}
   \caption{\label{fig:test3} Left panel Class\_C MACHO\_82.8283.41 and right panel Class\_C MACHO\_62.7361.30. }

\end{minipage}

   \vspace{-.2cm}
\begin{minipage}{\textwidth}
  \centering
 \includegraphics[width=.4\linewidth,height=3cm]{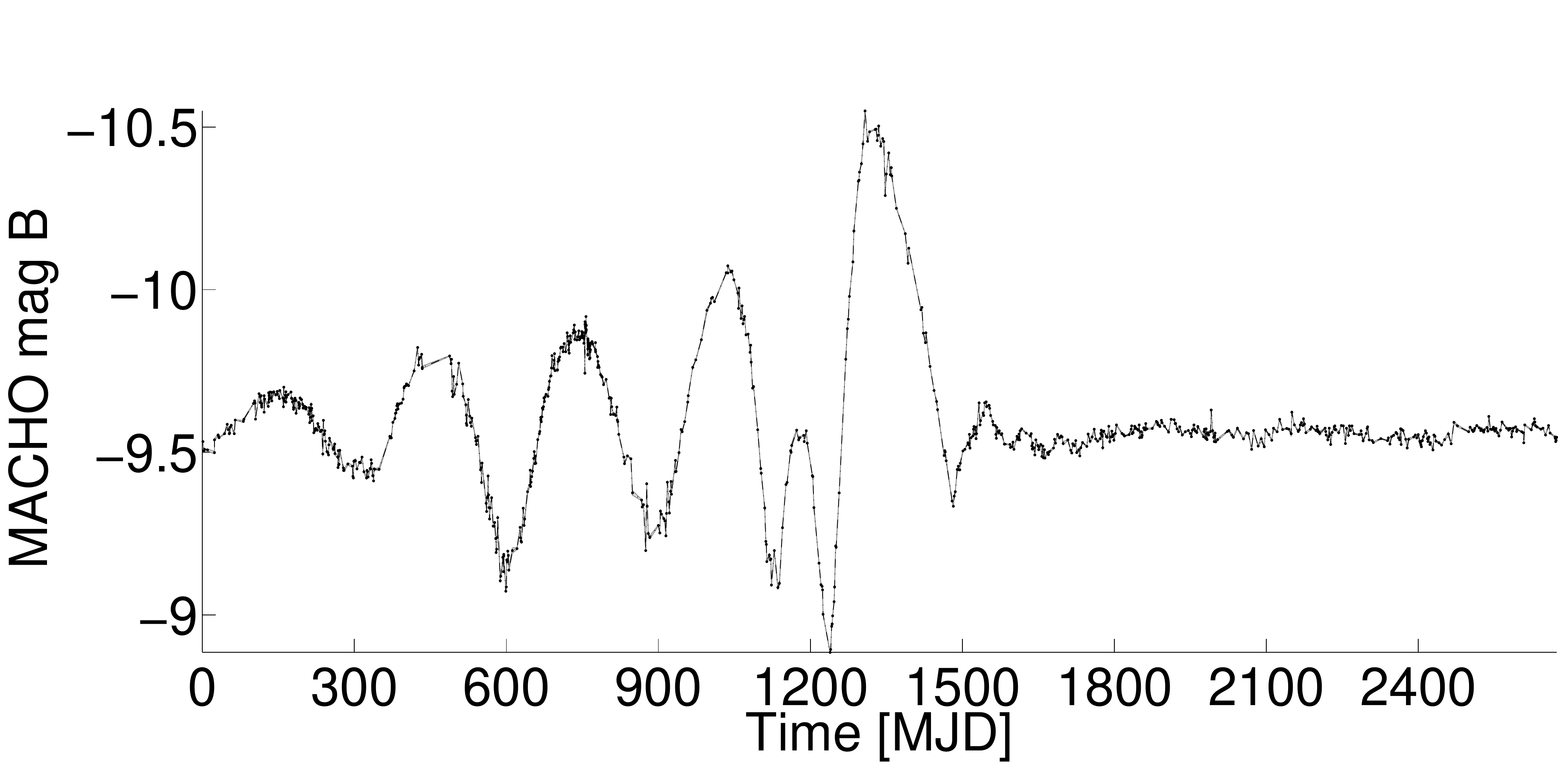}
    \includegraphics[width=.4\linewidth,height=3cm]{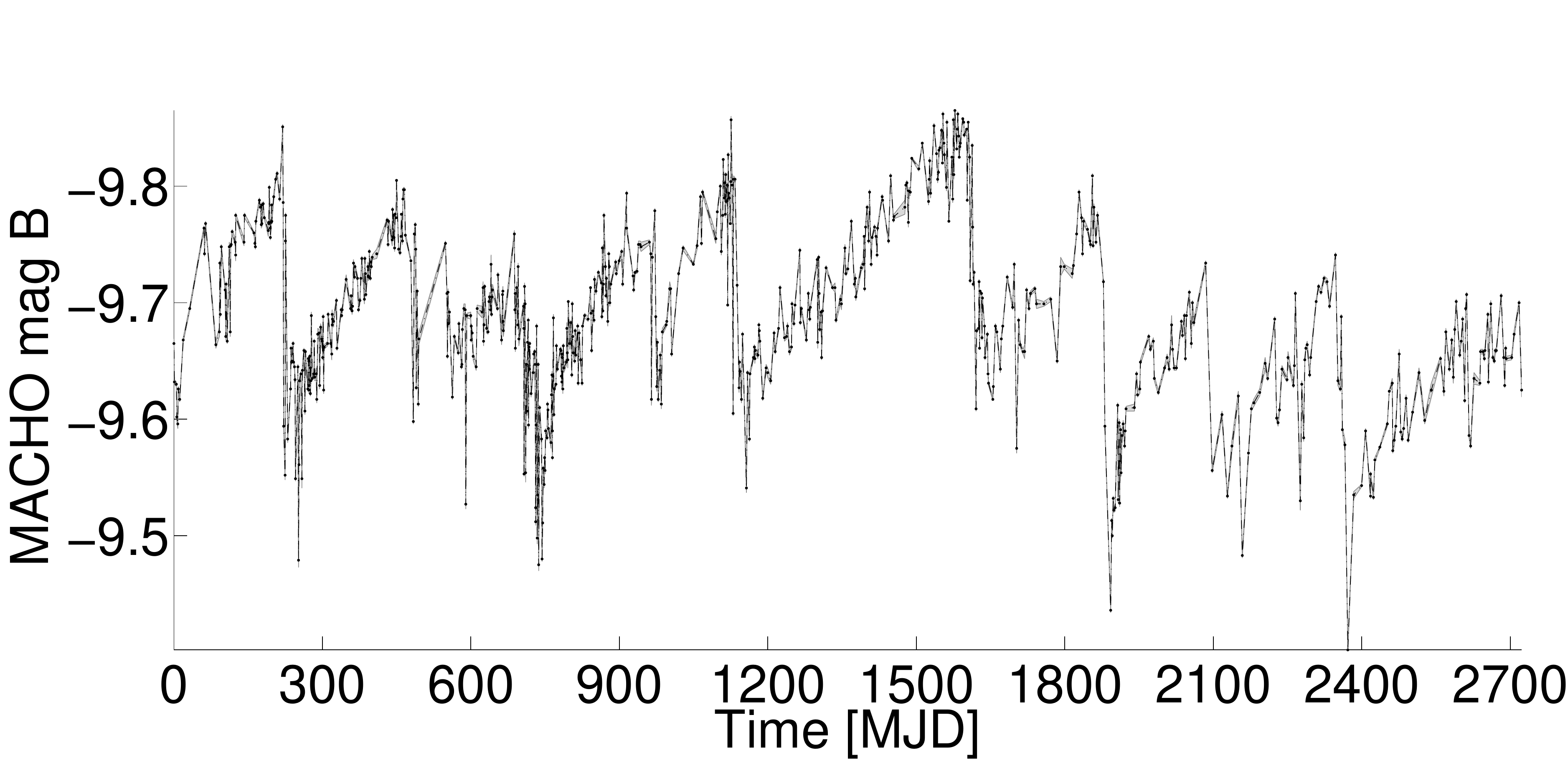}
   \includegraphics[width=.4\linewidth,height=3cm]{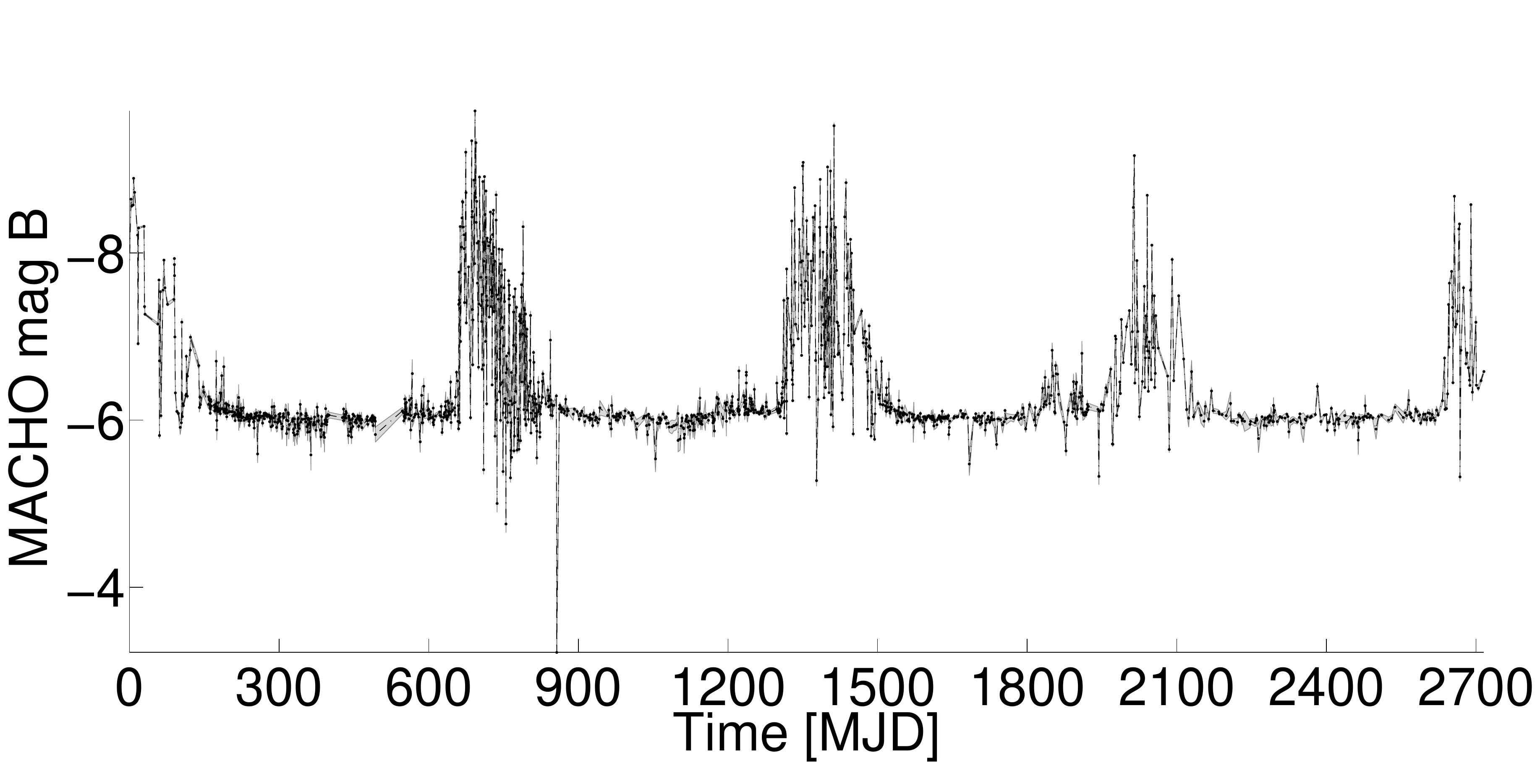}
    \includegraphics[width=.4\linewidth,height=3cm]{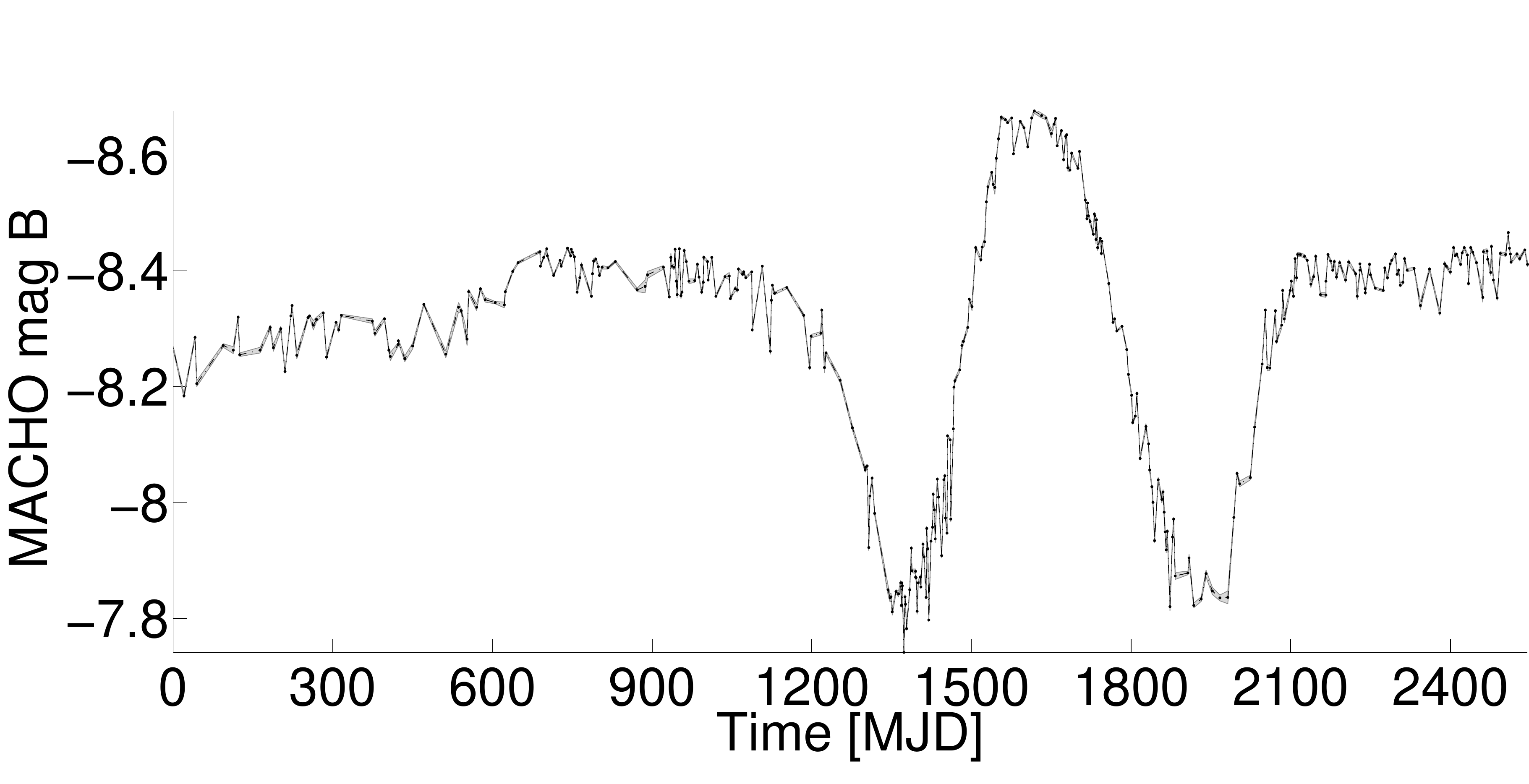}
 \caption{\label{fig:test4} Top left panel Outlier MACHO\_13.5835.11, top right panel Outlier MACHO\_18.2478.9, bottom left panel Outlier MACHO\_78.6462.561 and bottom right panel Outlier MACHO\_62.7241.19.}

\end{minipage}
\end{figure*}



\section{Conclusions}
\label{sec:conclusions}
The generation of precise, large and complete sky surveys in the last years has increased the need of developing automated analysis tools to process this tremendous amount of data. These tools should help astronomers to classify stars, characterize objects and detect anomaly among other applications.  In this paper we presented an algorithm based on a supervised classifier mechanism that enables us to discover outliers in catalogs of light-curves. To do so we trained a random forest classifier and used a Bayesian Network to obtain the joint probability distribution, which was used for our outlierness score. Different from existing methods, our work comprises a supervised algorithm where all the available information is used to our advantage.

Since the amount of data to be processed is huge, one could have expected a high computational complexity and the overtake of the resources. Nevertheless, our algorithm is only expensive in the training stage and extremely fast in the unknown light-curves analysis, allowing us to explore a very large datasets. Furthermore, our method is not only restricted to astronomical problems and could be applied to any data base where anomaly detection is necessary.

The results from the application of our work on catalogs of classified periodic stars from MACHO project are encouraging, and establish that our method correctly identifies light-curves that do not belong to these catalogs as outliers.

We have identified light-curves that were artifacts because of instrumental, mechanical, electronic or human errors and about \NumOutliers{} light-curves that emerged as intrinsic. After cross-matching these candidates with the available catalogs we found known but rare objects among our outliers and also objects that did not have previously information. By performing a clustering we classified some of them as new variability classes and others as intriguing unique outliers. As future work these objects will be followed up using spectroscopy, in order to characterize them and identify them with new observations. We hope that by doing this analysis we would be able to find more of these objects and turn our isolated outliers into new known variability classes.

On the other hand, we are planning to improve our algorithm in the future by creating new robust features and by constructing a more complete and large training set. Furthermore,  we aim to apply our algorithm to different large sky surveys as EROS \citep{Ansari2004}, Pan-Starrs \citep{Hodapp2004} and when finished LSST \citep{Tyson}. 

Finally, in order to help astronomers, we are planning a full release of a software which will include feature calculation of the light-curves and the application of our algorithm as a downloadable software and as an on-line tool and web services in the near future.

\section{acknowledgments}
We thank M. Stockle and D. Acu\~na  for helpful comments and advices. The analysis in this paper has been done using the Odyssey cluster offered by the FAS Research Computing Group at Harvard.  This work is supported by Vicerrector\'ia de Investigaci\'on(VRI) from Pontificia Universidad Cat\'olica de Chile and by the Ministry of Economy, Development, and Tourism's Millennium Science Initiative through grant
IC\,12009, awarded to The Millennium Institute of Astrophysics.

\clearpage
\bibliographystyle{apj}
\bibliography{Paper}

\begin{thebibliography}{}
\expandafter\ifx\csname natexlab\endcsname\relax\def\natexlab#1{#1}\fi

\bibitem[{Agarwal(2005)}]{Agarwal2005}
Agarwal, D. 2005, in Proceedings of the 5th {IEEE} International Conference on
  Data Mining. {IEEE} Computer Society, 26--33

\bibitem[{Aggarwal \& Yu(2001)}]{Aggarwal2001}
Aggarwal, C.~C., \& Yu, P.~S. 2001, {ACM SIGMOD Record}, 30, 37

\bibitem[{Alcock {et~al.}(2001)Alcock, Allsman, Alves, Axelrod, Becker, \&
  Bennett}]{Alcock2001}
Alcock, C., Allsman, R.~A., Alves, D.~R., {et~al.} 2001, The Astrophysical
  \ldots, 20

\bibitem[{{Alcock} {et~al.}(1996){Alcock}, {Allsman}, {Axelrod}, {Bennett},
  {Cook}, {Freeman}, {Griest}, {Marshall}, {Peterson}, {Pratt}, {Quinn},
  {Rodgers}, {Stubbs}, {Sutherland}, \& {Welch}}]{Alcock1996AJ}
{Alcock}, C., {Allsman}, R.~A., {Axelrod}, T.~S., {et~al.} 1996, The
  Astronomical Journal, 111, 1146

\bibitem[{{Alcock} {et~al.}(1997{\natexlab{a}}){Alcock}, {Allsman}, {Alves},
  {Axelrod}, {Becker}, {Bennett}, {Cook}, {Freeman}, {Griest}, {Keane},
  {Lehner}, {Marshall}, {Minniti}, {Peterson}, {Pratt}, {Quinn}, {Rodgers},
  {Stubbs}, {Sutherland}, {Tomaney}, {Vandehei}, \&
  {Welch}}]{Alcock1997ApJ...491L..11A}
{Alcock}, C., {Allsman}, R.~A., {Alves}, D., {et~al.} 1997{\natexlab{a}}, The
  Astrophysical Journal Letters, 491, L11

\bibitem[{{Alcock} {et~al.}(1997{\natexlab{b}}){Alcock}, {Allsman}, {Alves},
  {Axelrod}, {Bennett}, {Cook}, {Freeman}, {Griest}, {Guern}, {Lehner},
  {Marshall}, {Park}, {Perlmutter}, {Peterson}, {Pratt}, {Quinn}, {Rodgers},
  {Stubbs}, \& {Sutherland}}]{Alcock1997ApJ...479..119A}
---. 1997{\natexlab{b}}, The Astrophysical Journal, 479, 119

\bibitem[{{Alcock} {et~al.}(1997{\natexlab{c}}){Alcock}, {Allsman}, {Alves},
  {Axelrod}, {Becker}, {Bennett}, {Cook}, {Freeman}, {Griest}, {Guern},
  {Lehner}, {Marshall}, {Peterson}, {Pratt}, {Quinn}, {Rodgers}, {Stubbs},
  {Sutherland}, {Welch}, \& {MACHO Collaboration}}]{Alcock1997ApJ...486..697A}
---. 1997{\natexlab{c}}, The Astrophysical Journal, 486, 697

\bibitem[{{Alcock} {et~al.}(1997{\natexlab{d}}){Alcock}, {Allsman}, {Alves},
  {Axelrod}, {Becker}, {Bennett}, {Cook}, {Freeman}, {Griest}, {Guern},
  {Lehner}, {Marshall}, {Minniti}, {Peterson}, {Pratt}, {Quinn}, {Rodgers},
  {Sutherland}, \& {Welch}}]{Alcock1997ApJ...482...89A}
---. 1997{\natexlab{d}}, The Astrophysical Journal, 482, 89

\bibitem[{{Alcock} {et~al.}(1997{\natexlab{e}}){Alcock}, {Allsman}, {Alves},
  {Axelrod}, {Becker}, {Bennett}, {Cook}, {Freeman}, {Griest}, {Lacy},
  {Lehner}, {Marshall}, {Minniti}, {Peterson}, {Pratt}, {Quinn}, {Rodgers},
  {Stubbs}, {Sutherland}, \& {Welch}}]{Alcock1997AJ}
---. 1997{\natexlab{e}}, The Astronomical Journal, 114, 326

\bibitem[{{Alcock} {et~al.}(1999){Alcock}, {Allsman}, {Alves}, {Axelrod},
  {Becker}, {Bennett}, {Bersier}, {Cook}, {Freeman}, {Griest}, {Guern},
  {Lehner}, {Marshall}, {Minniti}, {Peterson}, {Pratt}, {Quinn}, {Rodgers},
  {Stubbs}, {Sutherland}, {Tomaney}, {Vandehei}, \&
  {Welch}}]{Alcock1999AJ....117..920A}
{Alcock}, C., {Allsman}, R.~A., {Alves}, D.~R., {et~al.} 1999, The Astronomical
  Journal, 117, 920

\bibitem[{{Alcock} {et~al.}(2002){Alcock}, {Allsman}, {Alves}, {Becker},
  {Bennett}, {Cook}, {Drake}, {Freeman}, {Griest}, {Hawley}, {Keller},
  {Lehner}, {Lepischak}, {Marshall}, {Minniti}, {Nelson}, {Peterson},
  {Popowski}, {Pratt}, {Quinn}, {Rodgers}, {Suntzeff}, {Sutherland},
  {Vandehei}, \& {Welch}}]{Alcock2002}
---. 2002, The Astrophysical Journal, 573, 338

\bibitem[{Alcock {et~al.}(2004)Alcock, Alves, Axelrod, Becker, Bennett,
  Clement, Cook, Drake, Freeman, Geha, Griest, Lehner, Marshall, Minniti,
  Muzzin, Nelson, Peterson, Popowski, Quinn, Rodgers, Rowe, Sutherland,
  Vandehei, \& Welch}]{Alcock2004}
Alcock, C., Alves, D.~R., Axelrod, T.~S., {et~al.} 2004, The Astronomical
  Journal, 127, 334

\bibitem[{Ansari(2004)}]{Ansari2004}
Ansari, R. 2004, in International Conference on Cosmic Rays and Dark Matter,
  1--9

\bibitem[{Arning {et~al.}(1996)Arning, Agrawal, \& Raghavan}]{Arning1996}
Arning, A., Agrawal, R., \& Raghavan, P. 1996, in {Proceedings of the ACM
  SIGKDD International Conference on Knowledge Discovery and Data Mining},
  164--169

\bibitem[{Artyukhina {et~al.}(1996)Artyukhina, Durlevich, Frolov, Goranskij,
  Gorynya, Karitskaya, Kazarovets, Kholopov, Kireeva, Kurochkin,
  {et~al.}}]{artyukhina1996gcvs}
Artyukhina, N., Durlevich, O., Frolov, M., {et~al.} 1996, VizieR Online Data
  Catalog, 2205, 0

\bibitem[{Bernard {et~al.}(2008)Bernard, Heutte, \& Adam}]{Bernard2008}
Bernard, S., Heutte, L., \& Adam, S. 2008, in Advanced Intelligent Computing
  Theories and Applications. With Aspects of Artificial Intelligence
  (Springer), 430--437

\bibitem[{Bhattacharyya {et~al.}(2012)Bhattacharyya, Richards, Rice, Starr,
  Butler, \& Bloom}]{Bhattacharyya}
Bhattacharyya, S., Richards, J.~W., Rice, J., {et~al.} 2012, 1

\bibitem[{Bishop(1994)}]{Bishop1994}
Bishop, C. 1994, in Proceedings of IEE Conference on Vision, Image and Signal
  Processing., 217--222

\bibitem[{{Blanco} \& {Heathcote}(1986)}]{Blanco1986PASP...98..635B}
{Blanco}, V.~M., \& {Heathcote}, S. 1986, PASP, 98, 635

\bibitem[{Breiman(2001)}]{Breiman2001}
Breiman, L. 2001, Machine learning, 45, 5

\bibitem[{Breunig {et~al.}(2000)Breunig, Kriegel, Ng, \& Sander}]{Breunig2000}
Breunig, M., Kriegel, H., Ng, R., \& Sander, J. 2000, in {Proceedings of 2000
  ACM SIGMOD International Conference on Management of Data}, 93--104

\bibitem[{{Cassisi} \& {Salaris}(2011)}]{Cassisi2011}
{Cassisi}, S., \& {Salaris}, M. 2011, The Astrophysical Journal Letters, 728,
  L43

\bibitem[{Chandola {et~al.}(2009)Chandola, Banerjee, \& Kumar}]{Chandola2009}
Chandola, V., Banerjee, A., \& Kumar, V. 2009, {ACM Computing Surveys}, 41, 1

\bibitem[{{Chen} {et~al.}(2006){Chen}, {Sanchawala}, \&
  {Chiu}}]{Chen2006AJ....131..990C}
{Chen}, W.~P., {Sanchawala}, K., \& {Chiu}, M.~C. 2006, The Astronomical
  Journal, 131, 990

\bibitem[{Clayton(1996)}]{Clayton1996}
Clayton, G.~C. 1996, Publications of the Astronomical Society of the Pacific,
  225

\bibitem[{Cooper \& Herskovits(1992)}]{Cooper:Herskovits:1992}
Cooper, G., \& Herskovits, E. 1992, Machine Learning, 9, 309

\bibitem[{{Dobrzycki} {et~al.}(2002){Dobrzycki}, {Groot}, {Macri}, \&
  {Stanek}}]{Dobrzycki2002ApJ...569L..15D}
{Dobrzycki}, A., {Groot}, P.~J., {Macri}, L.~M., \& {Stanek}, K.~Z. 2002, The
  Astrophysical Journal Letters, 569, L15

\bibitem[{{Done} {et~al.}(2007){Done}, {Gierli{\'n}ski}, \&
  {Kubota}}]{Done2007AARv..15....1D}
{Done}, C., {Gierli{\'n}ski}, M., \& {Kubota}, A. 2007, The Astronomy and
  Astrophysics Review, 15, 1

\bibitem[{Eskin(2000)}]{Eskin2000}
Eskin, E. 2000, in {Proceedings of the Seventeenth International Conference on
  Machine Learning.}, 255--262

\bibitem[{{Fender} {et~al.}(2004){Fender}, {Belloni}, \&
  {Gallo}}]{Fender2004MNRAS.355.1105F}
{Fender}, R.~P., {Belloni}, T.~M., \& {Gallo}, E. 2004, Monthly Notices of the
  Royal Astronomical Society, 355, 1105

\bibitem[{Fraser {et~al.}(2008)Fraser, Hawley, \& Cook}]{Fraser2008}
Fraser, O.~J., Hawley, S.~L., \& Cook, K.~H. 2008, The Astronomical Journal,
  136, 1242

\bibitem[{{Geha} {et~al.}(2003){Geha}, {Alcock}, {Allsman}, {Alves}, {Axelrod},
  {Becker}, {Bennett}, {Cook}, {Drake}, {Freeman}, {Griest}, {Keller},
  {Lehner}, {Marshall}, {Minniti}, {Nelson}, {Peterson}, {Popowski}, {Pratt},
  {Quinn}, {Stubbs}, {Sutherland}, {Tomaney}, {Vandehei}, \&
  {Welch}}]{Geha2003AJ....125....1G}
{Geha}, M., {Alcock}, C., {Allsman}, R.~A., {et~al.} 2003, The Astronomical
  Journal, 125, 1

\bibitem[{Geurts {et~al.}(2006)Geurts, Ernst, \& Wehenkel}]{Geurts2006}
Geurts, P., Ernst, D., \& Wehenkel, L. 2006, Machine Learning, 63, 3

\bibitem[{Gibbons \& Matias(1998)}]{Gibbons1998}
Gibbons, P.~B., \& Matias, Y. 1998in , ACM, 331--342

\bibitem[{Groenewegen(2004)}]{Groenewegen2004}
Groenewegen, M. 2004, Astronomy and Astrophysics, 425, 595

\bibitem[{Grubb \& Frank(1969)}]{Grubb1969}
Grubb, \& Frank, E. 1969, Technometrics, 11, 1

\bibitem[{He \& Carbonell(2006)}]{He2006}
He, J., \& Carbonell, J. 2006

\bibitem[{Henrion {et~al.}(2013)Henrion, Hand, Gandy, \&
  Mortlock}]{Henrion2012}
Henrion, M., Hand, D.~J., Gandy, A., \& Mortlock, D.~J. 2013, Statistical
  Analysis and Data Mining, 6, 53

\bibitem[{Herschel(1857)}]{Herschel1857}
Herschel, J. 1857, {Outlines of astronomy} (Blanchard and Lea), 268

\bibitem[{Hodapp {et~al.}(2004)Hodapp, Kaiser, Aussel, Burgett, Chambers, Chun,
  Dombeck, Douglas, Hafner, Heasley, Hoblitt, Hude, Isani, Jedicke, Jewitt,
  Laux, Luppino, Lupton, Maberry, Magnier, Mannery, Monet, Morgan, Onaka,
  Price, Ryan, Siegmund, Szapudi, Tonry, Wainscoat, \& Waterson}]{Hodapp2004}
Hodapp, K.~W., Kaiser, N., Aussel, H., {et~al.} 2004, Astronomische
  Nachrichten, 325, 636

\bibitem[{Hodge \& Austin(2004)}]{HodgeAustin2004}
Hodge, V., \& Austin, J. 2004, Artificial Intelligence Review, 22, 85

\bibitem[{Ita {et~al.}(2004)Ita, Tanab\'{e}, Matsunaga, Nakajima, Nagashima,
  Nagayama, Kato, Kurita, Nagata, Sato, Tamura, Nakaya, \& Nakada}]{Ita2004}
Ita, Y., Tanab\'{e}, T., Matsunaga, N., {et~al.} 2004, Monthly Notices of the
  Royal Astronomical Society, 353, 705

\bibitem[{Jin {et~al.}(2001)Jin, Tung, \& Han}]{Jin2001}
Jin, W., Tung, A., \& Han, J. 2001, in {Proceedings of the seventh ACM SIGKDD
  International Conference on Knowledge discovery and data mining}, 293--298

\bibitem[{John(1995)}]{John1995}
John, G. 1995, in Proceedings of the first International Conference on
  Knowledge Discovery and Data Mining, 174--179

\bibitem[{Keller {et~al.}(2002)Keller, Bessell, Cook, Geha, \&
  Syphers}]{Keller2002}
Keller, S.~C., Bessell, M.~S., Cook, K.~H., Geha, M., \& Syphers, D. 2002, The
  Astronomical Journal, 124, 2039

\bibitem[{{Keller} \& {Wood}(2002)}]{Keller2002a}
{Keller}, S.~C., \& {Wood}, P.~R. 2002, The Astrophysical Journal, 578, 144

\bibitem[{Keller {et~al.}(2007)Keller, Schmidt, Bessell, Conroy, Francis,
  Granlund, Kowald, Oates, Martin-Jones, Preston, {et~al.}}]{Keller2007}
Keller, S.~C., Schmidt, B.~P., Bessell, M.~S., {et~al.} 2007, Publications of
  the Astronomical Society of Australia, 24, 1

\bibitem[{Kim {et~al.}(2014)Kim, Protopapas, Bailer-Jones, Byun, Chang,
  Marquette, \& Shin}]{Kim2014}
Kim, D.-W., Protopapas, P., Bailer-Jones, C., {et~al.} 2014, submitted to
  Astronomy and Astrophysics

\bibitem[{Kim {et~al.}(2011)Kim, Protopapas, Byun, Alcock, Khardon, \&
  Trichas}]{Kim2011}
Kim, D.-W., Protopapas, P., Byun, Y.-I., {et~al.} 2011, The Astrophysical
  Journal, 735, 68

\bibitem[{Kim {et~al.}(2012)Kim, Protopapas, Trichas, Rowan-Robinson, Khardon,
  Alcock, \& Byun}]{Kim2012}
Kim, D.-W., Protopapas, P., Trichas, M., {et~al.} 2012, The Astrophysical
  Journal, 747, 107

\bibitem[{{Knigge}(2011)}]{Knigge2011}
{Knigge}, C. 2011, in Astronomical Society of the Pacific Conference Series,
  Vol. 447, Evolution of Compact Binaries, ed. L.~{Schmidtobreick}, M.~R.
  {Schreiber}, \& C.~{Tappert}, 3

\bibitem[{Knorr \& Ng(1998)}]{Knorr1998}
Knorr, E., \& Ng, R. 1998, in {Proceedings of the VLDB Conference, New York,
  USA}, 392--403

\bibitem[{Koller \& Friedman(2009)}]{Koller2009}
Koller, D., \& Friedman, N. I.~R. 2009, {Probabilistic graphical models}

\bibitem[{Kou {et~al.}(2004)Kou, Lu, Sirwongwattana, \& Huang}]{survey2}
Kou, Y., Lu, C., Sirwongwattana, S., \& Huang, Y. 2004, in Proceedings of the
  {IEEE} International Conference on Networking, Sensing and Control, 749--754

\bibitem[{{Liu} {et~al.}(2005){Liu}, {van Paradijs}, \& {van den
  Heuvel}}]{Liu2005AA...442.1135L}
{Liu}, Q.~Z., {van Paradijs}, J., \& {van den Heuvel}, E.~P.~J. 2005, Astronomy
  and Astrophysics, 442, 1135

\bibitem[{{Liu} {et~al.}(2007){Liu}, {van Paradijs}, \& {van den
  Heuvel}}]{Liu2007AA...469..807L}
---. 2007, Astronomy and Astrophysics, 469, 807

\bibitem[{{Marconi} {et~al.}(2013){Marconi}, {Molinaro}, {Bono},
  {Pietrzy{\'n}ski}, {Gieren}, {Pilecki}, {Stellingwerf}, {Graczyk}, {Smolec},
  {Konorski}, {Suchomska}, {G{\'o}rski}, \& {Karczmarek}}]{Marconi2013}
{Marconi}, M., {Molinaro}, R., {Bono}, G., {et~al.} 2013, The Astrophysical
  Journal Letters, 768, L6

\bibitem[{Monti \& Cooper(1998)}]{Montit}
Monti, S., \& Cooper, G.~F. 1998, in {Proceedings of the Fourteenth conference
  on Uncertainty in artificial intelligence}, Morgan Kaufmann Publishers Inc.,
  404--413

\bibitem[{Nairac {et~al.}(1999)Nairac, Towsend, Carr, King, Cowley, \&
  Tarassenko}]{Nairac1999}
Nairac, A., Towsend, N., Carr, R., {et~al.} 1999, Integrated ComputerAided
  Engineering, 6, 53

\bibitem[{{Ness} {et~al.}(2002){Ness}, {Schmitt}, {Burwitz}, {Mewe}, \&
  {Predehl}}]{Ness2002AA...387.1032N}
{Ness}, J.-U., {Schmitt}, J.~H.~M.~M., {Burwitz}, V., {Mewe}, R., \& {Predehl},
  P. 2002, Astronomy and Astrophysics, 387, 1032

\bibitem[{Papadimitriou {et~al.}(2003)Papadimitriou, Kitagawa, Gibbons, \&
  Faloutsos}]{Papadimitriou2002}
Papadimitriou, S., Kitagawa, H., Gibbons, P.~B., \& Faloutsos, C. 2003, in Data
  Engineering, 2003. Proceedings. 19th International Conference on, {IEEE},
  315--326

\bibitem[{Penzias \& Wilson(1965)}]{Penzias1965}
Penzias, A., \& Wilson, R. 1965, The Astrophysical Journal, 419

\bibitem[{Pichara \& Protopapas(2013)}]{Pichara2013}
Pichara, K., \& Protopapas, P. 2013, The Astrophysical Journal, 777, 83

\bibitem[{Pichara {et~al.}(2012)Pichara, Protopapas, Kim, Marquette, \&
  Tisserand}]{Pichara2012}
Pichara, K., Protopapas, P., Kim, D.-W., Marquette, J.-B., \& Tisserand, P.
  2012, Monthly Notices of the Royal Astronomical Society, 427, 1284

\bibitem[{Pichara \& Soto(2011)}]{Pichara2011}
Pichara, K., \& Soto, A. 2011, Intelligent Data Analysis, 15, 151

\bibitem[{Pichara {et~al.}(2008)Pichara, Soto, \& Araneda}]{Pichara}
Pichara, K., Soto, A., \& Araneda, A. 2008, in Advances in Artificial
  Intelligence-IBERAMIA 2008 (Springer), 163--172

\bibitem[{{Pietrzy{\'n}ski} {et~al.}(2010){Pietrzy{\'n}ski}, {Thompson},
  {Gieren}, {Graczyk}, {Bono}, {Udalski}, {Soszy{\'n}ski}, {Minniti}, \&
  {Pilecki}}]{Pietrzynski2010}
{Pietrzy{\'n}ski}, G., {Thompson}, I.~B., {Gieren}, W., {et~al.} 2010, Nature,
  468, 542

\bibitem[{Poleski(2008)}]{Poleski2009}
Poleski, R. 2008, Acta Astronomica, 58, 313

\bibitem[{Protopapas {et~al.}(2006)Protopapas, Giammarco, Faccioli, Struble,
  Dave, \& Alcock}]{Protopapas2008}
Protopapas, P., Giammarco, J., Faccioli, L., {et~al.} 2006, Monthly Notices of
  the Royal Astronomical Society, 369, 677

\bibitem[{{Quimby} {et~al.}(2011){Quimby}, {Kulkarni}, {Kasliwal}, {Gal-Yam},
  {Arcavi}, {Sullivan}, {Nugent}, {Thomas}, {Howell}, {Nakar}, {Bildsten},
  {Theissen}, {Law}, {Dekany}, {Rahmer}, {Hale}, {Smith}, {Ofek}, {Zolkower},
  {Velur}, {Walters}, {Henning}, {Bui}, {McKenna}, {Poznanski}, {Cenko}, \&
  {Levitan}}]{Quimby2011}
{Quimby}, R.~M., {Kulkarni}, S.~R., {Kasliwal}, M.~M., {et~al.} 2011, Nature,
  474, 487

\bibitem[{Ramaswamy {et~al.}(2000)Ramaswamy, Rastogi, \& Shim}]{Ramaswamy2000}
Ramaswamy, S., Rastogi, R., \& Shim, K. 2000, in {Proceedings of the ACM SIGMOD
  Conference on Management of Data, Dallas, TX}, 427--438

\bibitem[{Rebbapragada {et~al.}(2008)Rebbapragada, Protopapas, Brodley, \&
  Alcock}]{Rebbapragada2008}
Rebbapragada, U., Protopapas, P., Brodley, C.~E., \& Alcock, C. 2008, Machine
  Learning, 74, 281

\bibitem[{Richards {et~al.}(2012)Richards, Starr, Miller, Bloom, Butler, Brink,
  \& Crellin-Quick}]{Richards2012}
Richards, J.~W., Starr, D.~L., Miller, A.~A., {et~al.} 2012, The Astrophysical
  Journal Supplement Series, 203, 32

\bibitem[{{Ridley} {et~al.}(2013){Ridley}, {Crawford}, {Lorimer}, {Bailey},
  {Madden}, {Anella}, \& {Chennamangalam}}]{Ridley2013MNRAS.433..138R}
{Ridley}, J.~P., {Crawford}, F., {Lorimer}, D.~R., {et~al.} 2013, Monthly
  Notices of the Royal Astronomical Society, 433, 138

\bibitem[{{Schaefer}(2010)}]{Schaefer2010}
{Schaefer}, B.~E. 2010, The Astrophysical Journal Supplement Series, 187, 275

\bibitem[{{Schmidtke} {et~al.}(1999){Schmidtke}, {Cowley}, {Crane}, {Taylor},
  {McGrath}, {Hutchings}, \& {Crampton}}]{Schmidtke1999AJ....117..927S}
{Schmidtke}, P.~C., {Cowley}, A.~P., {Crane}, J.~D., {et~al.} 1999, The
  Astronomical Journal, 117, 927

\bibitem[{Seidl {et~al.}(2009)Seidl, M{\"u}ller, Assent, \&
  Steinhausen}]{Seidl2009}
Seidl, T., M{\"u}ller, E., Assent, I., \& Steinhausen, U. 2009, in Uncertainty
  Management in Information Systems

\bibitem[{Serio {et~al.}(2002)Serio, Manara, Sicoli, \& Bottke}]{Serio2001}
Serio, G.~F., Manara, A., Sicoli, P., \& Bottke, W.~F. 2002, Giuseppe Piazzi
  and the discovery of Ceres (University of Arizona Press)

\bibitem[{{Shafter}(2013)}]{Shafter2013}
{Shafter}, A.~W. 2013, The Astronomical Journal, 145, 117

\bibitem[{Son {et~al.}(2009)Son, Cho, \& Yoo}]{SonChoYoo2009}
Son, C., Cho, S., \& Yoo, J. 2009, in Management Enabling the Future Internet
  for Changing Business and New Computing Services, Vol. 5787 (Springer Berlin,
  Heidelberg), 291--300

\bibitem[{Soszynski {et~al.}(2003)Soszynski, Udalski, Kubiak, Zebrun, \&
  Szewczyk}]{Alamos2003}
Soszynski, I., Udalski, A., Kubiak, M., Zebrun, K., \& Szewczyk, O. 2003, Acta
  Astronomica, 53, 93

\bibitem[{Soszynski {et~al.}(2008)Soszynski, Udalski, Szymanski, Kubiak,
  Pietrzynski, Wyrzykowski, Ulaczyk, \& Poleski}]{Soszynski2008}
Soszynski, I., Udalski, A., Szymanski, M., {et~al.} 2008, Acta Astronomica, 58,
  293

\bibitem[{{Thomas} {et~al.}(2005){Thomas}, {Griest}, {Popowski}, {Cook},
  {Drake}, {Minniti}, {Myer}, {Alcock}, {Allsman}, {Alves}, {Axelrod},
  {Becker}, {Bennett}, {Freeman}, {Geha}, {Lehner}, {Marshall}, {Nelson},
  {Peterson}, {Quinn}, {Stubbs}, {Sutherland}, {Vandehei}, {Welch}, \& {MACHO
  Collaboration}}]{Thomas2005ApJ...631..906T}
{Thomas}, C.~L., {Griest}, K., {Popowski}, P., {et~al.} 2005, The Astrophysical
  Journal, 631, 906

\bibitem[{Tyson {et~al.}(2002)Tyson, Collaboration, Labs, Technologies, \&
  Hill}]{Tyson}
Tyson, J.~A., Collaboration, L., Labs, B., Technologies, L., \& Hill, M. 2002,
  Astronomical Telescopes and Instrumentation, 10

\bibitem[{{van den Heuvel} {et~al.}(1992){van den Heuvel}, {Bhattacharya},
  {Nomoto}, \& {Rappaport}}]{Heuvel1992AA...262...97V}
{van den Heuvel}, E.~P.~J., {Bhattacharya}, D., {Nomoto}, K., \& {Rappaport},
  S.~A. 1992, Astronomy and Astrophysics, 262, 97

\bibitem[{{van der Klis}(2000)}]{Klis2000ARAA..38..717V}
{van der Klis}, M. 2000, Astronomy and Astrophysics, 38, 717

\bibitem[{Voges {et~al.}(1999{\natexlab{a}})Voges, Aschenbach, Boller,
  Br{\"a}uninger, Briel, Burkert, Dennerl, Englhauser, Gruber, Haberl,
  {et~al.}}]{Voges1999}
Voges, W., Aschenbach, B., Boller, T., {et~al.} 1999{\natexlab{a}}, Astronomy
  and Astrophysics, 349, 389

\bibitem[{Voges {et~al.}(1999{\natexlab{b}})Voges, Aschenbach, Boller,
  Br{\"a}uninger, Briel, Burkert, Dennerl, Englhauser, Gruber, Haberl,
  {et~al.}}]{Voges1999AA...349..389V}
---. 1999{\natexlab{b}}, Astronomy and Astrophysics, 349, 389

\bibitem[{Watson {et~al.}(2009)Watson, Schr{\"o}der, Fyfe, Page, Mateos, Pye,
  Sakano, Rosen, Denby, Denkinson, {et~al.}}]{Watson2009}
Watson, M., Schr{\"o}der, A., Fyfe, D., {et~al.} 2009, Astronomy and
  Astrophysics, 493, 339

\bibitem[{{Watson} {et~al.}(2009){Watson}, {Schr{\"o}der}, {Fyfe}, {Page},
  {Lamer}, {Mateos}, {Pye}, {Sakano}, {Rosen}, {Ballet}, {Barcons}, {Barret},
  {Boller}, {Brunner}, {Brusa}, {Caccianiga}, {Carrera}, {Ceballos}, {Della
  Ceca}, {Denby}, {Denkinson}, {Dupuy}, {Farrell}, {Fraschetti}, {Freyberg},
  {Guillout}, {Hambaryan}, {Maccacaro}, {Mathiesen}, {McMahon}, {Michel},
  {Motch}, {Osborne}, {Page}, {Pakull}, {Pietsch}, {Saxton}, {Schwope},
  {Severgnini}, {Simpson}, {Sironi}, {Stewart}, {Stewart}, {Stobbart}, {Tedds},
  {Warwick}, {Webb}, {West}, {Worrall}, \& {Yuan}}]{Watson2009AA...493..339W}
{Watson}, M.~G., {Schr{\"o}der}, A.~C., {Fyfe}, D., {et~al.} 2009, Astronomy
  and Astrophysics, 493, 339

\bibitem[{{Wood}(2000)}]{Wood2000PASA...17...18W}
{Wood}, P.~R. 2000, PASP, 17, 18

\bibitem[{Wyrzykowski {et~al.}(2003)Wyrzykowski, Udalski, Kubiak, Szymanski,
  Zebrun, Soszynski, Wozniak, Pietrzynski, \& Szewczyk}]{Observatory2003}
Wyrzykowski, L., Udalski, A., Kubiak, M., {et~al.} 2003, Acta Astronomica, 53,
  1

\bibitem[{Xiong {et~al.}(2010)Xiong, Poczos, Connolly, \&
  Schneider}]{Connolly2010}
Xiong, L., Poczos, B., Connolly, A., \& Schneider, J. 2010, Anomaly detection
  for astronomical data

\bibitem[{Yang {et~al.}(2006)Yang, Xie, \& Lu}]{YangXieLu2006}
Yang, H., Xie, F., \& Lu, Y. 2006, in Fuzzy Systems and Knowledge Discovery,
  Vol. 4223, 1082--1091

\bibitem[{Zhang {et~al.}(1996)Zhang, Ramakrishnan, \& Livny}]{Zhang1996}
Zhang, T., Ramakrishnan, R., \& Livny, M. 1996, in {Proceedings of the ACM
  SIGMOD International Conference on Management of Data}, 103--114

\end{thebibliography}

\end{document}